\documentclass[universe,article,accept,pdftex,oneauthor]{Definitions/mdpi}   

\usepackage[normalem]{ulem} 

\firstpage{1} 
\pubvolume{1}
\issuenum{1}
\articlenumber{0}
\pubyear{2022} 
\copyrightyear{2022}
\externaleditor{Academic Editors: Robert H. Bernstein, Bertrand Echenard} 
\datereceived{15 December 2021} 
\dateaccepted{26 February 2022} 
\datepublished{} 
\hreflink{https://doi.org/} 

\Title{Charged Lepton Flavor Violation at the High-Energy Colliders: Neutrino Mass Relevant Particles
} 

\TitleCitation{Charged Lepton Flavor Violation at the High-Energy Colliders: Neutrino Mass Relevant Particles}

\Author{Yongchao Zhang}

\AuthorNames{Yongchao Zhang}

\AuthorCitation{Zhang, Y.} 

\address[1]{%
School of Physics, Southeast University, Nanjing 211189, China; zhangyongchao@seu.edu.cn}

\abstract{ 
We summarize the potential charged lepton flavor violation (LFV) from neutrino mass relevant models, for instance the seesaw mechanisms. In particular, we study, in a model-dependent way, the LFV signals at the high-energy hadron and lepton colliders originating from the beyond standard model (BSM) neutral scalar $H$, doubly charged scalar $H^{\pm\pm}$, heavy neutrino $N$, heavy $W_R$ boson, and the $Z'$ boson. For the neutral scalar, doubly charged scalar and $Z'$ boson, the LFV signals originate from the (effective) LFV couplings of these particles to the charged leptons, while for the heavy neutrino $N$ and $W_R$ boson, the LFV effects are from flavor mixing in the neutrino sector. We consider current limits on these BSM particles and estimate their prospects at future high-energy hadron and lepton colliders.} 

\keyword{lepton flavor violation; BSM neutral scalar; doubly charged scalar; heavy neutrino; heavy $W_R$ boson; $Z'$ boson }  
\RequirePackage[normalem]{ulem} 
\RequirePackage{color}\definecolor{RED}{rgb}{1,0,0}\definecolor{BLUE}{rgb}{0,0,1} 
\providecommand{\DIFaddbegin}{} 
\providecommand{\DIFaddend}{} 
\providecommand{\DIFdelbegin}{} 
\providecommand{\DIFdelend}{} 

\begin{document}

\section{Introduction}

In the standard model (SM) of particle physics, charged lepton flavor violation (LFV) processes are vanishing, such as the decays $\ell_\alpha \to \ell_\beta \gamma$, $\ell_\alpha \to \ell_\beta \ell_\gamma \ell_\delta$ (with $\alpha,\,\beta,\, \gamma,\, \delta = e,\, \mu,\, \tau$ the lepton flavor indices) and muonium-antimuonium oscillation~\cite{ParticleDataGroup:2020ssz}. Other observables, for instance the anomalous magnetic dipole moments of electron and muon, might also receive sizable contributions from the beyond SM (BSM) particles and their interactions~\cite{Lindner:2016bgg}. Therefore, these precision measurements at the high-intensity frontier are one of the primary probes of the BSM physics~\mbox{
\cite{Jaeckel:2010ni, Proceedings:2012ulb, Essig:2013lka, Beacham:2019nyx}}. LFV in the charged lepton sector can be, to some extent, relevant to BSM in the neutrino sector, for instance the various seesaw mechanisms to generate the tiny neutrino masses. Depending on model details and specific UV completions, the BSM particles from the seesaw frameworks can be of spin zero, half, or one, i.e., the  CP-even scalars or pseudoscalars, fermions, and vector bosons. In principle, these particles can be either heavy at the TeV scale or relatively light at the GeV scale or even lighter, if allowed by experimental data.  Furthermore, some of the LFV signals induced by these particles can also be lepton number violating (LNV)~\mbox{\cite{FileviezPerez:2015mlm, Cai:2017mow}} and/or lepton flavor universaility violating (LFUV)~\mbox{\cite{Alguero:2018nvb, Datta:2019zca, Kumar:2019qbv, Hurth:2021nsi, London:2021lfn}}, which are closely related to neutrino mass generation and the $B$-anomalies~\cite{LHCb:2021trn}, respectively. For simplicity, we will not consider LNV or LFUV, however focus here only on LFV~\mbox{\cite{Marciano:2008zz, Bernstein:2013hba, Lindner:2016bgg, Calibbi:2017uvl, Heeck:2016xwg}}. 

In this paper, we summarize the LFV effects from a couple of representative particles in the seesaw mechanisms, i.e., the BSM neutral scalar $H$, doubly charged scalar $H^{\pm\pm}$, heavy neutrino $N$, heavy $W_R$ boson, and the $Z'$ boson. It is worked as much in a model independent way as possible, and we do not consider too much dependence on the model details. With given LFV couplings, we will check the relevant LFV limits from current data, and  estimate the prospects of these particles at future high-energy lepton and hadron colliders. More details can be found in the original papers.


\section{BSM Neutral Scalar \boldmath{$H$}}
\label{sec:scalar}

If a BSM neutral scalar couples directly to the SM quarks or mixes with the SM Higgs, it may induce flavor-changing neutral currents (FCNCs) in the quark sector, which are stringently constrained by the $K$ and $B$ meson data~\mbox{\cite{ParticleDataGroup:2020ssz}}. Therefore, for simplicity, we assume the neutral scalar $H$ does not couple directly to the SM quarks and its mixing with the SM Higgs $h$ is sufficiently small. In a large variety of BSM scenarios, the scalar $H$ can couple to the SM-charged leptons in a flavor-violating way~\mbox{\cite{Hou:1995dg, Dev:2017ftk, Li:2018cod, Arganda:2019gnv}},\footnote{See e.g., Refs.~\mbox{\cite{Brignole:2004ah,Harnik:2012pb, Blankenburg:2012ex, Banerjee:2016foh, Herrero-Garcia:2016uab, Chakraborty:2016gff, Chakraborty:2017tyb, Qin:2017aju, Lu:2020dkx, Arganda:2014dta, Arganda:2015naa, Arganda:2016zvc,Marcano:2019rmk}} for the rich phenomenologies of LFV couplings of the SM Higgs.} e.g., in the left-right symmetric model (LRSM)~\cite{Dev:2016dja,Maiezza:2016ybz,BhupalDev:2016nfr,Dev:2017dui}, two Higgs doublet model~\mbox{\cite{Branco:2011iw, Crivellin:2013wna, Crivellin:2015hha}}
, $R$-parity violating supersymmetric theories~\cite{Aulakh:1982yn,Hall:1983id,Ross:1984yg,Barbier:2004ez}, and mirror models~\cite{Hung:2006ap,Bu:2008fx,Chang:2016ave,Hung:2017voe}. The LFV couplings of $H$ can arise at the tree-level or 1-loop level, depending on model details. The scalar $H$ can be from a BSM singlet, doublet, triplet, or other multiplets, and thus has different quantum numbers (such as the isospin), gauge couplings, which will provide different production channels of $H$ at the high-energy colliders. In this paper, we focus only on the production of $H$ at future lepton colliders via the LFV couplings in a model independent way. In the perspective of effective theory, the couplings of $H$ to the SM-charged leptons can be written in the following way.\footnote{Note that without any details on the couplings of $H$ to the active neutrinos the couplings in Equation~(\ref{eqn:Lagrangian:H}) are not invariant with respect to the SM $SU(2)_L$ group.}

\begin{equation}
\DIFaddbegin \label{eqn:Lagrangian:H}
    \DIFaddend {\cal L}_Y = h_{\alpha\beta} \overline{\ell}_\alpha H \ell_\beta ~+~ {\rm H.c.} \,,
\end{equation}
where $\ell$ is the charged lepton, 
and $h_{\alpha\beta}$ is the Yukawa coupling. For simplicity, we have assumed $H$ to be a real field and CP to be even. As a result, the matrix $h_{\alpha\beta}$ is symmetric.

The couplings $h_{\alpha\beta}$ of the scalar $H$ can induce very rich LFV signals at the high-energy colliders. In light of the clean backgrounds, the future $e^+ e^-$ colliders are the primary facilities to search for such smoking-gun signals, such as at the International Linear Collider (ILC)~\cite{Baer:2013cma}, Circular Electron-Positron Collider (CEPC)~\cite{CEPC-SPPCStudyGroup:2015csa}, Future Circular Collider (FCC-ee)~\cite{TLEPDesignStudyWorkingGroup:2013myl},  and Compact Linear Collider
(CLIC)~\cite{CLICPhysicsWorkingGroup:2004qvu}. 
Given a single LFV coupling $h_{\alpha\beta}$ with $\alpha\neq \beta$, if kinematically allowed, the scalar $H$ can be on-shell produced at high-energy $e^+ e^-$ colliders via the process:
\begin{eqnarray}
\label{eqn:H}
e^+ e^- \to \ell_\alpha^\pm \ell_\beta^\mp H \,.
\end{eqnarray}

\begin{figure}[H] \includegraphics[width=0.25\textwidth]{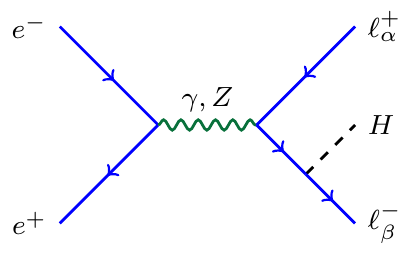}
  \includegraphics[width=0.25\textwidth]{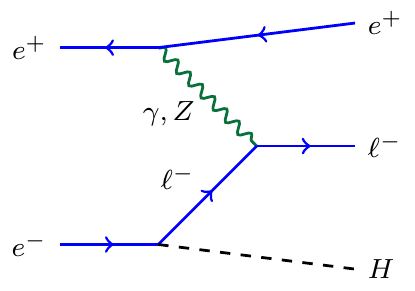}
  \caption{Representative Feynman diagrams for on-shell production of $H$ at future lepton colliders via the process in Equation~(\ref{eqn:H}). Figure from Ref.~\mbox{\cite{Dev:2017ftk}}.}
  \label{fig:diagram1}
\end{figure}

The Feynman diagrams can be found in Figure~\ref{fig:diagram1}. For simplicity, it is assumed that all other Yukawa couplings in the matrix $h_{\alpha\beta}$ are vanishing. As a result, a single coupling $h_{\alpha\beta}$ can not induce the LFV decays, such as $\ell_\beta \to \ell_\alpha\gamma$ and $\ell_\beta \to 3\ell_\alpha$, which depend on the combinations of $h_{\alpha\beta}$ with other Yukawa couplings, e.g., $h_{\alpha\alpha}h_{\alpha\beta}$. It turns out that only a few precise LFV measurements can be used to set limits on the single coupling $h_{\alpha\beta}$.

\begin{figure}[H]
  \includegraphics[width=0.25\textwidth]{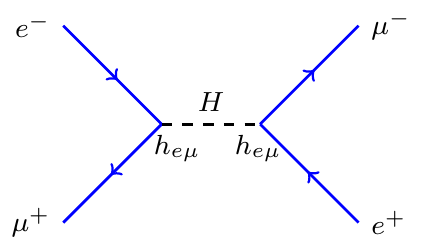}
  \includegraphics[width=0.25\textwidth]{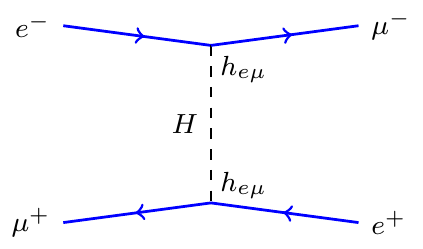} \\
  \includegraphics[width=0.25\textwidth]{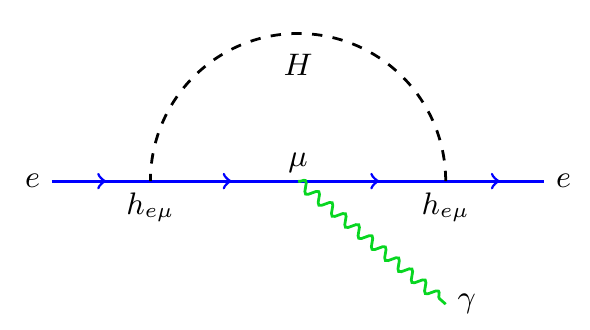}
  \includegraphics[width=0.25\textwidth]{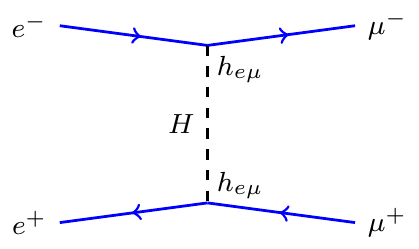}
  \caption{Feynman diagrams for the $H$ contribution to muonium-antimuonium oscillation (\textbf{upper left and right}), electron $g-2$ (\textbf{lower left}), and the LEP 
   $e^+ e^- \to \mu^+ \mu^-$ process (\textbf{lower right}) with the LFV coupling $h_{e\mu}$. Figure from Ref.~\mbox{\cite{Dev:2017ftk}}.}
  \label{fig:diagram:limit}
\end{figure}

\begin{figure}[H] \includegraphics[width=0.31\textwidth]{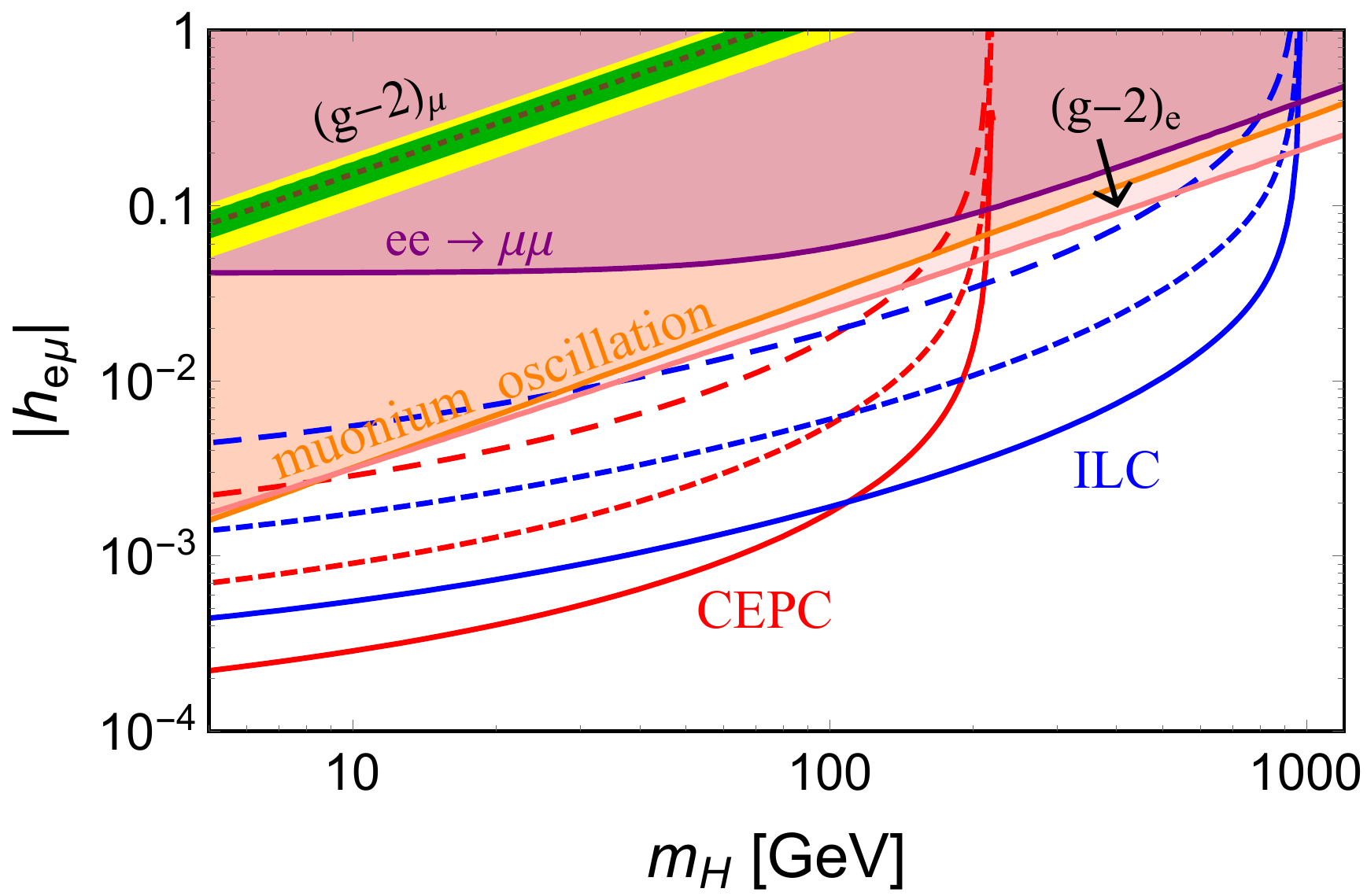}
  \includegraphics[width=0.31\textwidth]{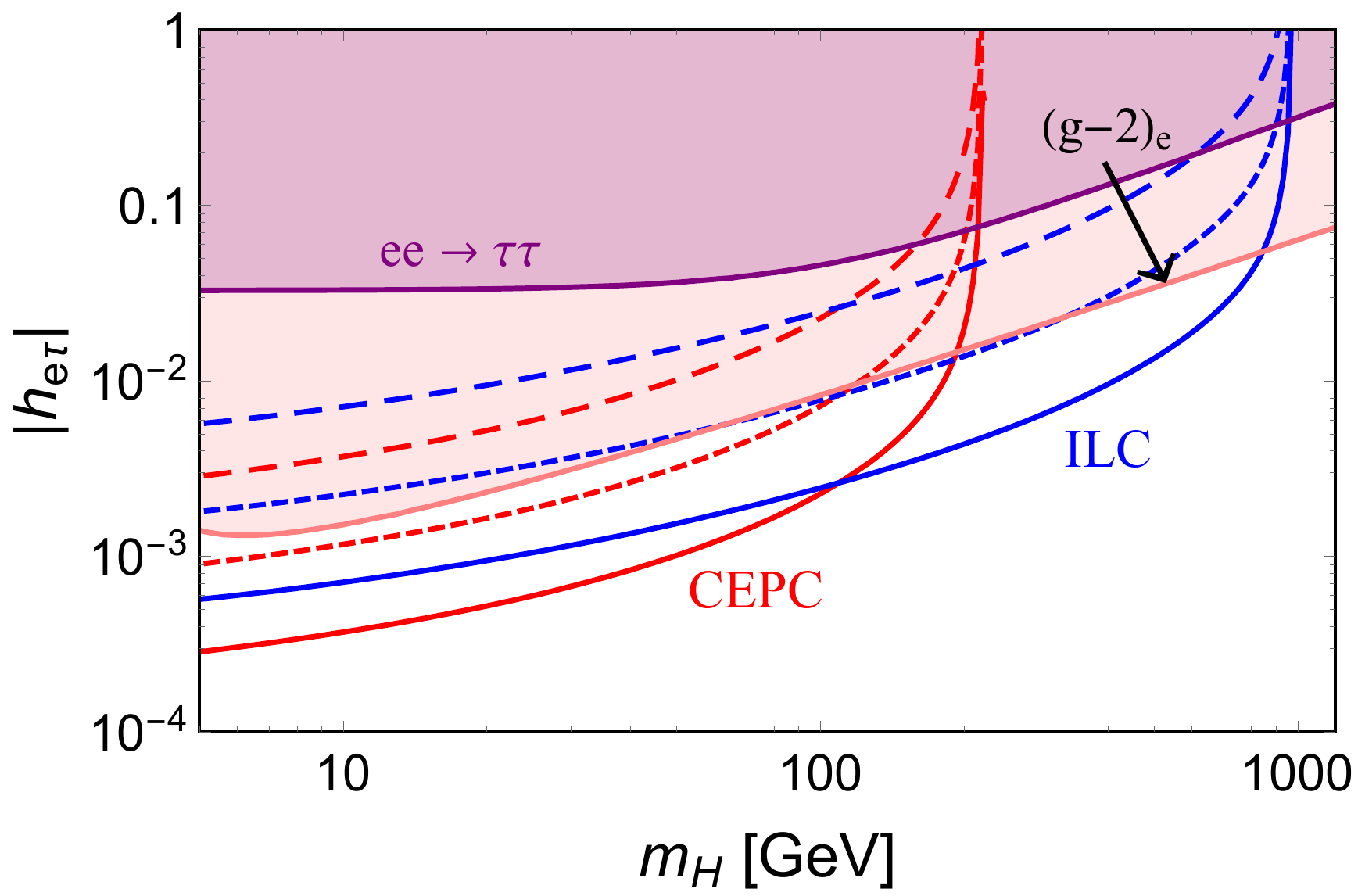}
  \includegraphics[width=0.31\textwidth]{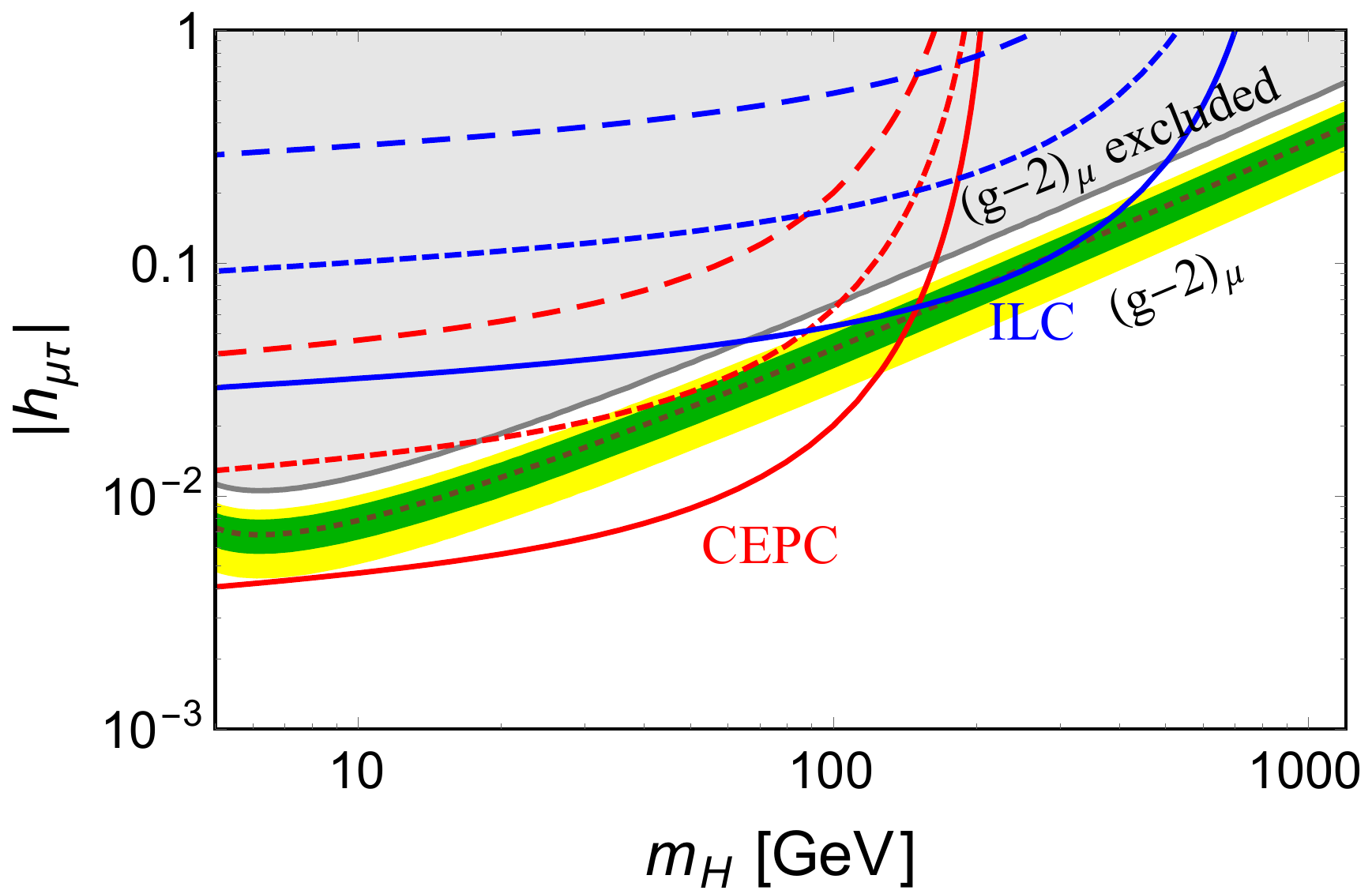}
  \caption{Prospects of LFV couplings $|h_{e\mu}|$ (\textbf{left}), $|h_{e\tau}|$ (\textbf{middle}), and $|h_{\mu\tau}|$ (\textbf{right}) from the process (\ref{eqn:H}) at CEPC 240 GeV with 5 ab$^{-1}$ (red) and ILC 1 TeV with 1 ab$^{-1}$ (blue). A branching ratio of 1\% (long-dashed), 10\% (short-dashed), or 100\% (solid) from $H$ decay is assumed to be visible. The shaded regions are excluded by the corresponding limits. In the left and right panels, the brown line could fit the central value of $\Delta a_\mu$, and the green and yellow bands cover the $1\sigma$ and $2\sigma$ ranges of $\Delta a_\mu$. See text for more details. Figure from Ref.~\cite{Dev:2017ftk}.}
  \label{fig:H:onshell}
\end{figure}

The coupling $h_{e\mu}$ can induce muonium-antimuonium oscillation at the tree-level, 
contributes to the anomalous magnetic moment $a_e$ of electron via the $H - \mu$ loop, 
and affects also the measurements of $e^+ e^- \to \mu^+ \mu^-$ at the LEP. The corresponding Feynman diagrams are shown in Figure~\ref{fig:diagram:limit}. The precise measurements of muonium oscillation by MACS~\cite{Willmann:1998gd}, the current value of $a_e$~\cite{Mohr:2015ccw} and the LEP data~\cite{DELPHI:2005wxt} exclude a large region in the parameter space of the scalar mass $m_H$ and the coupling $h_{e\mu}$, as depicted by the shaded regions in Figure~\ref{fig:H:onshell}. More calculation details can be found in Refs.~\cite{Dev:2017ftk,BhupalDev:2018vpr}. The coupling $h_{e\mu}$ contributes also to the magnetic moment $a_\mu$ of muon via the $H - e$ loop. However, this contribution is highly suppressed by the electron mass, and can not explain the current discrepancy of the muon $g-2$ anomaly $\Delta a_\mu$~\cite{Muong-2:2006rrc, Muong-2:2021ojo} (cf. the left panel of Figure~\ref{fig:H:onshell}).

\vspace{-10pt}

The coupling $h_{e\mu}$ can induce the signal $e^+ e^- \to e^\pm \mu^\mp H$ at future high-energy $e^+ e^-$ colliders. After being produced, the scalar $H$ can decay back into a pair of opposite-sign different flavor-charged leptons, i.e., $H \to e^\pm \mu^\mp$. In the SM, such processes are absent, and the most important backgrounds at $e^+ e^-$ colliders are mainly from electrons mis-identified as muons or vice versa. Therefore, the LFV signal of $e\mu$ resonance can be easily distinguished from the SM backgrounds. Taking CEPC 240 GeV with an integrated luminosity of 5 ab$^{-1}$ and ILC 1 TeV with 1 ab$^{-1}$ as benchmark facilities, 
the resultant prospects of $m_H$ and $h_{e\mu}$ 
are shown as the red and blue lines in the left panel of Figure~\ref{fig:H:onshell}, assuming at least \mbox{10 signal events}. The long-dashed, short-dashed, and solid lines correspond respectively to the branching ratio (BR) of $H \to e^\pm \mu^\mp$ to be 1\%, 10\%, and 100\%. 
Benefiting from the larger luminosity, the CEPC can probe a smaller $h_{e\mu}$ than ILC, down to ${\cal O} (10^{-4})$ in the limit of massless $H$. On the other end, when $H$ is heavy, ILC can probe a heavier $H$ than CEPC due to the kinematics reason.

If the LFV coupling, $h_{e\tau}$ is present, it will contribute to the electron $g-2$ via the $H - \tau$ loop and the $e^+ e^- \to \tau^+ \tau^-$ process at LEP. The corresponding constraints are shown as shaded regions in the middle panel of Figure~\ref{fig:H:onshell}. With the coupling $h_{\mu\tau}$, the neutral scalar $H$ provides a simple explanation for the muon $g-2$ anomaly via the $H - \tau$ loop. The corresponding $1\sigma$ and $2\sigma$ ranges of $\Delta a_\mu$ in the parameter space of $m_H$ and $h_{\mu\tau}$ are presented in the right panel of Figure~\ref{fig:H:onshell}, respectively, as the green and yellow bands, while the gray shaded region is excluded by the muon $g-2$ discrepancy at the $5\sigma$ C.L. As for the case of $h_{e\mu}$, the prospects of $h_{e\tau}$ and $h_{\mu\tau}$ are shown respectively in the middle and right panels of Figure~\ref{fig:H:onshell}, with again at least 10 signal events. The reconstruction efficiency of $\tau$ lepton has taken conservatively to be 60\% at both CEPC and ILC~\cite{Baer:2013cma}. For all the three LFV couplings $h_{e\mu,\, e\tau,\, \mu\tau}$, even if all the current stringent lepton flavor constraints are taken into account, there are yet large regions of parameter space of $m_H$ and $h_{\alpha\beta}$ (with $\alpha\neq\beta$) that can be probed at future high-energy $e^+ e^-$ colliders, as seen in Figure~\ref{fig:H:onshell}.

\begin{figure}[H] \includegraphics[width=0.25\textwidth]{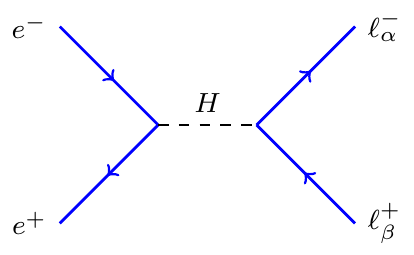}
  \includegraphics[width=0.25\textwidth]{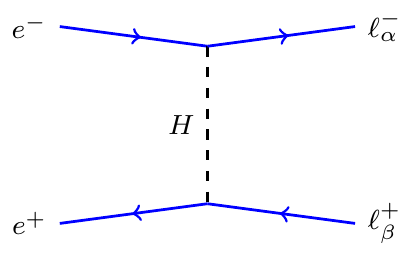}
  \caption{Feynman diagrams for off-shell production of $H$ at future lepton colliders (cf. the process in Equation~(\ref{eqn:H:2})). Figure from Ref.~\mbox{\cite{Dev:2017ftk}}.}
  \label{fig:diagram2}
\end{figure}

\begin{figure}[H] \includegraphics[width=0.31\textwidth]{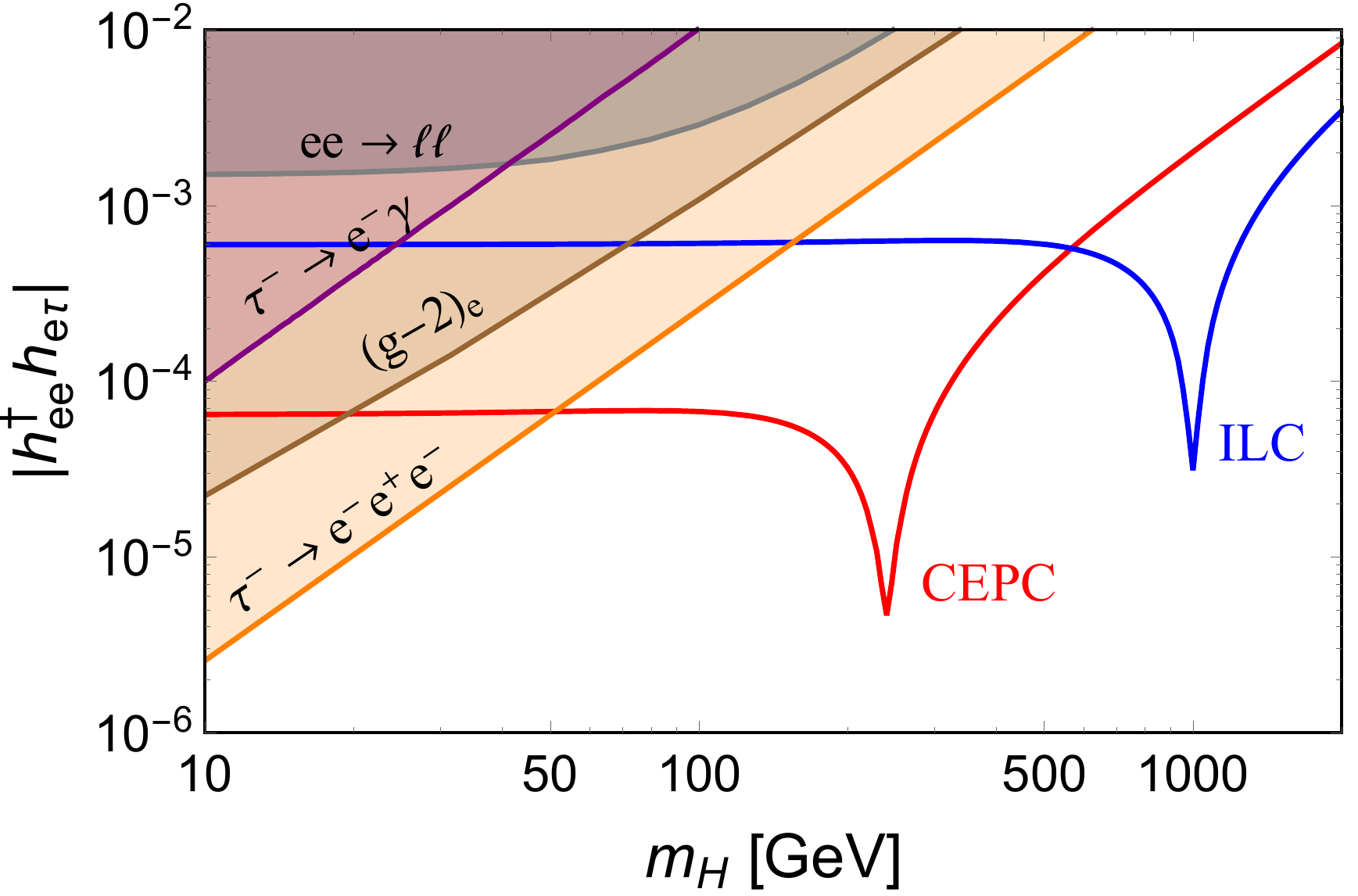}
  \includegraphics[width=0.31\textwidth]{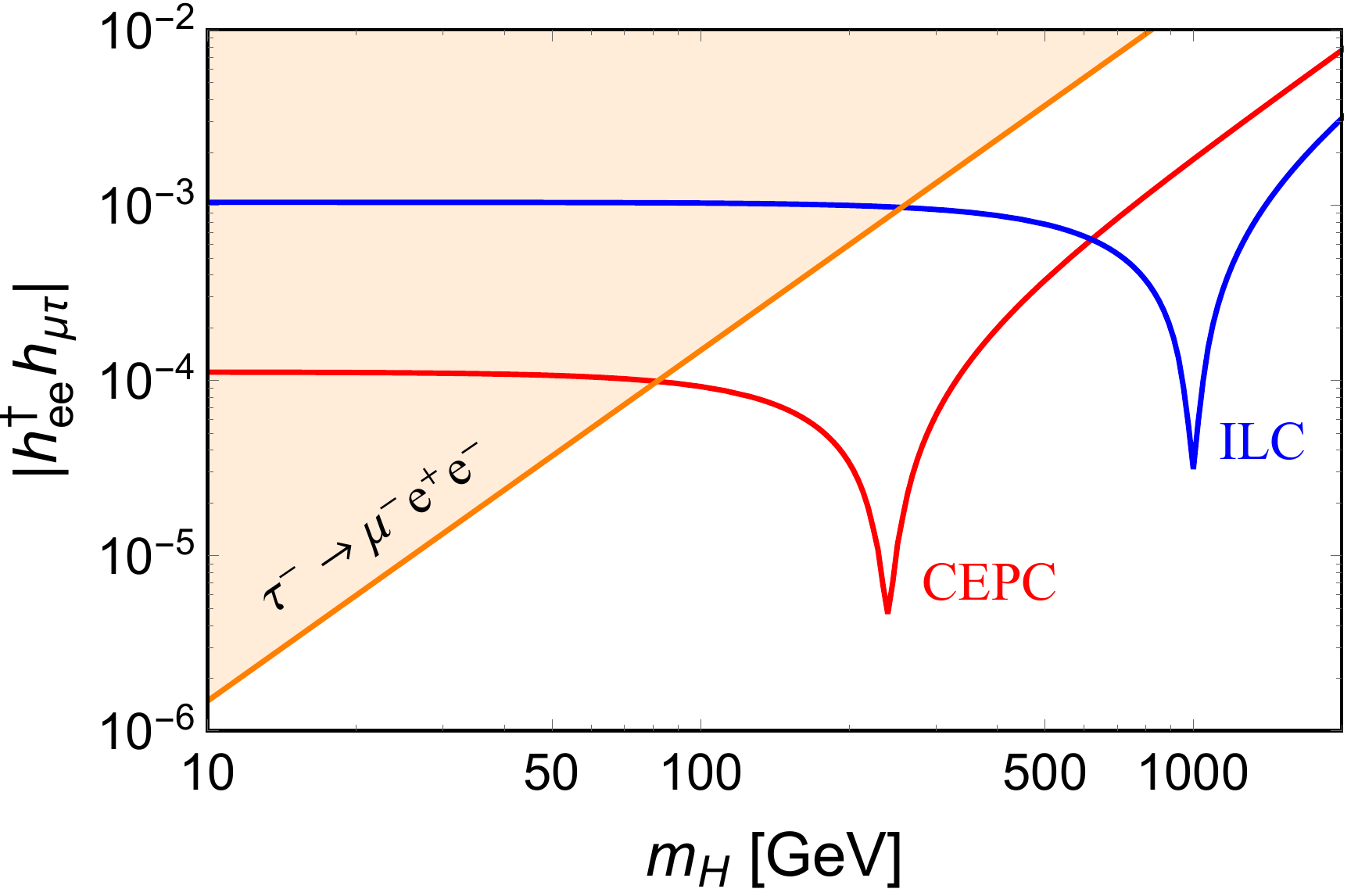}
  \includegraphics[width=0.31\textwidth]{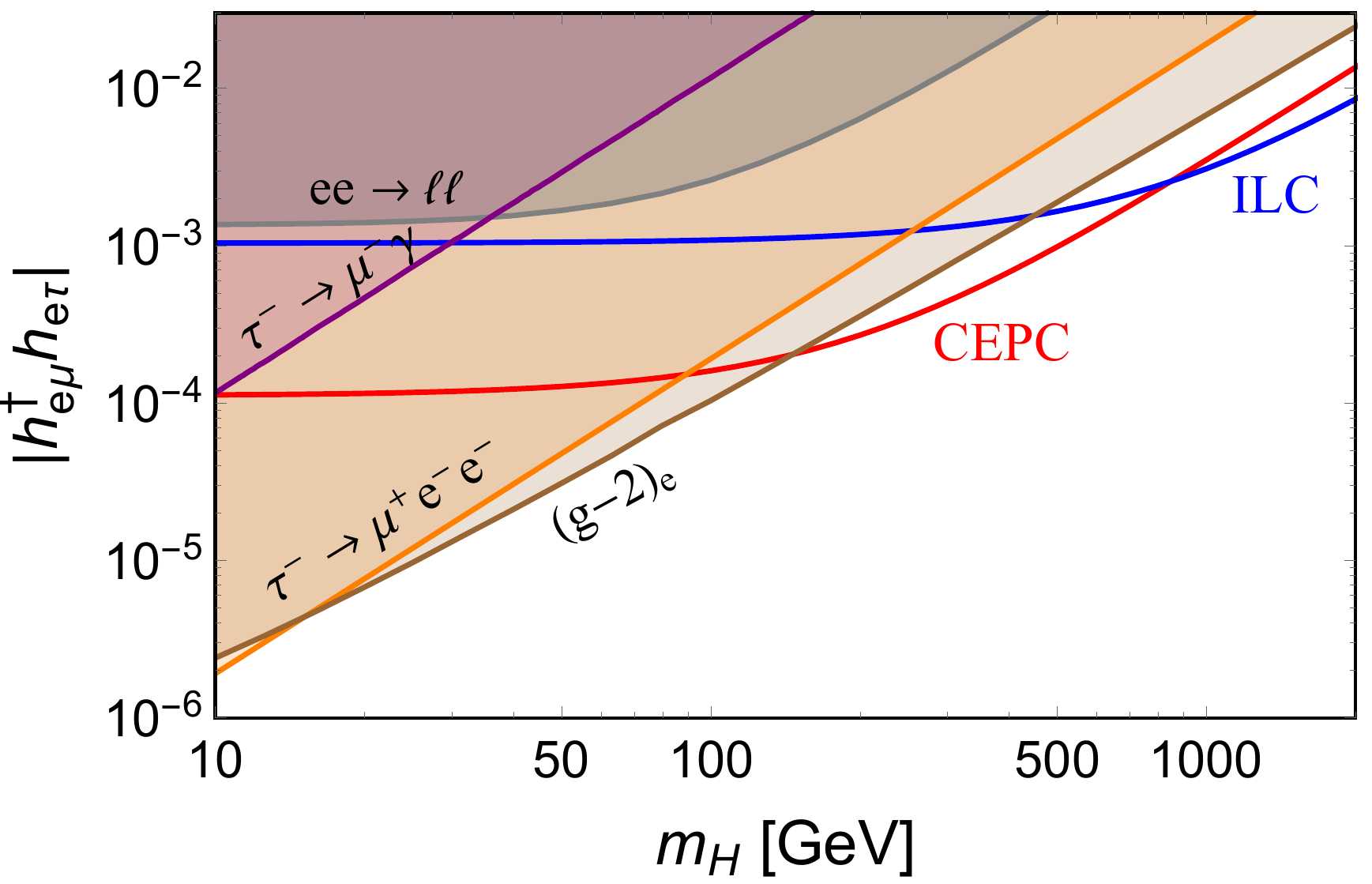}
  \caption{Prospects of $|h^\dagger_{ee} h_{e\tau}|$ (\textbf{left}), $|h^\dagger_{ee} h_{\mu\tau}|$ (\textbf{middle}), and $|h^\dagger_{e\mu} h_{e\tau}|$ (\textbf{right}) from searches of $e^+ e^- \to e^\pm \tau^\mp,\, \mu^\pm \tau^\mp$ at CEPC 240 GeV with 5 ab$^{-1}$ (red) and ILC 1 TeV with 1 ab$^{-1}$ (blue). The shaded regions are excluded by the corresponding limits. See text for more details. Figure from Ref.~\cite{Dev:2017ftk}.}
  \label{fig:H:offshell}
\end{figure}

If the couplings $h_{ee}$ and $h_{e\mu}$ are both nonzero, as an $s$-channel mediator the scalar $H$ will induce the process: 
\begin{eqnarray}
\DIFaddbegin \label{eqn:H:2}
\DIFaddend e^+ e^- \to e^\pm \mu^\mp
\end{eqnarray}
at high-energy colliders, which is also apparently LFV. The Feynman diagram is shown in the left panel of Figure~\ref{fig:diagram2} (with $\ell_\alpha \ell_\beta = e\mu$). However, the coupling $h_{ee}h_{e\mu}$ is severely constrained by the limit from $\mu \to eee$~\cite{SINDRUM:1987nra}, which precludes all the prospects of $e^\pm \mu^\mp$ at future high-energy lepton colliders. In the $\tau$ sector, the limits are much weaker. Let us consider the process $e^+ e^- \to e^+ \tau^-$, which can be induced by an $s$-channel $H$ mediator and the combination $h_{ee}^\dagger h_{e\tau}$ of Yukawa couplings (cf. the left panel of Figure~\ref{fig:diagram2} for the Feynman diagram). Such couplings are subject to the limits from the decays $\tau \to eee$, $\tau \to e \gamma$, electron $g-2$, and the LEP data $e^+ e^- \to e^+ e^- ,\, \tau^+ \tau^-$ (the corresponding Feynman diagrams are similar to those in Figure~\ref{fig:diagram:limit}), which are shown as the shaded regions in the left panel of Figure~\ref{fig:H:offshell}. 
Requiring at least 10 signal events, the prospects of $m_H$ and $|h_{ee}^\dagger h_{e\tau}|$ at CEPC 240 GeV and ILC 1 TeV are presented in the left panel of Figure~\ref{fig:H:offshell}, respectively, as the solid and blue lines. The dips are due to the resonant production of $H$ at CEPC and ILC with $m_{H} \simeq \sqrt{s}$. As $H$ is only in the mediator for the LFV processes, its mass can go well above the center-of-mass energy $\sqrt{s}$ at the lepton colliders, as seen in the three panels of Figure~\ref{fig:H:offshell}. In the limit of $m_H \gg \sqrt{s}$, the high-energy lepton colliders probe the effective four-fermion interaction $(\overline{e}e)(\overline{e}\tau)/\Lambda^2$ with the cut-off scale $\Lambda \simeq m_H/\sqrt{|h_{ee}^\dagger h_{e\tau}|}$~\mbox{\cite{Kabachenko:1997aw, Cho:2016zqo, Ferreira:2006dg, Aranda:2009kz, Murakami:2014tna, Calibbi:2021pyh}}.

\vspace{-10pt}

To obtain the signatures of $e^+ e^- \to \mu^+ \tau^-$, there are two different combinations of Yukawa couplings, i.e., $h_{ee}^\dagger h_{\mu\tau}$ and $h_{e\mu}^\dagger h_{e\tau}$, which are mediated respectively by an $s$- and $t$-channel $H$ (cf. the Feynman diagrams in the left and right panels of Figure~\ref{fig:diagram2}). For the case of $h_{ee}^\dagger h_{\mu\tau}$, there is only constraint from the decay $\tau^- \to \mu^- e^+ e^-$~\cite{Hayasaka:2010np}, while for $h_{e\mu}^\dagger h_{e\tau}$ the limits are mainly from the rare decays $\tau^- \to \mu^+ e^- e^-$, $\tau^- \to \mu^- \gamma$, electron $g-2$, and the LEP data $e^+ e^- \to \mu^+ \mu^-,\, \tau^+ \tau^-$. These constraints are shown respectively in the middle and right panels of Figure~\ref{fig:H:offshell} as the shaded regions. The prospects of $h_{ee}^\dagger h_{\mu\tau}$ and $h_{e\mu}^\dagger h_{e\tau}$ at the CEPC and ILC are also shown in these two figures as the red and blue lines. As the coupling combination $h_{e\mu}^\dagger h_{e\tau}$ can not induce any $s$-channel process at colliders, the prospects in the right panel of Figure~\ref{fig:H:offshell} do not have any resonance structure. For all these $H$-induced LFV processes $e^+ e^- \to e^\pm \tau^\mp,\, \mu^\pm \tau^\mp$, the future high-energy $e^+ e^-$ colliders can probe a large unconstrained parameter space of $m_H$ and $h_{\alpha\beta}^\dagger h_{\delta\gamma}$, as seen in Figure~\ref{fig:H:offshell}.

\begin{figure}[H]
  \includegraphics[width=0.25\textwidth]{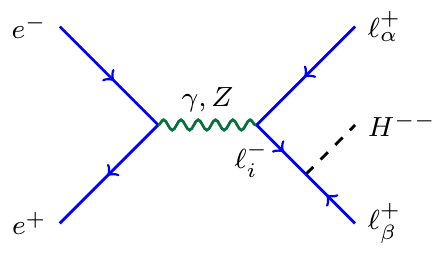}
  \includegraphics[width=0.25\textwidth]{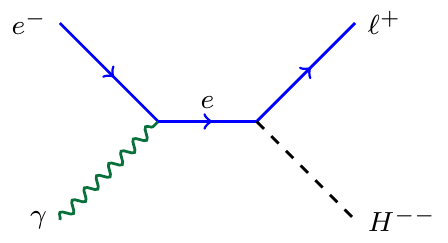}
  \includegraphics[width=0.25\textwidth]{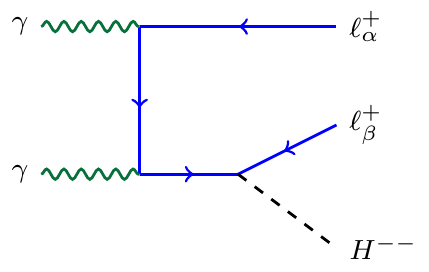}
  \caption{{Representative Feynman diagrams for on-shell production of $H^{\pm\pm}$ at future lepton colliders (cf. the processes in Equation~(\ref{eqn:Hpp:channels})). Figure from Ref.~\mbox{\cite{BhupalDev:2018vpr}}}\hspace{0pt}.}
  \label{fig:diagram3}
\end{figure}

Before switching to the doubly-charged scalar in the next section, here are a few more comments on $H$:
\begin{itemize}
    \item Analogous to the neutral current process in Equation~(\ref{eqn:H}), we can also have charged current process $e^+ e^- \to \overline{\nu}_\alpha \nu_\beta H$, which is mediated by a $t$-channel $W$ boson. See Ref.~\cite{BhupalDev:2018vpr} for more details. 
    \item At future high-energy $e^+ e^-$ colliders, high-energy photon beams can be obtained from the back scattering of high-intensity low-energy laser beam off high-energy electron beams~\cite{Ginzburg:1981vm, Ginzburg:1982yr, Telnov:1989sd},  which provides more production channels for the neutral scalar $H$, e.g., via the process $\gamma\gamma \to e^\pm \mu^\mp H$ (the corresponding Feynman diagram is similar to the third diagram in Figure~\ref{fig:diagram3}). These channels are largely complementary to the $e^+ e^-$ channels above~\cite{BhupalDev:2018vpr}. 

    \item Future high-energy muon colliders can probe larger regions of parameter space  for the muon flavor couplings, in particular for the explanation of muon $g-2$ anomaly (see e.g., Ref.~\cite{Capdevilla:2020qel}).
       \item The LFV signal due to the scalar $H$ can also be searched for at the hadron colliders, i.e., in the channel $q\bar{q} \to \ell_\alpha^\pm \ell_\beta^\mp H$ at the parton level. As a result of the relatively ``dirty'' backgrounds at the hadron colliders in particular for the tau leptons, the searches of LFV signals from $H$ is more challenging at the hadron colliders than at the lepton colliders. However, the scalar $H$ can be probed to a larger mass at the hadron colliders (cf. Ref.~\mbox{\cite{Iguro:2020qbc}}). \end{itemize}

\vspace{-10pt}

\section{Doubly Charged Scalar \boldmath{$H^{\pm\pm}$}}

The doubly charged  scalar $H^{\pm\pm}$ exists in a large variety of BSM scenarios, such as the type-II seesaw~\cite{Magg:1980ut, Schechter:1980gr, Cheng:1980qt, Lazarides:1980nt, Mohapatra:1980yp}, LRSM~\cite{Pati:1974yy, Mohapatra:1974gc, Senjanovic:1975rk}, and the Zee--Babu model~\cite{Babu:1988ki}. The doubly charged scalars can couple either to left-handed or right-handed charged fermions in the SM, and the Yukawa couplings can be written in the form of, from the perspective of effective Lagrangian:\footnote{As in Equation~(\ref{eqn:Lagrangian:H}), the couplings in Equation~(\ref{eqn:Lagrangian:Hpp}) may not be invariant under the SM $SU(2)_L$ if the charged leptons involved are left-handed.}
\vspace{-10pt}

 \begin{eqnarray}
 \label{eqn:Lagrangian:Hpp}
 {\cal L}_Y = f_{\alpha\beta} H^{++} \overline{\ell_\alpha^C} \ell_\beta ~+~ {\rm H.c.}
\end{eqnarray}
which is not only LNV  {if the doubly charged scalar does not carry any lepton number~\mbox{\cite{Crivellin:2018ahj}},}\footnote{{The lepton number of the doubly charged scalar is model dependent: for instance, $H^{++}$ has lepton number $L = -2$ in the type-II seesaw, while in some other models its lepton number can be zero.}}  but also potentially LFV. As a result of the gauge couplings of $H^{\pm\pm}$ to photon as well as the $Z$   {boson } and the Yukawa couplings $f_{\alpha\beta}$, $H^{\pm\pm}$ can be pair produced at the high-energy lepton and hadron colliders, i.e., $e^+ e^-,\, pp \to H^{++} H^{--}$. The usual gauge interaction inducing the Drell--Yan process can not be used to directly probe the Yukawa couplings $f_{\alpha\beta}$, unless the Yukawa couplings are sufficiently small such that $H^{\pm\pm}$ is long-lived at the high-energy colliders. After being produced, the   {doubly charged } scalars decay into pairs of same-sign dileptons, i.e., $H^{\pm\pm} \to \ell_\alpha^\pm \ell_\beta^\pm$, which is apparently   {signatures } beyond the SM. In some models, the {doubly charged } scalar may decay also into other particles, e.g., $H^{\pm\pm} \to W^\pm W^\pm$ in the type-II seesaw. 
The most stringent limit on the {doubly charged } scalar mass $M_{\pm\pm}$ is from the direct searches of same-sign dilepton pairs from $H^{\pm\pm}$ decay at the LHC~{\mbox{\cite{ATLAS:2017xqs, CMS:2017pet, ATLAS:2018ceg,
ATLAS:2021jol} }(see e.g., Refs.~\mbox{\cite{Rizzo:1981dla, Akeroyd:2005gt, FileviezPerez:2008jbu} }
for classic phenomenological studies and Refs.~\mbox{
\cite{Ferreira:2019qpf, deMelo:2019asm, Padhan:2019jlc, Fuks:2019clu, Gluza:2020qrt, Ashanujjaman:2021txz} }\hspace{0pt}
for projections of $H^{\pm\pm}$ at future hadron colliders)}. At the low-energy high-precision frontier, the coupling $f_{\alpha\beta}$ is also constrained by the LFV decays $\ell_\alpha \to \ell_\beta \gamma$, $\ell_\alpha \to \ell_\beta \ell_\gamma \ell_\delta$, the electron and muon $g-2$, muonium oscillation, and the LEP $e^+ e^- \to \ell^+ \ell^-$ data (see e.g., Ref.~\cite{BhupalDev:2018vpr}).

\textls[-5]{As for the neutral scalar $H$ case above, the {doubly charged } scalar $H^{\pm\pm}$ can also be singly produced at the high-energy lepton colliders via the Yukawa couplings $f_{\alpha\beta}$}~\mbox{\cite{Rizzo:1981dla, Lusignoli:1989tr, Barenboim:1996pt, Kuze:2002vb, Yue:2007kv, Yue:2007ym}}, e.g., $e^+ e^- \to H^{\pm\pm} \ell_\alpha^\mp \ell_\beta^\mp$ {(cf. the left panel of Figure~\ref{fig:diagram3})}. The corresponding production cross section of this channel is proportional to  the Yukawa coupling $|f_{\alpha\beta}|^2$, therefore the Yukawa coupling $f_{\alpha\beta}$ can be directly measured in such processes. Take $f_{e\mu}$ as an explicit example, with the $e^\pm$ beams and the the high-energy photon beams, we can have the following single production processes:
\begin{eqnarray}
 \label{eqn:Hpp:channels}
 e^+ e^- ,\, \gamma\gamma \to H^{\pm\pm} e^\mp \mu^\mp  \,, \quad
e^\pm \gamma \to H^{\pm\pm} \mu^\mp \,.
\end{eqnarray}

\begin{figure}[H]
  \includegraphics[width=0.4\textwidth]{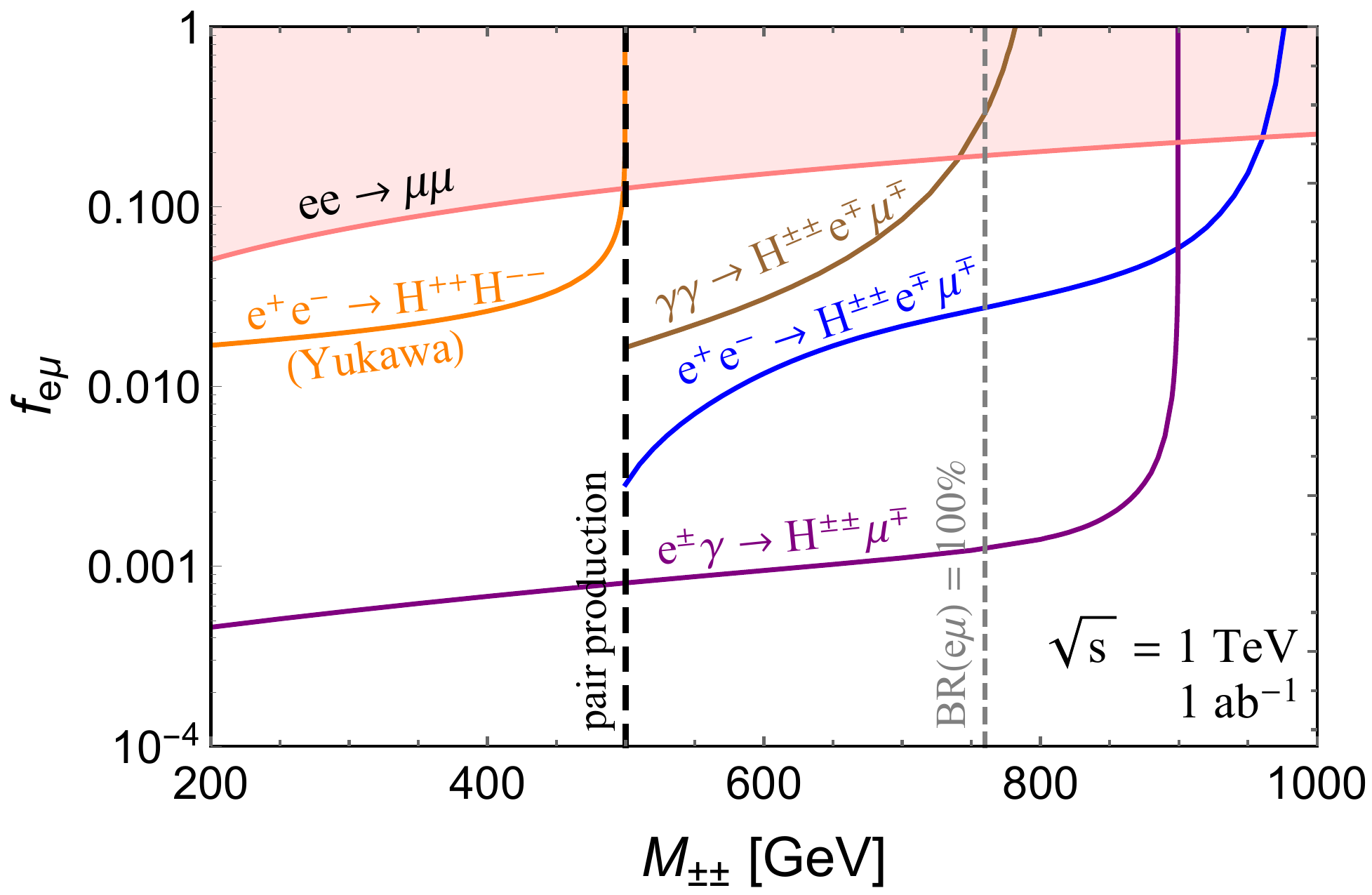} 
  \includegraphics[width=0.4\textwidth]{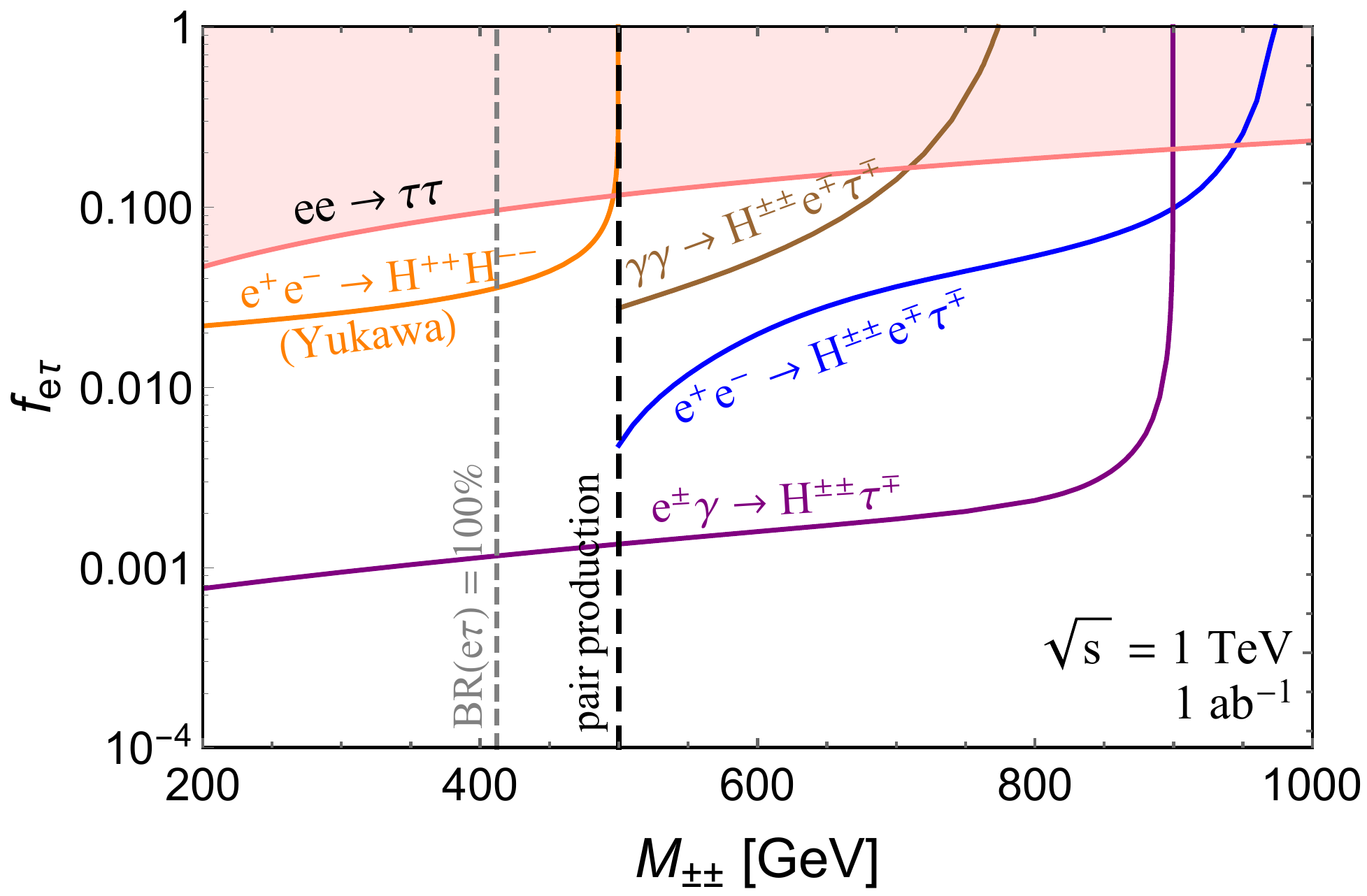}
  \caption{Prospects of the {doubly charged } scalar mass $M_{\pm\pm}$ and the LFV coupling $f_{e\mu}$  {(\textbf{left}) and $f_{e\tau}$ (\textbf{right}) } at ILC $1$ TeV with 1 ab$^{-1}$, in the Yukawa pair (orange) and single production modes of the $e^+ e^-$ (blue), $e\gamma$ (purple) and $\gamma\gamma$ (brown) processes. The pink shaded {regions are } excluded by the LEP {$e^+ e^- \to \ell^+ \ell^-$ } data. The vertical dashed gray lines indicate the current same-sign dilepton limits on the {doubly charged } scalar mass from LHC, assuming {respectively, ${\rm BR} (H^{\pm\pm} \to e^{\pm} \mu^{\pm}) = 100\%$ and ${\rm BR} (H^{\pm\pm} \to e^{\pm} \tau^{\pm}) = 100\%$ for the left and right panels} . See text for more details. Figure from Ref.~\cite{BhupalDev:2018vpr}.}
  \label{fig:Hpp:1}
\end{figure}

 {The corresponding Feynman diagrams can be found in Figure~\ref{fig:diagram3}.} For simplicity, we assume the {doubly charged } scalar decays predominately into $e^\pm \mu^\pm$,\footnote{{In the type-II seesaw, this is not a good approximation, as the leptonic branching fractions are largely determined by the PMNS neutrino mixing matrix, see e.g., Ref.~\mbox{\cite{FileviezPerez:2008jbu}}.}}  which can be ensured if the coupling $f_{e\mu}$ is much larger than other Yukawa couplings and other decay channels of the {doubly charged } scalar, such as $H^{\pm\pm} \to W^\pm W^\pm$ are subdominant. 
With ${\rm BR} (H^{\pm\pm} \to e^\pm \mu^\pm)=100\%$, the corresponding LHC limit on $M_{\pm\pm}$ is shown as the vertical dashed line in  {the left panel of } Figure~\ref{fig:Hpp:1}~\cite{ATLAS:2017xqs,CMS:2017pet}. The coupling $f_{e\mu}$ will also induce extra contribution to $e^+ e^- \to \mu^+ \mu^-$ at the LEP, and the corresponding limit is presented as the pink shaded region in  {the left panel of } Figure~\ref{fig:Hpp:1}. The contribution of {doubly charged} scalars to the muon $g-2$ is always negative, and therefore can not explain the muon $g-2$ anomaly. Furthermore,  the {doubly charged } scalar contribution to muon $g-2$ is highly suppressed by the charged lepton mass~\cite{Lindner:2016bgg}, thus not shown in Figure~\ref{fig:Hpp:1}.

The prospects of $H^{\pm\pm}$  {and $f_{e\mu}$ } at the ILC 1 TeV with luminosity of 1 ab$^{-1}$ are shown in  {the left panel of } Figure~\ref{fig:Hpp:1}. For the photon beam, we take the effective photon luminosity distribution from Refs.~\cite{Ginzburg:1981vm, Ginzburg:1982yr, Telnov:1989sd}. In the $e^+ e^-$ and $\gamma\gamma$ channels, for {doubly charged scalar mass $M_{\pm\pm} \lesssim \sqrt{s}/2 =500$ } GeV, the production of $H^{\pm\pm}$ will be dominated by pair production via both  {the } gauge and Yukawa interactions, while for $M_{\pm\pm}\gtrsim 500$ GeV only the single production of $H^{\pm\pm}$ is kinematically allowed. The $e^\pm \gamma$ process has only two particles in the final state, thus it can probe a smaller Yukawa coupling $f_{e\mu}$. 
For the LFV coupling $f_{e\tau}$, the prospects at the ILC 1 TeV is to some extent similar to the $f_{e\mu}$ case  {(cf. the right panel of Figure~\ref{fig:Hpp:1})} , with a smaller reconstruction efficiency for $\tau$ than muons. For the LFV coupling $f_{\mu\tau}$, the production cross section for $e^+ e^-,\, \gamma\gamma \to H^{\pm\pm}\mu^\mp \tau^\mp$ is much smaller than that for $H^{\pm\pm} e^\mp \mu^\mp$ and $H^{\pm\pm} e^\mp \tau^\mp$. As a result, the prospects of $f_{\mu\tau}$ are only at the order of 0.1~\cite{BhupalDev:2018vpr}.

\begin{figure}[H]
  \includegraphics[width=0.25\textwidth]{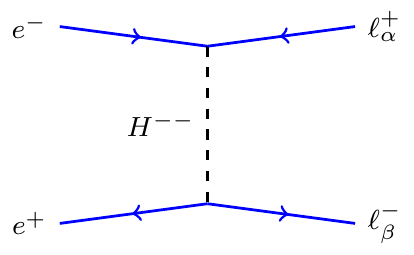}
  \includegraphics[width=0.25\textwidth]{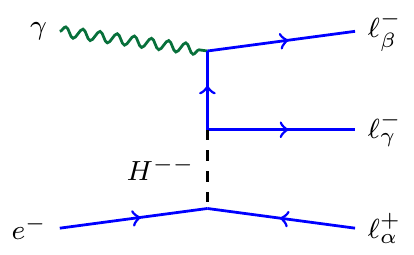}
  \caption{{Feynman diagrams for off-shell production of $H^{\pm\pm}$ at future lepton colliders (cf. the processes in Equation~(\ref{eqn:dcs:offshell})). Figure from Ref.~\mbox{\cite{BhupalDev:2018vpr}}.}}
  \label{fig:diagram4}
\end{figure}

Analogous to the neutral scalar $H$ above, the {doubly charged } scalar $H^{\pm\pm}$ can also induce LFV processes $e^+ e^- \to \ell_\alpha^\pm \ell_\beta^\mp$ at the high-energy lepton colliders by playing the role of the {$t$-channel mediator (cf. the left panel of Figure~\ref{fig:diagram4})}. The limit from $\mu \to eee$ is stringent, and it precludes the prospects of $H^{\pm\pm}$ in the channel of $e^\pm \mu^\mp$. In the $\tau$ flavor sector, with the coupling $f_{ee}^\dagger f_{e\tau}$, the {doubly charged } scalar can induce the processes 
{(see Figure~\ref{fig:diagram4} for diagrams):
} \begin{eqnarray}
\label{eqn:dcs:offshell}
e^+ e^- \to e^\pm \tau^\mp \,, \quad
e^\pm \gamma \to e^\pm e^\pm \tau^\mp,\, \tau^\pm e^\pm e^\mp \,.
\end{eqnarray}

\begin{figure}[H]
  \includegraphics[width=0.4\textwidth]{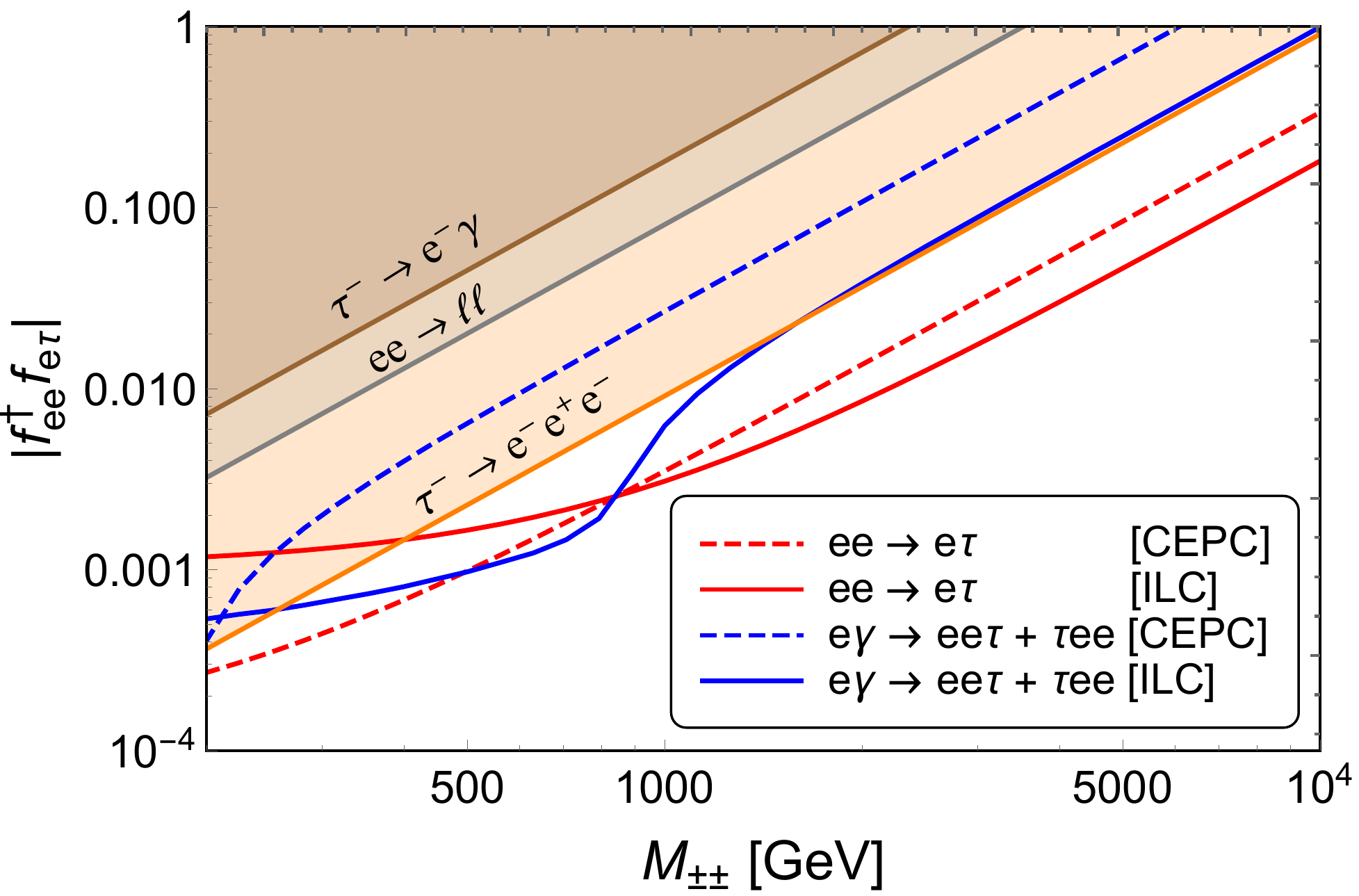}
  \includegraphics[width=0.4\textwidth]{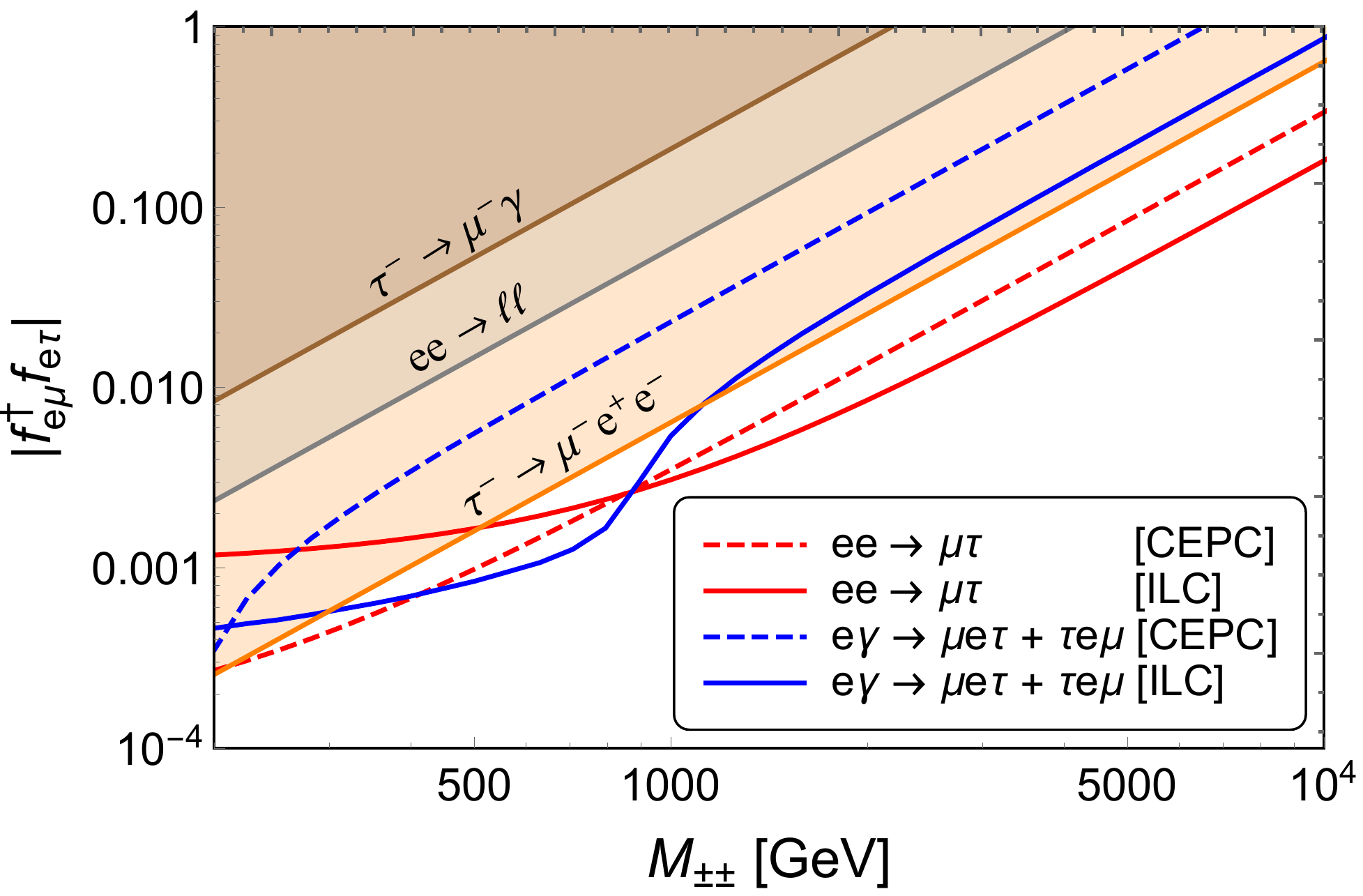}
  \caption{Prospects of the Yukawa couplings $|f_{ee}^\dagger f_{e\tau}|$ (\textbf{left}) and $|f_{e\mu}^\dagger f_{e\tau}|$ (\textbf{right}) for the {doubly charged } scalar $H^{\pm\pm}$ production via the $ee \to \ell_\alpha \ell_\beta$ (red) and $e\gamma \to \ell_\alpha \ell_\beta \ell_\gamma$ (blue) processes, at CEPC  240 GeV with 5 ab$^{-1}$ (dashed) and ILC 1 TeV with 1 ab$^{-1}$ (solid). The shaded regions are excluded by the corresponding limits. See text for more details. Figure from Ref.~\cite{BhupalDev:2018vpr}.}
  \label{fig:Hpp:2}
\end{figure}

Similar to the neutral scalar case, the relevant limits for the {doubly charged } scalar are from the LFV decays $\tau \to e \gamma$, $\tau \to eee$, and the LEP data $\ell^+ \ell^-$ data, which are presented as the shaded regions in the left panel of Figure~\ref{fig:Hpp:2}. The prospects of $M_{\pm\pm}$ and $|f_{ee}^\dagger f_{e\tau}|$ at the CEPC 240 GeV with the luminosity of 5 ab$^{-1}$ and ILC 1 TeV with 1 ab$^{-1}$ are shown in the left panel of Figure~\ref{fig:Hpp:2}, with red lines for the channel $e^+ e^- \to e^\pm \tau^\pm$ and the blue lines for $e\gamma \to ee\tau$ (combing the two possible final state in Equation~(\ref{eqn:dcs:offshell})). The dashed lines denote the CEPC sensitivities, and  the solid lines are for ILC. As in the neutral scalar case in Section~\ref{sec:scalar}, both CEPC and ILC can probe $H^{\pm\pm}$ with a mass larger than the center-of-mass energy in the $e^+ e^-$ channel, which corresponds to detecting the effective four-fermion interaction $(\overline{e}e)(\overline{e}\tau)$ with the cut-off scale $\Lambda \simeq M_{\pm\pm}/\sqrt{|f_{ee}^\dagger f_{e\tau}|}$. With three particles in the final state, the cross section for $e\gamma$ processes are not as competitive as the $e^+ e^-$ channel, as seen in the left panel of Figure~\ref{fig:Hpp:2}.

\vspace{-10pt}

Given the coupling $f_{e\mu}^\dagger f_{e\tau}$, the {doubly charged } scalar $H^{\pm\pm}$ can induce the processes $e^+ e^- \to \mu^\pm \tau^\mp$ and $e^\pm \gamma \to \mu^\pm e^\pm \tau^\mp,\, \tau^\pm e^\pm \mu^\mp${. The } corresponding limits from $\tau^- \to \mu^-\gamma$, $\tau^-\to \mu^- e^+ e^-$, the LEP $\ell^+ \ell^-$ data, and prospects  {of $H^{\pm\pm}$ } at the CEPC 240 GeV and ILC 1 TeV are presented in the right panel of Figure~\ref{fig:Hpp:2}. As for the case of $f_{ee}^\dagger f_{e\tau}$, the $e^+ e^-$ processes have better sensitivities than the $e\gamma$ collisions. These processes can also be induced by the coupling $f_{ee}^\dagger f_{\mu\tau}$. However, the corresponding production cross sections are smaller, which weakens the detectability of $f_{ee}^\dagger f_{\mu\tau}$ at future high-energy lepton colliders. More $H^{\pm\pm}$ induced LFV processes, such as $e\gamma \to \mu\mu\mu,\, \mu\mu\tau$ are also possible, however the corresponding prospects are suppressed by the small cross sections. More details can be found in Ref.~\cite{BhupalDev:2018vpr}.

Here are more comments on the {doubly charged } scalar $H^{\pm\pm}$:
\begin{itemize}
    \item The CLIC energy can go up to 3 TeV, and this will improve significantly the prospects of $M_{\pm\pm}$ in both the on-shell and off-shell searches at future lepton colliders.
    \item The future muon collider will provide more channels for searches of LFV due to the {doubly charged scalar, for instance in the processes $\mu^+ \mu^- \to e^\pm \mu^\mp, \; \mu^\pm \tau^\mp$. Furthermore, the muon collider can explore higher energy scales than the $e^+ e^-$ colliders~\mbox{
\cite{Delahaye:2019omf}}}. 
    \item For sufficiently small couplings, the {doubly charged } scalar can be long-lived at the high-energy colliders, which is  {however largely model dependent. For instance, in some regions of parameter space in the type-II seesaw, the decays $H^{\pm\pm} \to \ell_\alpha^\pm \ell_\beta^\pm,$ $W^{\pm(\ast)} W^{\pm(\ast)}$ are suppressed, respectively, by the tiny active neutrinos and the small vacuum expectation value $v_\Delta$ of the triplet (and the off-shell $W$ bosons), which makes $H^{\pm\pm}$ potentially long-lived~\mbox{
\cite{FileviezPerez:2008jbu}}. The searches of long-lived doubly charged scalar is } largely complementary to the searches of prompt signals from $H^{\pm\pm}$ decay~\cite{BhupalDev:2018tox}.
     \item {As mentioned above, the doubly charged scalar can be produced at the high-energy colliders via the gauge interactions, e.g., the Drell--Yan process $pp \to H^{++} H^{--}$. At future high-energy hadron colliders, the doubly charged scalar can be probed to a higher mass than at the lepton colliders~\mbox{
\cite{Arkani-Hamed:2015vfh}}. Though the LFV couplings can not be directly measured in these channels (unless the doubly charged scalar is long-lived at the colliders), the LFV signals at the hadron colliders are largely complementary to the signals in this section at the lepton colliders.
} \end{itemize}

\section{Heavy Neutrino \boldmath{$N$}}

In the type-I seesaw~\cite{Minkowski:1977sc,Mohapatra:1979ia, Yanagida:1979as,Gell-Mann:1979vob,Glashow:1979nm}, heavy neutrinos are introduced to generate the tiny neutrino masses. With the mass term ${\cal M}_{N,ij} \overline{N_i^C} N_j$ ($i,\,j$ are the mass indices), the heavy neutrinos $N_i$ are Majorana particles.  In some other seesaw models such as the inverse seesaw~\cite{Mohapatra:1986aw, Mohapatra:1986bd, Bernabeu:1987gr}, by adding three more neutral singlet fermions, the heavy neutrinos form pseudo-Dirac states. Whether they are Majorana or Dirac fermions, the heavy neutrinos could mix with the active neutrinos $\nu_\alpha$ ($\alpha = e,\,\mu,\,\tau$), and thus couple to the SM  {Higgs,} as well as the $W$ and $Z$ bosons through  heavy-light neutrino mixing $V_{\alpha N}$. If the heavy neutrinos are at or below the TeV-scale, they can be produced at the high-energy lepton and hadron colliders. For simplicity, let us focus here only on the LFV signals from the heavy neutrinos. To this end, let us neglect the specific UV completions of seesaw models and consider only one heavy neutrino $N = N_1$, with the other two states $N_{2,3}$ much heavier and not contributing significantly to the LFV signals. 

The heavy neutrino $N$ can be produced at $pp$ colliders via the charged current Drell--Yan and vector-boson fusion (VBF) process~{\mbox{\cite{Keung:1983uu, Datta:1993nm, Dev:2013wba, Alva:2014gxa, Degrande:2016aje}}}:\footnote{{The heavy neutrino $N$ can also be produced from the $W^+ W^-$ and $W^\pm W^\pm$ scattering processes~\mbox{\cite{Dicus:1991fk, Fuks:2020att}}.}}

\vspace{-10pt}

 \begin{eqnarray}
\label{eqn:N}
&&q\bar{q}^\prime \to W^{\pm\ast} \to \ell_\alpha^\pm N \to \ell_\alpha^\pm \ell_\beta{^{\mp} } W{^{\pm},\;
\ell_\alpha^\pm \ell_\beta^{\pm} W^{\mp} } \,, \nonumber \\
&& W^{\pm\ast} \gamma \to \ell_\alpha^\pm N \to \ell_\alpha^\pm \ell_\beta{^{\mp} W^\pm,\; 
\ell_\alpha^\pm \ell_\beta^{\pm} } W^\mp \,{.
} \end{eqnarray}

\begin{figure}[H]
  \includegraphics[width=0.25\textwidth]{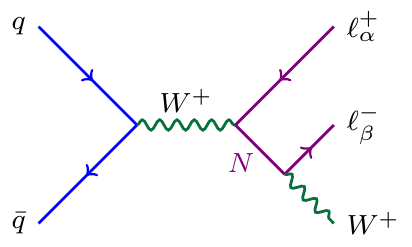}
  \includegraphics[width=0.25\textwidth]{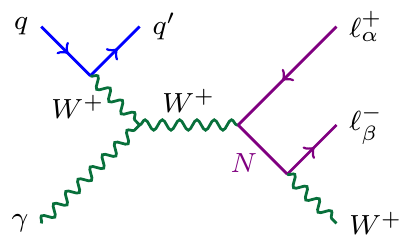}
  \caption{{Feynman diagrams for production of heavy neutrino $N$ at future hadron colliders (cf. the processes in Equation~(\ref{eqn:N})). Figure from Ref.~\mbox{
\cite{Pascoli:2018heg}}.}}
  \label{fig:diagram5}
\end{figure}

{The representative diagrams are presented in Figure~\ref{fig:diagram5}, and } we have assumed the heavy neutrino mass $m_N > m_W$ (with $m_W$ the $W$ boson mass). If the heavy neutrino $N$ is a Dirac fermion, there will only be opposite-sign charged leptons $\ell_\alpha^\pm \ell_\beta^\mp$ in the final state. For the case of Majorana $N$, there will also be the same-sign charged leptons $\ell_\alpha^\pm \ell_\beta^\pm$, which is  undoubtedly LNV signals beyond the SM. 
For some mass ranges of $m_N$, the gluon-fusion process $gg\to Z^\ast,\, h^\ast \to \nu N$ is also very important~{\mbox{\cite{Willenbrock:1985tj,Dicus:1991wj, Hessler:2014ssa, Ruiz:2017yyf}}}.  More details about the production channels of $N$ can be found e.g., in Ref.~\cite{Pascoli:2018heg}. If the heavy neutrino $N$ mixes with two neutrino flavors $\nu_{\alpha,\,\beta}$ (with $\alpha\neq \beta$) in the SM, the production and decay of $N$ at the high-energy colliders will produce LFV signals {via the charged currents} , i.e., $\ell_\alpha\neq\ell_\beta$ in the final state of the process in Equation~(\ref{eqn:N}){~\mbox{\cite{Arganda:2015ija}}}.

\begin{figure}[H]
   \includegraphics[width=0.45\textwidth]{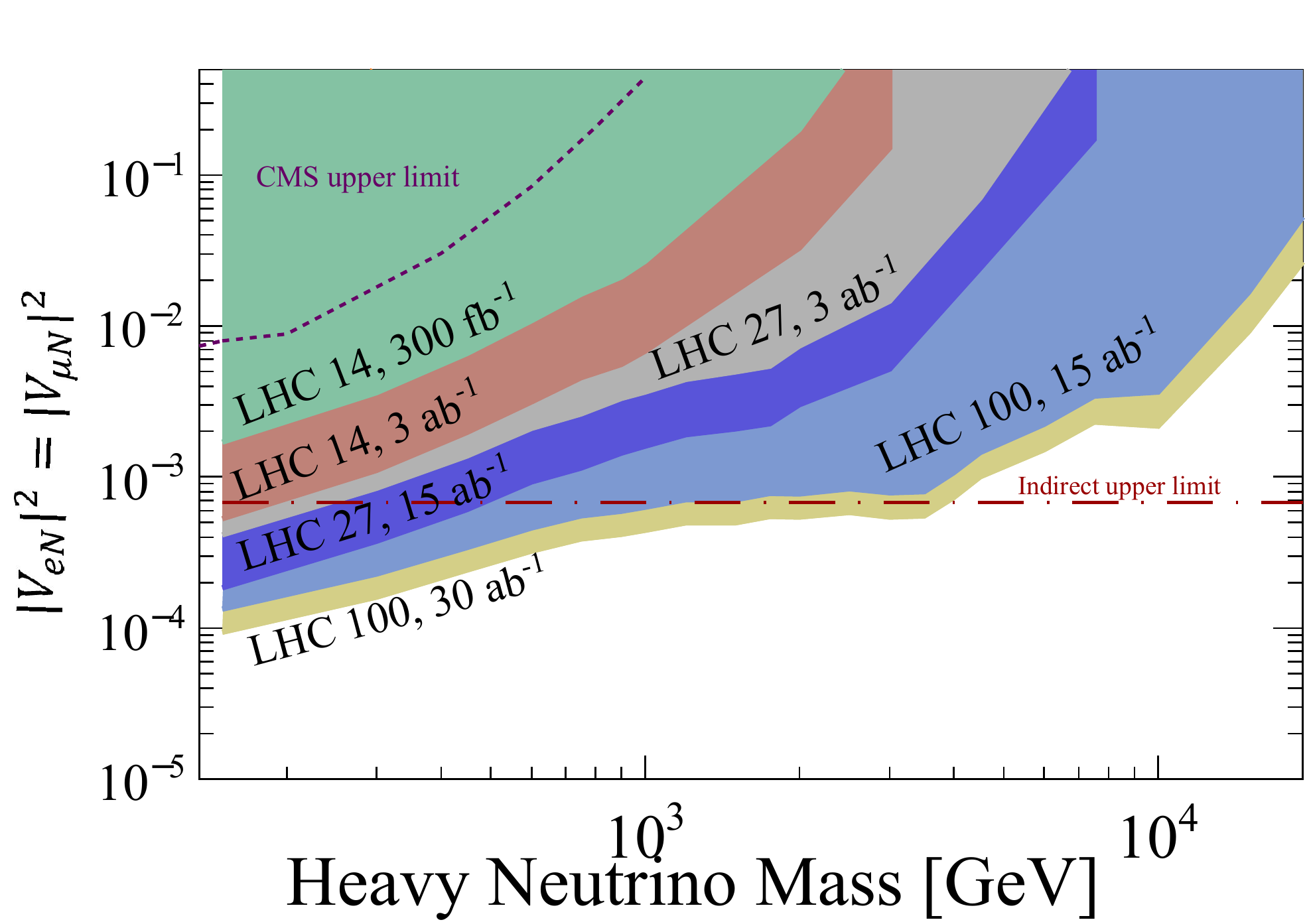} 
  \includegraphics[width=0.45\textwidth]{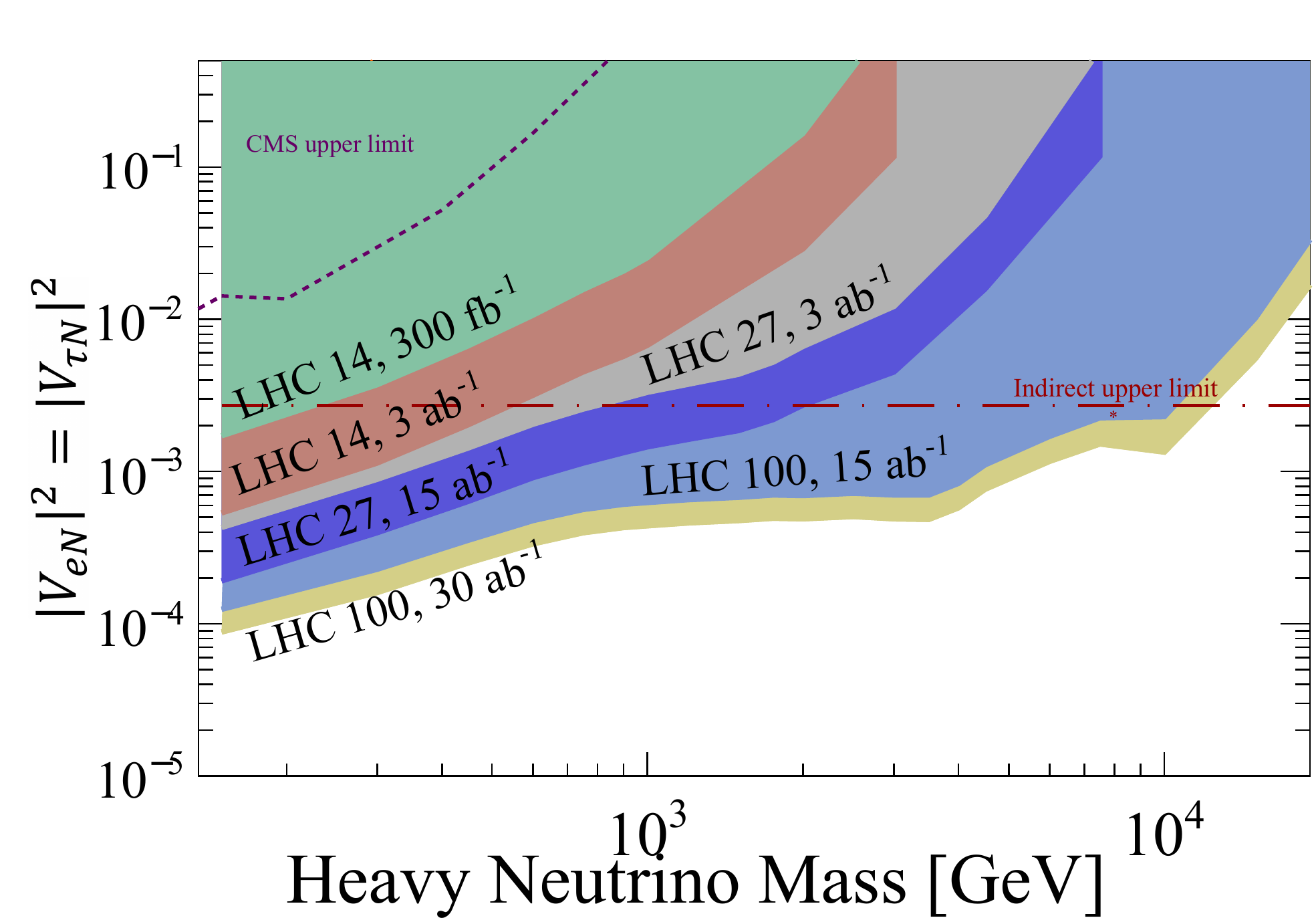} 
  \caption{Sensitivities of heavy neutrino mass $m_N$ and the heavy-light neutrino mixing\linebreak $|V_{eN}|^2 = |V_{\mu N}|^2$ {(\textbf{left}) or $|V_{eN}|^2 = |V_{\mu N}|^2$ (\textbf{right})}, at future LHC 14 TeV with luminosities of 300 fb$^{-1}$ and 3 ab$^{-1}$, LHC 27 TeV with 3 ab$^{-1}$ and 15 ab$^{-1}$, and 100 TeV collider with 15 ab$^{-1}$ and 30 ab$^{-1}$. Regions above the short dashed {lines } are excluded by CMS data, and the dot-dashed horizontal {lines indicate} the indirect limit from current EWPD
   data. See text for more details. Figure from Ref.~\cite{Pascoli:2018heg}.}
  \label{fig:N}
\end{figure}

 Considering both the charged current Drell--Yan and VBF processes in Equation~(\ref{eqn:N}), if the $W$ boson from $N$ decays leptonically, we will have three charged leptons plus significant missing transverse energy (MET) in the final state, with potentially extra (VBF) jets. With $|V_{eN}| = |V_{\mu N}|$ {and $V_{\tau N} = 0$}, we can have the processes $pp \to ee\ell_X + e\mu\tau_h$, with $\tau_h$ referring to the hadronic decaying tauons, $\ell_X = e,\, \mu,\,\tau_h$, and all the possible lepton charges are included. Adopting dynamic jet vetoes, the sensitivities of $m_N$ and the heavy-light neutrino mixing $|V_{eN}| = |V_{\mu N}|$ at future hadron colliders are presented in {the left panel of } Figure~\ref{fig:N}~\cite{Pascoli:2018heg}. {For the case of $|V_{eN}| = |V_{\tau N}|$ with $V_{\mu N} = 0$, the charged leptons in the final state can be $eee$, $ee\mu$, and $e\tau_h \ell_X$, and the corresponding prospects of $m_N$ and $|V_{eN}| = |V_{\tau N}|$ are shown in the right panel of Figure~\ref{fig:N}.} It is found that the prospects here are not sensitive to the Majorana or Dirac nature of $N$. In the near future, the LHC 14 TeV data with a luminosity of 300 fb$^{-1}$ can probe {heavy-light neutrino mixing angles $|V_{eN}|^2 = |V_{\mu N}|^2$ and $|V_{eN}|^2 = |V_{\tau N}|^2$ up to the order of $10^{-3}$ } for a heavy neutrino at the scale of 100 GeV. At  future 100 TeV collider, the sensitivity of heavy-light neutrino mixing can even go down to ${\cal O}(10^{-4})$. 
More simulation details can be found in Ref.~\cite{Pascoli:2018heg}. In Figure~\ref{fig:N}, the regions above the short dashed {lines } are excluded by the direct trilepton searches of $N$ at CMS~\cite{CMS:2018iaf}, and horizontal dot-dashed {lines represent} the current electroweak precision data (EWPD) constraints on $|V_{\alpha N}|$~\cite{Fernandez-Martinez:2016lgt}. More limits on heavy-light neutrino mixing can be found e.g., in {Refs.~\mbox{\cite{Bolton:2019pcu, ATLAS:2019kpx, CMS:2022fut}}}.

At the GeV-scale, the limits on the neutrino mass $m_N$ and $|V_{\alpha N}|^2$ will be mostly from the high-precision LFV measurements below~{\mbox{\cite{Atre:2009rg, Cottin:2018nms, Coloma:2020lgy, DeVries:2020jbs, Zhou:2021ylt}}}:

\begin{itemize}
    \item At lower energies, the charged current Drell--Yan process in Equation~(\ref{eqn:N}) will ``hadronize'' to be semileptonic weak decays of charged mesons, i.e.,
\begin{eqnarray}
\label{eqn:meson}
{\cal P}_2^\pm \to \ell_\alpha N^{(\ast)} \to \ell_\alpha^\pm \ell_\beta^{\mp} {\cal P}_1^\pm \,, 
\end{eqnarray}
with ${\cal P}_{1,\,2}^\pm$ the charged mesons. As at the high energy {scale}, the heavy neutrino $N$ will induce LFV in meson decays if $\ell_\alpha\neq \ell_\beta$. LNV decays ${\cal P}_2^\pm \to \ell_\alpha^\pm \ell_\beta^{\pm} {\cal P}_1^\mp$ are also possible if $N$ is a Majorana fermion. 
It is found that for the mass range of $m_\pi < m_N < m_K$ (with $m_{\pi,\,K}$ the masses of pions and Kaons), the most stringent limit on $|V_{eN}V_{\mu N}|$ is from the decay $K^+ \to \pi^+ e^\pm \mu^\mp$ leading to $|V_{eN}V_{\mu N}|\lesssim 10^{-9}$~\cite{Hu:2019zan}. For $m_K < m_N < m_B$ (with $m_{B}$ the $B$ meson mass), the strongest limit is from $B^+ \to \pi^+ e^\pm \mu^\mp$.  In the process (\ref{eqn:meson}), if $N$ is on-shell, it can be directly searched via the two-body meson decays, e.g., $K^+ \to \ell^+ N$. The NA62 data have excluded $|V_{eN}|^2 ,\, |V_{\mu N}|^2 \lesssim 10^{-8}$ to $10^{-9}$ for $170\, {\rm MeV} < m_N < 450$ MeV~\cite{NA62:2017qcd}. More processes can be found e.g., in Refs.~\cite{Bolton:2019pcu,Coloma:2020lgy}. 
\item For neutral mesons ${\cal P}^0$, the heavy neutrino will induce LFV leptonic decays at the 1-loop level, i.e., {~\mbox{\cite{Ilakovac:1999md}}}: 
\begin{eqnarray}
{\cal P}^0 \to \ell_\alpha^\pm \ell_\beta^\mp \,,
\end{eqnarray}
which can be applied to the LFV leptonic decays of $K_L$, $D^0$, $B^0$, and $B_s^0$. Such LFV meson decays are all highly suppressed in the SM, and the mass $m_N$ and heavy-light neutrino mixing $|V_{\alpha N} V_{\beta N}|$ are tightly constrained by precision meson data. 
More details can be found e.g., in Refs.~{\mbox{\cite{Ilakovac:1999md,Hu:2019zan}}}. 
\item In the charge lepton sector, if $N$ mixes with two SM neutrino flavors, it will induce extra contribution to the LFV radiative decays of charged leptons $\ell_\alpha \to \ell_\beta\gamma$ and $\mu - e$ conversion in nuclei. It is found that the most stringent limit is from $\mu \to e \gamma$, which leads to $|V_{eN} V_{\mu N}| \lesssim 10^{-3}$ for a 10 GeV $N$~\cite{Bolton:2019pcu}.
 \item  {For sufficiently light $N$, it could be long-lived at the high-energy colliders. The corresponding rich phenomenologies can be found e.g., in Refs.~\mbox{\cite{Helo:2013esa, Izaguirre:2015pga, Cottin:2018nms, Dube:2017jgo, Dib:2018iyr, Gago:2015vma, Accomando:2016rpc, Caputo:2017pit, Deppisch:2018eth, Liu:2018wte, Antusch:2017hhu, Kling:2018wct, Helo:2018qej, Jana:2018rdf, Caputo:2016ojx, Blondel:2014bra, Antusch:2016vyf, Bonivento:2013jag, Abada:2018sfh,Abada:2014cca, Antusch:2017pkq, Hernandez:2018cgc, Abada:2018oly, Boiarska:2019jcw, Lavignac:2020yld, Dib:2019ztn, Drewes:2019fou, Liu:2019ayx, Das:2019fee, Drewes:2019vjy, Chiang:2019ajm, Dib:2019tuj, Jones-Perez:2019plk, Coloma:2020lgy, Barducci:2020icf, Borsato:2021aum, Tastet:2021vwp, Liu:2022kid}}.
}\end{itemize}

If two (or more) heavy neutrino $N_{1,\,2}$ are involved, there could be more phenomenological implications, such as heavy neutrino mixing and CP violation due to the TeV-scale $N$~\cite{Bray:2007ru} or in the decays of mesons and the SM Higgs~\cite{Tapia:2021gne, Cvetic:2021lmm, Abada:2019bac}. All these processes might involve LFV in some way.

\section{Heavy \boldmath{$W_R$} Boson}

\begin{figure}[H]
  \includegraphics[width=0.28\textwidth]{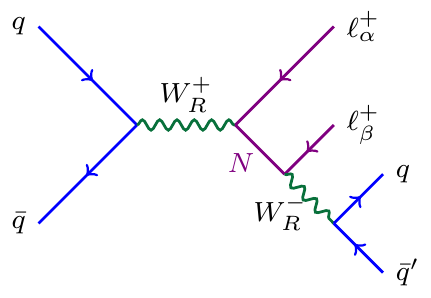}
  \caption{{Feynman diagram for production of heavy $W_R$ boson at future hadron colliders (cf. the process in Equation~(\ref{eqn:WR})). }}
  \label{fig:diagram6}
\end{figure}

In the LRSM, the heavy neutrinos are states from the right-handed isospin doublets, therefore $N$ couples to the heavy right-handed $W_R$ boson. As a result of the Majorana nature of $N$, the production and decay of $W_R$ at the high-energy hadron colliders will produce same-sign dileptons via the process:
\begin{eqnarray}
\DIFaddbegin \label{eqn:WR}
\DIFaddend q\bar{q}^\prime \to W_R^{\pm} \to \ell_\alpha^\pm N \to \ell_\alpha^\pm \ell_\beta^{\pm} W_R^{\mp\ast} \to 
\ell_\alpha^\pm \ell_\beta^\pm jj \,, 
\end{eqnarray}
which constitutes the smoking-gun signal of LRSM.  {The Feynman diagram is shown in Figure~\ref{fig:diagram6}.} In such a process, the heavy neutrino mass eigenstate $N$ may be a state of {a pure flavor}, or combinations of two {or more flavors. Take the two flavor case as an explicit example, the heavy neutrino mass eigenstate can be decomposed as } $N = \cos\theta N_\alpha + \sin\theta N_\beta${,} with $\theta$ as the mixing angle of heavy neutrinos $N_{\alpha,\,\beta}${, which leads } to LFV signatures $\ell_\alpha^\pm \ell_\beta^\pm$. The most stringent  LHC limit on $W_R$ and $N$ is from the recent searches by CMS with a luminosity of 138 fb$^{-1}$ at 13 TeV, which covers both the lepton flavor conserving cases $ee$, $\mu\mu$, and LFV case $e\mu$~\cite{CMS:2021dzb}. No significant excess is found above the backgrounds, and the $W_R$ mass is excluded up to 5.4 TeV. With 3000 fb$^{-1}$ data at 14 TeV, the $W_R$ mass can be improved up to roughly 6.5 TeV~\cite{Nemevsek:2018bbt, Chauhan:2018uuy}. At future 100 TeV colliders, the $W_R$ mass can be probed up to roughly 38 TeV with a luminosity of 30 fb$^{-1}$, assuming the heavy neutrino decays promptly~\cite{Mitra:2016kov}. If the neutrino $N$ is {relatively light, say at the 100 GeV scale, } the parameter space of three-body decay $N \to \ell jj$ will be highly compressed. {The highly compressed signature leads to a boosted topology~\mbox{
\cite{Ferrari:2000sp, Mitra:2016kov}}\hspace{0pt}
, which has been searched at the LHC~\mbox{
\cite{ATLAS:2018dcj, ATLAS:2019isd}}\hspace{0pt}
.  Projections at future colliders can be found in e.g., Ref.~\mbox{
\cite{Nemevsek:2018bbt}}\hspace{0pt}
. If sufficiently light, the heavy neutrino } $N$ {will be} long-lived at the high-energy colliders, with {the decay length } up to the meter level for $m_N \sim$ 10 GeV~{\mbox{
\cite{Helo:2013esa, Maiezza:2015lza, Nemevsek:2016enw, Nemevsek:2018bbt, Cottin:2018kmq, Cottin:2019drg}}\hspace{0pt}
}. The displaced vertex signals from $N$ decay can be used to search for the heavy $W_R$ boson, and the  prospect of $W_R$ mass can go up to 33 TeV at the future 100 TeV collider, with $m_N \sim 100$ GeV~\cite{Nemevsek:2018bbt}. The prompt and displaced vertex signals from $W_R$ and $N$ can both be sources for LFV at future high-energy colliders.

\section{\boldmath{$Z'$} Boson}

\begin{figure}[H]
  \includegraphics[width=0.4\textwidth]{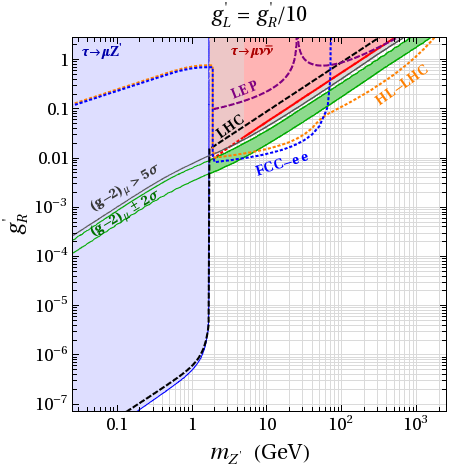} 
  \caption{Prospect of $m_{Z'}$ and $g'_R$ at {HL-LHC}
  (dotted orange line) and FCC-ee (dotted blue line) with $g'_L = g'_R/10$. The green band corresponds the muon $g-2$ anomaly at the $2\sigma$ C.L.. The blue, gray, and red shaded regions are excluded by $\tau \to \mu Z'$, muon $g-2$ discrepancy at the $5\sigma$ C.L., and $\tau \to \mu \nu\bar{\nu}$. The purple and black dashed lines are, respectively, the limits from the LEP $Z$-pole data and the LHC $W$ data. See text for more details. Figure from Ref.~\cite{Altmannshofer:2016brv}. }
  \label{fig:Zprime}
\end{figure}

The {$Z'$ gauge boson is associated with a BSM $U(1)'$ gauge symmetry. There have been  intensive studies on the rich phenomenologies of $Z'$, see e.g., Refs.~\mbox{
\cite{Langacker:2000ju, Langacker:2008yv, delAguila:2010mx, deBlas:2012qp, Accomando:2013sfa, Jezo:2014wra, Accomando:2016sge, Deppisch:2019kvs, Buras:2021btx, Bossi:2020yne,Iguro:2020qbc}}\hspace{0pt}
. The $Z'$ boson is intimately related to the neutrino mass generation, e.g., in the $U(1)_{B-L}$ gauge group~\mbox{
\cite{Wetterich:1981bx, Buchmuller:1991ce, Emam:2007dy, Basso:2008iv, FileviezPerez:2009hdc, Heeck:2014zfa, Khalil:2006yi, Huitu:2008gf, Accomando:2016sge, Accomando:2017qcs, Dev:2017xry}}\hspace{0pt}
. In some cases, the } heavy neutral $Z'$ boson can have flavor changing neutral currents, \footnote{{See e.g., Refs.~\mbox{
\cite{Langacker:2000ju,Davidson:2012wn,Abada:2015zea,DeRomeri:2016gum,Brignole:2004ah, Herrero:2018luu,Abada:2014cca,Calibbi:2021pyh} }\hspace{0pt}
for the rich phenomenologies of LFV couplings of the SM $Z$ boson.}}  e.g., from the mixing of SM fermions with heavy exotic fermions~\cite{Langacker:1988ur}.  Another important motivation of LFV couplings of $Z'$ is the muon $g-2$ anomaly, where the $Z'$ boson can be either heavy or light. For instance, the effective LFV couplings of $Z'$ to $\mu$ and $\tau$ can be written as:
\begin{eqnarray}
\label{eqn:Zprime}
{\cal L} \supset Z_\mu^\prime \left( g'_{L} \bar{\mu}_L \gamma^\mu \tau_L + g'_{R} \bar{\mu}_R \gamma^\mu \tau_R \right) ~+~ {\rm H.c.} \,.
\end{eqnarray}
with $\tau$ running in the loop, the $Z'$ boson with LFV couplings can explain the muon $g-2$ discrepancy, which is in some sense similar to the neutral scalar case above~\cite{Altmannshofer:2016brv, Lindner:2016bgg}. The corresponding $2\sigma$ regions for the muon $g-2$ anomaly is shown as the green band in  Figure~\ref{fig:Zprime}, where we have assumed  that the gauge couplings $g'_L = g'_R/10$. 

The direct searches of heavy $Z' \to e\mu,\, e\tau,\, \mu\tau$ have been performed at the LHC~\cite{CMS:2021tau}. Assuming the LFV couplings of $Z'$ are the same as the couplings of $Z$ boson to charged leptons, $Z'$ mass has been excluded up to 5.0 TeV in the $e\mu$ channel, 4.3 TeV in the $e\tau$ channel, and 4.1 TeV in the $\mu\tau$ channel~\cite{CMS:2021tau}. 
The LFV couplings $g'_{L,\,R}$ in Equation~(\ref{eqn:Zprime}) are also severely constrained by the decay {$\tau \to \mu Z'$} with $Z' \to \nu_\mu \bar\nu_\tau$, the lepton flavor universality violation in tauon decays, the high-precision LHC $W$ data,  the LEP $Z$-pole data, and the muon $g-2$ anomaly at the $5\sigma$ C.L.. As seen in Figure~\ref{fig:Zprime}, when all these constraints are taken into consideration, a $Z'$ with mass $m_{Z'} \gtrsim2$ GeV can provide viable interpretation of the muon $g-2$ anomaly. 

\begin{figure}[H]
  \includegraphics[width=0.25\textwidth]{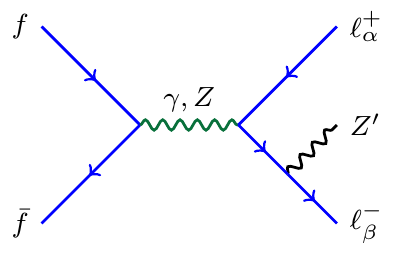}
  \caption{{Representative Feynman diagram for on-shell production of $Z'$ at future high-energy colliders via the process in Equation~(\ref{eqn:Zprime2}). Figure from Ref.~\mbox{
\cite{Altmannshofer:2016brv}}\hspace{0pt}
. }}
  \label{fig:diagram:Zprime}
\end{figure}

The LFV coupling $g'_{L,\,R}$ of the light $Z'$ boson can be directly measured at the high-energy hadron and lepton colliders via the process: 
\begin{eqnarray}
\DIFdelbegin 
\DIFdelend \DIFaddbegin \label{eqn:Zprime2}
\DIFaddend pp,\, e^+ e^- \to \mu^\pm \tau^\mp Z' \,,
\end{eqnarray}
which is quite similar to the process (\ref{eqn:H}) for the neutral scalar $H${, and the Feynman diagram is shown in Figure~\ref{fig:diagram:Zprime}}. With $Z' \to \mu^\pm \tau^\mp$, the LFV coupling $g'_{L,\,R}$ will generate same-sign dilepton pairs  $\mu^\pm \mu^\pm \tau^\mp \tau^\mp$. The prospects at the high-luminosity LHC (HL-LHC) 14 TeV with an integrated luminosity of 3 ab$^{-1}$ and the FCC-ee 91 GeV with 2.6 ab$^{-1}$ are presented, respectively, as the dotted orange and blue curves in Figure~\ref{fig:Zprime}. It is  clear that both machines can probe large regions of the $Z'$ interpretation of the muon $g-2$ anomaly. More details can be found in Ref.~\cite{Altmannshofer:2016brv}. It should be noted that the prospects can be further improved at a future muon collider.

\vspace{-10pt}

With the LFV coupling $g'_{e\mu}$  {to electron and muon}, the $Z'$ boson could contribute to $\mu -e$ scattering~\cite{Dev:2020drf,Masiero:2020vxk}, e.g., in the MUonE experiment, which is proposed to determine the contribution of hadronic vacuum polarization to muon $g-2$~\cite{CarloniCalame:2015obs, Abbiendi:2016xup}. However, the LFV coupling $g'_{e\mu}$ of $Z'$ boson is tightly constrained by muonium-antimuonium oscillation and electron $g-2$, which has precluded the sensitivity of the MUonE experiment. More details can be found in Ref.~\cite{Dev:2020drf}.


\section{Conclusions}

In this paper, we summarized briefly the LFV signals from {some representative particles in the seesaw frameworks, i.e., } the BSM neutral scalar $H$, {doubly charged } scalar $H^{\pm\pm}$, heavy neutrino $N$, heavy $W_R$ boson, and the $Z'$ boson. Constrained by current data, some of these particles are required to be heavy, {i.e., the doubly charged } scalar and the $W_R$ boson, while others can be either heavy at the TeV-scale or light down to the GeV-scale or even lighter{, i.e., the scalar $H$, the neutrino $N$ and the $Z'$ boson}. The LFV signals induced by these particles can originate from the {direct } (effective) LFV couplings in the charged lepton sector, i.e., the neutral scalar $H$, the {doubly charged } scalar, and the $Z'$ boson.  {Some of these couplings might be relevant to the neutrino sector, for instance the couplings of doubly charged scalar to charged leptons are largely determined by the neutrino oscillation data in the type-II seesaw. } For the heavy $N$ and $W_R$ boson, the LFV signals are intimately related to flavor mixing in the neutrino sector, which is transited to the charged lepton sector via the charged currents by the $W$ and $W_R$ bosons. {This paper  focused mainly on the LFV signals induced by these neutrino mass relevant particles } at the high-energy {lepton and hadron colliders. The LFV signatures can also be searched for at the high-intensity frontier}. In particular, if the muon $g-2$ anomaly is revealed to be true, or any LFV signal is found in future experiments, we need to understand better these particles, as well as other well-motivated particles, such as leptoquarks and supersymmetric particles. 
\vspace{6pt}

\funding{The author is supported by the National Natural Science Foundation of China under grant No.\  12175039, the 2021 Jiangsu Shuangchuang (Mass Innovation and Entrepreneurship)
Talent Program No.\ JSSCBS20210144, and the ``Fundamental Research Funds for the Central Universities''.}

\acknowledgments{The author would like to thank the anonymous referees for their valuable comments and suggestions.} 

\begin{adjustwidth}{-4.6cm}{0cm}


\end{adjustwidth}

\begin{adjustwidth}{-\extralength}{0cm}
\reftitle{References}

\end{adjustwidth}




\begin{thebibliography}{999}

\bibitem[Zyla \em{et~al.}(2020)Zyla et~al.]{ParticleDataGroup:2020ssz}
Zyla, P.A.;  Barnett, R.M.;  Beringer, J.;  Dahl, O.;  Dwyer, D.A.;  Groom, D.E.;  Lin, C.-J.;  Lugovsky, K.S.;  Pianori, E.;  Robinson, D.J.; et~al.
\newblock {Review of Particle Physics}.
\newblock {\em Prog. Theor. Exp. Phys.} {\bf 2020}, {\em 2020},~083C01.
\newblock
  doi:{\changeurlcolor{black}\href{https://doi.org/10.1093/ptep/ptaa104}{\detokenize{10.1093/ptep/ptaa104}}}.

\bibitem[Lindner \em{et~al.}(2018)Lindner, Platscher, and
  Queiroz]{Lindner:2016bgg}
Lindner, M.; Platscher, M.; Queiroz, F.S.
\newblock {A Call for New Physics: The Muon Anomalous Magnetic Moment and
  Lepton Flavor Violation}.
\newblock {\em Phys. Rept.} {\bf 2018}, {\em 731},~1--82,
\newblock
  doi:{\changeurlcolor{black}\href{https://doi.org/10.1016/j.physrep.2017.12.001}{\detokenize{10.1016/j.physrep.2017.12.001}}}.

\bibitem[Jaeckel and Ringwald(2010)]{Jaeckel:2010ni}
Jaeckel, J.; Ringwald, A.
\newblock {The Low-Energy Frontier of Particle Physics}.
\newblock {\em Ann. Rev. Nucl. Part. Sci.} {\bf 2010}, {\em 60},~405--437,
\newblock
  doi:{\changeurlcolor{black}\href{https://doi.org/10.1146/annurev.nucl.012809.104433}{\detokenize{10.1146/annurev.nucl.012809.104433}}}.

\bibitem[Pro(2012)]{Proceedings:2012ulb}
Hewett, J.L.;  Weerts, H.;  Brock, R.;  Butler, J.N.;  Casey, B.C.K.;  Collar, J.;  de Gouvea, A.;  Essig, R.;  Grossman, Y.;  Haxton, W.;  et al. {{Fundamental Physics at the Intensity Frontier}}. \emph{arXiv} \textbf{2012}, arXiv:1205.2671.

\bibitem[Essig \em{et~al.}(2013)Essig et~al.]{Essig:2013lka}
Essig, R.; Jaros, J.A.; Wester, W.;  Hansson Adrian, P.; Andreas, S.; Averett, T.; Baker, O.; Batell, B.; Battaglieri,  M.; Beacham, J.; et~al.
\newblock {Working Group Report: New Light Weakly Coupled Particles}. \emph{arXiv} \textbf{2013}, arXiv:1311.0029.

\bibitem[Beacham \em{et~al.}(2020)Beacham et~al.]{Beacham:2019nyx}
Beacham, J.;  Burrage, C.; Curtin, D.; De Roeck, A.; Evans, J.; Feng, J.L.; Gatto, C.; Gninenko, S.; Hartin, A.; Irastorza, I.; et~al.
\newblock {Physics Beyond Colliders at CERN: Beyond the Standard Model Working
  Group Report}.
\newblock {\em J. Phys. G} {\bf 2020}, {\em 47},~010501,
\newblock
  doi:{\changeurlcolor{black}\href{https://doi.org/10.1088/1361-6471/ab4cd2}{\detokenize{10.1088/1361-6471/ab4cd2}}}.

\bibitem[Fileviez~Perez(2015)]{FileviezPerez:2015mlm}
Fileviez~Perez, P.
\newblock {New Paradigm for Baryon and Lepton Number Violation}.
\newblock {\em Phys. Rept.} {\bf 2015}, {\em 597},~1--30,  doi:10.1016/j.physrep.2015.\linebreak 09.001.

\bibitem[Cai \em{et~al.}(2018)Cai, Han, Li, and Ruiz]{Cai:2017mow}
Cai, Y.; Han, T.; Li, T.; Ruiz, R.
\newblock {Lepton Number Violation: Seesaw Models and Their Collider Tests}.
\newblock {\em Front. Phys.} {\bf 2018}, {\em 6},~40,
\newblock
  doi:{\changeurlcolor{black}\href{https://doi.org/10.3389/fphy.2018.00040}{\detokenize{10.3389/fphy.2018.00040}}}.

\bibitem[Alguer\'o \em{et~al.}(2019)Alguer\'o, Capdevila, Descotes-Genon,
  Masjuan, and Matias]{Alguero:2018nvb}
Alguer\'o, M.; Capdevila, B.; Descotes-Genon, S.; Masjuan, P.; Matias, J.
\newblock {Are we overlooking lepton flavour universal new physics in $b\to
  s\ell\ell$ ?}
\newblock {\em Phys. Rev. D} {\bf 2019}, {\em 99},~075017,
\newblock
  doi:{\changeurlcolor{black}\href{https://doi.org/10.1103/PhysRevD.99.075017}{\detokenize{10.1103/PhysRevD.99.075017}}}.

\bibitem[Datta \em{et~al.}(2019)Datta, Kumar, and London]{Datta:2019zca}
Datta, A.; Kumar, J.; London, D.
\newblock {The $B$ anomalies and new physics in $b \to s e^+ e^-$}.
\newblock {\em Phys. Lett. B} {\bf 2019}, {\em 797},~134858,
\newblock
  doi:{\changeurlcolor{black}\href{https://doi.org/10.1016/j.physletb.2019.134858}{\detokenize{10.1016/j.physletb.2019.134858}}}.

\bibitem[Kumar and London(2019)]{Kumar:2019qbv}
Kumar, J.; London, D.
\newblock {New physics in $b \to s e^+ e^-$?}
\newblock {\em Phys. Rev. D} {\bf 2019}, {\em 99},~073008,
\newblock
  doi:{\changeurlcolor{black}\href{https://doi.org/10.1103/PhysRevD.99.073008}{\detokenize{10.1103/PhysRevD.99.073008}}}.

\bibitem[Hurth \em{et~al.}(2022)Hurth, Mahmoudi, Santos, and
  Neshatpour]{Hurth:2021nsi}
Hurth, T.; Mahmoudi, F.; Santos, D.M.; Neshatpour, S.
\newblock {More indications for lepton nonuniversality in $b\to s\ell^+
  \ell^-$}.
\newblock {\em Phys. Lett. B} {\bf 2022}, {\em 824},~136838,
\newblock
  doi:{\changeurlcolor{black}\href{https://doi.org/10.1016/j.physletb.2021.136838}{\detokenize{10.1016/j.physletb.2021.136838}}}.

\bibitem[London and Matias(2021)]{London:2021lfn}
London, D.; Matias, J.
\newblock {$B$ Flavour Anomalies: 2021 Theoretical Status Report}. \emph{arXiv} \textbf{2021}, arXiv:2110.13270.


\bibitem[Aaij \em{et~al.}(2021)Aaij et~al.]{LHCb:2021trn}
Aaij, R.; et~al. [LHCb Collaboration].
\newblock {Test of lepton universality in beauty-quark decays}. \emph{arXiv} {\bf 2021}, arXiv:hep-ex/2103.11769.

\bibitem[Marciano \em{et~al.}(2008)Marciano, Mori, and Roney]{Marciano:2008zz}
Marciano, W.J.; Mori, T.; Roney, J.M.
\newblock {Charged Lepton Flavor Violation Experiments}.
\newblock {\em Ann. Rev. Nucl. Part. Sci.} {\bf 2008}, {\em 58},~315--341.
\newblock
  doi:{\changeurlcolor{black}\href{https://doi.org/10.1146/annurev.nucl.58.110707.171126}{\detokenize{10.1146/annurev.nucl.58.110707.171126}}}.

\bibitem[Bernstein and Cooper(2013)]{Bernstein:2013hba}
Bernstein, R.H.; Cooper, P.S.
\newblock {Charged Lepton Flavor Violation: An Experimenter's Guide}.
\newblock {\em Phys. Rept.} {\bf 2013}, {\em 532},~27--64,
\newblock
  doi:{\changeurlcolor{black}\href{https://doi.org/10.1016/j.physrep.2013.07.002}{\detokenize{10.1016/j.physrep.2013.07.002}}}.

\bibitem[Calibbi and Signorelli(2018)]{Calibbi:2017uvl}
Calibbi, L.; Signorelli, G.
\newblock {Charged Lepton Flavour Violation: An Experimental and Theoretical
  Introduction}.
\newblock {\em Riv. Nuovo Cim.} {\bf 2018}, {\em 41},~71--174,
\newblock
  doi:{\changeurlcolor{black}\href{https://doi.org/10.1393/ncr/i2018-10144-0}{\detokenize{10.1393/ncr/i2018-10144-0}}}.

\bibitem[Heeck(2017)]{Heeck:2016xwg}
Heeck, J.
\newblock {Interpretation of Lepton Flavor Violation}.
\newblock {\em Phys. Rev. D} {\bf 2017}, {\em 95},~015022,
\newblock
  doi:{\changeurlcolor{black}\href{https://doi.org/10.1103/PhysRevD.95.015022}{\detokenize{10.1103/PhysRevD.95.015022}}}.

\bibitem[Hou and Wong(1996)]{Hou:1995dg}
Hou, W.S.; Wong, G.G.
\newblock {mu+ e- \ensuremath{<}---\ensuremath{>} mu- e+ transitions via
  neutral scalar bosons}.
\newblock {\em Phys. Rev. D} {\bf 1996}, {\em 53},~1537--1541,
\newblock
  doi:{\changeurlcolor{black}\href{https://doi.org/10.1103/PhysRevD.53.1537}{\detokenize{10.1103/PhysRevD.53.1537}}}.

\bibitem[Dev \em{et~al.}(2018)Dev, Mohapatra, and Zhang]{Dev:2017ftk}
Dev, P.S.B.; Mohapatra, R.N.; Zhang, Y.
\newblock {Lepton Flavor Violation Induced by a Neutral Scalar at Future Lepton
  Colliders}.
\newblock {\em Phys. Rev. Lett.} {\bf 2018}, {\em 120},~221804,
\newblock
  doi:{\changeurlcolor{black}\href{https://doi.org/10.1103/PhysRevLett.120.221804}{\detokenize{10.1103/PhysRevLett.120.221804}}}.

\bibitem[Li and Schmidt(2019)]{Li:2018cod}
Li, T.; Schmidt, M.A.
\newblock {Sensitivity of future lepton colliders to the search for charged
  lepton flavor violation}.
\newblock {\em Phys. Rev. D} {\bf 2019}, {\em 99},~055038,
\newblock
  doi:{\changeurlcolor{black}\href{https://doi.org/10.1103/PhysRevD.99.055038}{\detokenize{10.1103/PhysRevD.99.055038}}}.

\bibitem[Arganda \em{et~al.}(2019)Arganda, Marcano, Mileo, Morales, and
  Szynkman]{Arganda:2019gnv}
Arganda, E.; Marcano, X.; Mileo, N.I.; Morales, R.A.; Szynkman, A.
\newblock {Model-independent search strategy for the lepton-flavor-violating
  heavy Higgs boson decay to $\tau\mu$ at the LHC}.
\newblock {\em Eur. Phys. J. C} {\bf 2019}, {\em 79},~738,
\newblock
  doi:{\changeurlcolor{black}\href{https://doi.org/10.1140/epjc/s10052-019-7249-7}{\detokenize{10.1140/epjc/s10052-019-7249-7}}}.

\bibitem[Brignole and Rossi(2004)]{Brignole:2004ah}
Brignole, A.; Rossi, A.
\newblock {Anatomy and phenomenology of mu-tau lepton flavor violation in the
  MSSM}.
\newblock {\em Nucl. Phys. B} {\bf 2004}, {\em 701},~3--53,
\newblock
  doi:{\changeurlcolor{black}\href{https://doi.org/10.1016/j.nuclphysb.2004.08.037}{\detokenize{10.1016/j.nuclphysb.2004.08.037}}}.

\bibitem[Harnik \em{et~al.}(2013)Harnik, Kopp, and Zupan]{Harnik:2012pb}
Harnik, R.; Kopp, J.; Zupan, J.
\newblock {Flavor Violating Higgs Decays}.
\newblock {\em J. High Energy Phys.} {\bf 2013}, {\em 3},~026,
\newblock
  doi:{\changeurlcolor{black}\href{https://doi.org/10.1007/JHEP03(2013)026}{\detokenize{10.1007/JHEP03(2013)026}}}.

\bibitem[Blankenburg \em{et~al.}(2012)Blankenburg, Ellis, and
  Isidori]{Blankenburg:2012ex}
Blankenburg, G.; Ellis, J.; Isidori, G.
\newblock {Flavour-Changing Decays of a 125 GeV Higgs-like Particle}.
\newblock {\em Phys. Lett. B} {\bf 2012}, {\em 712},~386--390,
\newblock
  doi:{\changeurlcolor{black}\href{https://doi.org/10.1016/j.physletb.2012.05.007}{\detokenize{10.1016/j.physletb.2012.05.007}}}.

\bibitem[Banerjee \em{et~al.}(2016)Banerjee, Bhattacherjee, Mitra, and
  Spannowsky]{Banerjee:2016foh}
Banerjee, S.; Bhattacherjee, B.; Mitra, M.; Spannowsky, M.
\newblock {The Lepton Flavour Violating Higgs Decays at the HL-LHC and the
  ILC}.
\newblock {\em J. High Energy Phys.} {\bf 2016}, {\em 7},~059,
\newblock
  doi:{\changeurlcolor{black}\href{https://doi.org/10.1007/JHEP07(2016)059}{\detokenize{10.1007/JHEP07(2016)059}}}.

\bibitem[Herrero-Garcia \em{et~al.}(2016)Herrero-Garcia, Rius, and
  Santamaria]{Herrero-Garcia:2016uab}
Herrero-Garcia, J.; Rius, N.; Santamaria, A.
\newblock {Higgs lepton flavour violation: UV completions and connection to
  neutrino masses}.
\newblock {\em J. High Energy Phys.} {\bf 2016}, {\em 11},~084,
\newblock
  doi:{\changeurlcolor{black}\href{https://doi.org/10.1007/JHEP11(2016)084}{\detokenize{10.1007/JHEP11(2016)084}}}.

\bibitem[Chakraborty \em{et~al.}(2016)Chakraborty, Datta, and
  Kundu]{Chakraborty:2016gff}
Chakraborty, I.; Datta, A.; Kundu, A.
\newblock {Lepton flavor violating Higgs boson decay ${\boldsymbol{h}}
  \rightarrow \mu \tau $ at the ILC}.
\newblock {\em J. Phys. G} {\bf 2016}, {\em 43},~125001,
\newblock
  doi:{\changeurlcolor{black}\href{https://doi.org/10.1088/0954-3899/43/12/125001}{\detokenize{10.1088/0954-3899/43/12/125001}}}.

\bibitem[Chakraborty \em{et~al.}(2017)Chakraborty, Mondal, and
  Mukhopadhyaya]{Chakraborty:2017tyb}
Chakraborty, I.; Mondal, S.; Mukhopadhyaya, B.
\newblock {Lepton flavor violating Higgs boson decay at $e^+ e^-$ colliders}.
\newblock {\em Phy. Rev. D} {\bf 2017}, {\em 96},~115020,
\newblock
  doi:{\changeurlcolor{black}\href{https://doi.org/10.1103/PhysRevD.96.115020}{\detokenize{10.1103/PhysRevD.96.115020}}}.

\bibitem[Qin \em{et~al.}(2018)Qin, Li, L\"u, Yu, and Zhou]{Qin:2017aju}
Qin, Q.; Li, Q.; L\"u, C.D.; Yu, F.S.; Zhou, S.H.
\newblock {Charged lepton flavor violating Higgs decays at future $e^+e^-$
  colliders}.
\newblock {\em Eur. Phys. J. C} {\bf 2018}, {\em 78},~835,
\newblock
  doi:{\changeurlcolor{black}\href{https://doi.org/10.1140/epjc/s10052-018-6298-7}{\detokenize{10.1140/epjc/s10052-018-6298-7}}}.

\bibitem[Lu \em{et~al.}(2021)Lu, Levin, Li, Agapitos, Li, Meng, Qian, Xiao, and
  Yang]{Lu:2020dkx}
Lu, M.; Levin, A.M.; Li, C.; Agapitos, A.; Li, Q.; Meng, F.; Qian, S.; Xiao,
  J.; Yang, T.
\newblock {The physics case for an electron-muon collider}.
\newblock {\em Adv. High Energy Phys.} {\bf 2021}, {\em 2021},~6693618,
\newblock
  doi:{\changeurlcolor{black}\href{https://doi.org/10.1155/2021/6693618}{\detokenize{10.1155/2021/6693618}}}.

\bibitem[Arganda \em{et~al.}(2015)Arganda, Herrero, Marcano, and
  Weiland]{Arganda:2014dta}
Arganda, E.; Herrero, M.J.; Marcano, X.; Weiland, C.
\newblock {Imprints of massive inverse seesaw model neutrinos in lepton flavor
  violating Higgs boson decays}.
\newblock {\em Phys. Rev. D} {\bf 2015}, {\em 91},~015001,
\newblock
  doi:{\changeurlcolor{black}\href{https://doi.org/10.1103/PhysRevD.91.015001}{\detokenize{10.1103/PhysRevD.91.015001}}}.

\bibitem[Arganda \em{et~al.}(2016)Arganda, Herrero, Marcano, and
  Weiland]{Arganda:2015naa}
Arganda, E.; Herrero, M.J.; Marcano, X.; Weiland, C.
\newblock {Enhancement of the lepton flavor violating Higgs boson decay rates
  from SUSY loops in the inverse seesaw model}.
\newblock {\em Phys. Rev. D} {\bf 2016}, {\em 93},~055010,
\newblock
  doi:{\changeurlcolor{black}\href{https://doi.org/10.1103/PhysRevD.93.055010}{\detokenize{10.1103/PhysRevD.93.055010}}}.

\bibitem[Arganda \em{et~al.}(2017)Arganda, Herrero, Marcano, Morales, and
  Szynkman]{Arganda:2016zvc}
Arganda, E.; Herrero, M.J.; Marcano, X.; Morales, R.; Szynkman, A.
\newblock {Effective lepton flavor violating
  H\ensuremath{\ell}i\ensuremath{\ell}j vertex from right-handed neutrinos
  within the mass insertion approximation}.
\newblock {\em Phys. Rev. D} {\bf 2017}, {\em 95},~095029,
\newblock
  doi:{\changeurlcolor{black}\href{https://doi.org/10.1103/PhysRevD.95.095029}{\detokenize{10.1103/PhysRevD.95.095029}}}.

\bibitem[Marcano and Morales(2020)]{Marcano:2019rmk}
Marcano, X.; Morales, R.A.
\newblock {Flavor techniques for LFV processes: Higgs decays in a general
  seesaw model}.
\newblock {\em Front. Phys.} {\bf 2020}, {\em 7},~228,
\newblock
  doi:{\changeurlcolor{black}\href{https://doi.org/10.3389/fphy.2019.00228}{\detokenize{10.3389/fphy.2019.00228}}}.

\bibitem[Dev \em{et~al.}(2016)Dev, Mohapatra, and Zhang]{Dev:2016dja}
Dev, P.S.B.; Mohapatra, R.N.; Zhang, Y.
\newblock {Probing the Higgs Sector of the Minimal Left-Right Symmetric Model
  at Future Hadron Colliders}.
\newblock {\em J. High Energy Phys.} {\bf 2016}, {\em 05},~174,
\newblock
  doi:{\changeurlcolor{black}\href{https://doi.org/10.1007/JHEP05(2016)174}{\detokenize{10.1007/JHEP05(2016)174}}}.

\bibitem[Maiezza \em{et~al.}(2017)Maiezza, Senjanovi\'c, and
  Vasquez]{Maiezza:2016ybz}
Maiezza, A.; Senjanovi\'c, G.; Vasquez, J.C.
\newblock {Higgs sector of the minimal left-right symmetric theory}.
\newblock {\em Phys. Rev. D} {\bf 2017}, {\em 95},~095004,
\newblock
  doi:{\changeurlcolor{black}\href{https://doi.org/10.1103/PhysRevD.95.095004}{\detokenize{10.1103/PhysRevD.95.095004}}}.

\bibitem[Bhupal~Dev \em{et~al.}(2017)Bhupal~Dev, Mohapatra, and
  Zhang]{BhupalDev:2016nfr}
Bhupal~Dev, P.S.; Mohapatra, R.N.; Zhang, Y.
\newblock {Displaced photon signal from a possible light scalar in minimal
  left-right seesaw model}.
\newblock {\em Phys. Rev. D} {\bf 2017}, {\em 95},~115001,
\newblock
  doi:{\changeurlcolor{black}\href{https://doi.org/10.1103/PhysRevD.95.115001}{\detokenize{10.1103/PhysRevD.95.115001}}}.

\bibitem[Dev \em{et~al.}(2017)Dev, Mohapatra, and Zhang]{Dev:2017dui}
Dev, P.S.B.; Mohapatra, R.N.; Zhang, Y.
\newblock {Long Lived Light Scalars as Probe of Low Scale Seesaw Models}.
\newblock {\em Nucl. Phys. B} {\bf 2017}, {\em 923},~179--221,
\newblock
  doi:{\changeurlcolor{black}\href{https://doi.org/10.1016/j.nuclphysb.2017.07.021}{\detokenize{10.1016/j.nuclphysb.2017.07.021}}}.

\bibitem[Branco \em{et~al.}(2012)Branco, Ferreira, Lavoura, Rebelo, Sher, and
  Silva]{Branco:2011iw}
Branco, G.C.; Ferreira, P.M.; Lavoura, L.; Rebelo, M.N.; Sher, M.; Silva, J.P.
\newblock {Theory and phenomenology of two-Higgs-doublet models}.
\newblock {\em Phys. Rept.} {\bf 2012}, {\em 516},~1--102,
\newblock
  doi:{\changeurlcolor{black}\href{https://doi.org/10.1016/j.physrep.2012.02.002}{\detokenize{10.1016/j.physrep.2012.02.002}}}.

\bibitem[Crivellin \em{et~al.}(2013)Crivellin, Kokulu, and
  Greub]{Crivellin:2013wna}
Crivellin, A.; Kokulu, A.; Greub, C.
\newblock {Flavor-phenomenology of two-Higgs-doublet models with generic Yukawa
  structure}.
\newblock {\em Phys. Rev. D} {\bf 2013}, {\em 87},~094031,
\newblock
  doi:{\changeurlcolor{black}\href{https://doi.org/10.1103/PhysRevD.87.094031}{\detokenize{10.1103/PhysRevD.87.094031}}}.

\bibitem[Crivellin \em{et~al.}(2016)Crivellin, Heeck, and
  Stoffer]{Crivellin:2015hha}
Crivellin, A.; Heeck, J.; Stoffer, P.
\newblock {A perturbed lepton-specific two-Higgs-doublet model facing
  experimental hints for physics beyond the Standard Model}.
\newblock {\em Phys. Rev. Lett.} {\bf 2016}, {\em 116},~081801,
\newblock
  doi:{\changeurlcolor{black}\href{https://doi.org/10.1103/PhysRevLett.116.081801}{\detokenize{10.1103/PhysRevLett.116.081801}}}.

\bibitem[Aulakh and Mohapatra(1982)]{Aulakh:1982yn}
Aulakh, C.S.; Mohapatra, R.N.
\newblock {Neutrino as the Supersymmetric Partner of the Majoron}.
\newblock {\em Phys. Lett. B} {\bf 1982}, {\em 119},~136--140.
\newblock
  doi:{\changeurlcolor{black}\href{https://doi.org/10.1016/0370-2693(82)90262-3}{\detokenize{10.1016/0370-2693(82)90262-3}}}.

\bibitem[Hall and Suzuki(1984)]{Hall:1983id}
Hall, L.J.; Suzuki, M.
\newblock {Explicit R-Parity Breaking in Supersymmetric Models}.
\newblock {\em Nucl. Phys. B} {\bf 1984}, {\em 231},~419--444.
\newblock
  doi:{\changeurlcolor{black}\href{https://doi.org/10.1016/0550-3213(84)90513-3}{\detokenize{10.1016/0550-3213(84)90513-3}}}.

\bibitem[Ross and Valle(1985)]{Ross:1984yg}
Ross, G.G.; Valle, J.W.F.
\newblock {Supersymmetric Models Without R-Parity}.
\newblock {\em Phys. Lett. B} {\bf 1985}, {\em 151},~375--381.
\newblock
  doi:{\changeurlcolor{black}\href{https://doi.org/10.1016/0370-2693(85)91658-2}{\detokenize{10.1016/0370-2693(85)91658-2}}}.

\bibitem[Barbier \em{et~al.}(2005)Barbier et~al.]{Barbier:2004ez}
Barbier, R.;  Berat, C.;  Besancon, M.;  Chemtob, M.;  Deandrea, A.; Dudas, E.;  Fayet, P.; Lavignac, S.; Moreau, G.;  Perez, E.; et~al.
\newblock {R-parity violating supersymmetry}.
\newblock {\em Phys. Rept.} {\bf 2005}, {\em 420},~1--202,
\newblock
  doi:{\changeurlcolor{black}\href{https://doi.org/10.1016/j.physrep.2005.08.006}{\detokenize{10.1016/j.physrep.2005.08.006}}}.

\bibitem[Hung(2007)]{Hung:2006ap}
Hung, P.Q.
\newblock {A Model of electroweak-scale right-handed neutrino mass}.
\newblock {\em Phys. Lett. B} {\bf 2007}, {\em 649},~275--279, 10.1016/j.physletb.\linebreak 2007.03.067.

\bibitem[Bu \em{et~al.}(2008)Bu, Liao, and Liu]{Bu:2008fx}
Bu, J.P.; Liao, Y.; Liu, J.Y.
\newblock {Lepton Flavor Violating Muon Decays in a Model of Electroweak-Scale
  Right-Handed Neutrinos}.
\newblock {\em Phys. Lett. B} {\bf 2008}, {\em 665},~39--43,
\newblock
  doi:{\changeurlcolor{black}\href{https://doi.org/10.1016/j.physletb.2008.05.059}{\detokenize{10.1016/j.physletb.2008.05.059}}}.

\bibitem[Chang \em{et~al.}(2016)Chang, Chang, Nugroho, and Yuan]{Chang:2016ave}
Chang, C.F.; Chang, C.H.V.; Nugroho, C.S.; Yuan, T.C.
\newblock {Lepton Flavor Violating Decays of Neutral Higgses in Extended Mirror
  Fermion Model}.
\newblock {\em Nucl. Phys. B} {\bf 2016}, {\em 910},~293--308,
\newblock
  doi:{\changeurlcolor{black}\href{https://doi.org/10.1016/j.nuclphysb.2016.07.009}{\detokenize{10.1016/j.nuclphysb.2016.07.009}}}.

\bibitem[Hung \em{et~al.}(2018)Hung, Le, Tran, and Yuan]{Hung:2017voe}
Hung, P.Q.; Le, T.; Tran, V.Q.; Yuan, T.C.
\newblock {Muon-to-Electron Conversion in Mirror Fermion Model with Electroweak
  Scale Non-Sterile Right-handed Neutrinos}.
\newblock {\em Nucl. Phys. B} {\bf 2018}, {\em 932},~471--504,
\newblock
  doi:{\changeurlcolor{black}\href{https://doi.org/10.1016/j.nuclphysb.2018.05.020}{\detokenize{10.1016/j.nuclphysb.2018.05.020}}}.

\bibitem[Bae(2013)]{Baer:2013cma}
{The International Linear Collider Technical Design Report---Volume 2: Physics}. \emph{arXiv} {\bf 2013}, arXiv:hep-ph/1306.6352.


\bibitem[Ahmad \em{et~al.}(2015)Ahmad et~al.]{CEPC-SPPCStudyGroup:2015csa}
Ahmad, M.;  Alves, D.;  An, H.; An, Q.; Arhrib, A.; Arkani-Hamed, N.; Ahmed, I.; Bai, Y.;  Ferroli, R.B.; Ban, Y.; et~al.
\newblock {CEPC-SPPC Preliminary Conceptual Design Report. 1. Physics and
  Detector. 2015. Available online: \url{http://cepc.ihep.ac.cn/preCDR/main_preCDR.pdf} (accessed on {1 March 2015}). 
} 

\bibitem[Bicer \em{et~al.}(2014)Bicer
  et~al.]{TLEPDesignStudyWorkingGroup:2013myl}
Bicer, M.;  et~al. [TLEP Design Study Working Group] 
\newblock {First Look at the Physics Case of TLEP}.
\newblock {\em J. High Energy Phys.} {\bf 2014}, {\em 1},~164,
\newblock
  doi:{\changeurlcolor{black}\href{https://doi.org/10.1007/JHEP01(2014)164}{\detokenize{10.1007/JHEP01(2014)164}}}.

\bibitem[Accomando \em{et~al.}(2004)Accomando
  et~al.]{CLICPhysicsWorkingGroup:2004qvu}
Accomando, E.; et~al. [CLIC Physics Working Group]
\newblock {Physics at the CLIC multi-TeV linear collider}.
\newblock  {In Proceedings of the 11th International Conference on Hadron Spectroscopy},  Rio de Janeiro, Brazil, 10 June 2005; doi:{\changeurlcolor{black}\href{https://doi.org/10.5170/CERN-2004-005}{\detokenize{10.5170/CERN-2004-005}}}.

\bibitem[Willmann \em{et~al.}(1999)Willmann et~al.]{Willmann:1998gd}
Willmann, L.;  Schmidt, P.V.; Wirtz, H.P.; Abela, R.; Baranov, V.; Bagaturia, J.; Bertl, W.H.; Engfer, R.; Grossmann, A.; Hughes, V.W.; et~al.
\newblock {New bounds from searching for muonium to anti-muonium conversion}.
\newblock {\em Phys. Rev. Lett.} {\bf 1999}, {\em 82},~49--52,
\newblock
  doi:{\changeurlcolor{black}\href{https://doi.org/10.1103/PhysRevLett.82.49}{\detokenize{10.1103/PhysRevLett.82.49}}}.

\bibitem[Mohr \em{et~al.}(2016)Mohr, Newell, and Taylor]{Mohr:2015ccw}
Mohr, P.J.; Newell, D.B.; Taylor, B.N.
\newblock {CODATA Recommended Values of the Fundamental Physical Constants:
  2014}.
\newblock {\em Rev. Mod. Phys.} {\bf 2016}, {\em 88},~035009,
\newblock
  doi:{\changeurlcolor{black}\href{https://doi.org/10.1103/RevModPhys.88.035009}{\detokenize{10.1103/RevModPhys.88.035009}}}.

\bibitem[Abdallah \em{et~al.}(2006)Abdallah et~al.]{DELPHI:2005wxt}
Abdallah, J.;  et~al. [DELPHI Collaboration], 
\newblock {Measurement and interpretation of fermion-pair production at LEP
  energies above the Z resonance}.
\newblock {\em Eur. Phys. J. C} {\bf 2006}, {\em 45},~589--632,
\newblock
  doi:{\changeurlcolor{black}\href{https://doi.org/10.1140/epjc/s2005-02461-0}{\detokenize{10.1140/epjc/s2005-02461-0}}}.

\bibitem[Bhupal~Dev \em{et~al.}(2018)Bhupal~Dev, Mohapatra, and
  Zhang]{BhupalDev:2018vpr}
Bhupal~Dev, P.S.; Mohapatra, R.N.; Zhang, Y.
\newblock {Probing TeV scale origin of neutrino mass at future lepton colliders
  via neutral and doubly-charged scalars}.
\newblock {\em Phys. Rev. D} {\bf 2018}, {\em 98},~075028,
\newblock
  doi:{\changeurlcolor{black}\href{https://doi.org/10.1103/PhysRevD.98.075028}{\detokenize{10.1103/PhysRevD.98.075028}}}.

\bibitem[Bennett \em{et~al.}(2006)Bennett et~al.]{Muong-2:2006rrc}
Bennett, G.W.;  et~al. [Muon g-2 Collaboration]
\newblock {Final Report of the Muon E821 Anomalous Magnetic Moment Measurement
  at BNL}.
\newblock {\em Phys. Rev. D} {\bf 2006}, {\em 73},~072003,
\newblock
  doi:{\changeurlcolor{black}\href{https://doi.org/10.1103/PhysRevD.73.072003}{\detokenize{10.1103/PhysRevD.73.072003}}}.

\bibitem[Abi \em{et~al.}(2021)Abi et~al.]{Muong-2:2021ojo}
Abi, B.;  et~al. [Muon g-2 Collaboration] 
\newblock {Measurement of the Positive Muon Anomalous Magnetic Moment to 0.46
  ppm}.
\newblock {\em Phys. Rev. Lett.} {\bf 2021}, {\em 126},~141801,
\newblock
  doi:{\changeurlcolor{black}\href{https://doi.org/10.1103/PhysRevLett.126.141801}{\detokenize{10.1103/PhysRevLett.126.141801}}}.

\bibitem[Bellgardt \em{et~al.}(1988)Bellgardt et~al.]{SINDRUM:1987nra}
Bellgardt, U.;  et~al. [SINDRUM Collaboration]
\newblock {Search for the Decay mu+ ---\ensuremath{>} $e^{+} e^{+} e^{-}$}.
\newblock {\em Nucl. Phys. B} {\bf 1988}, {\em 299},~1--6.
\newblock
  doi:{\changeurlcolor{black}\href{https://doi.org/10.1016/0550-3213(88)90462-2}{\detokenize{10.1016/0550-3213(88)90462-2}}}.

\bibitem[Kabachenko and Pirogov(1998)]{Kabachenko:1997aw}
Kabachenko, V.V.; Pirogov, Y.F.
\newblock {Studying lepton family violation in lepton lepton collisions}.
\newblock {\em Eur. Phys. J. C} {\bf 1998}, {\em 4},~525--532,
\newblock
  doi:{\changeurlcolor{black}\href{https://doi.org/10.1007/s100520050225}{\detokenize{10.1007/s100520050225}}}.

\bibitem[Cho and Shimo(2017)]{Cho:2016zqo}
Cho, G.C.; Shimo, H.
\newblock {Search for lepton flavor violation at future lepton colliders}.
\newblock {\em Mod. Phys. A} {\bf 2017}, {\em 32},~1750127,
\newblock
  doi:{\changeurlcolor{black}\href{https://doi.org/10.1142/S0217732317501279}{\detokenize{10.1142/S0217732317501279}}}.

\bibitem[Ferreira \em{et~al.}(2007)Ferreira, Guedes, and
  Santos]{Ferreira:2006dg}
Ferreira, P.M.; Guedes, R.B.; Santos, R.
\newblock {Lepton flavour violating processes at the International Linear
  Collider}.
\newblock {\em Phys. Rev. D} {\bf 2007}, {\em 75},~055015,
\newblock
  doi:{\changeurlcolor{black}\href{https://doi.org/10.1103/PhysRevD.75.055015}{\detokenize{10.1103/PhysRevD.75.055015}}}.

\bibitem[Aranda \em{et~al.}(2009)Aranda, Flores-Tlalpa, Ramirez-Zavaleta,
  Tlachino, Toscano, and Tututi]{Aranda:2009kz}
Aranda, J.I.; Flores-Tlalpa, A.; Ramirez-Zavaleta, F.; Tlachino, F.J.; Toscano,
  J.J.; Tututi, E.S.
\newblock {Effective Lagrangian description of Higgs mediated flavor violating
  electromagnetic transitions: Implications on lepton flavor violation}.
\newblock {\em Phys. Rev. D} {\bf 2009}, {\em 79},~093009,
\newblock
  doi:{\changeurlcolor{black}\href{https://doi.org/10.1103/PhysRevD.79.093009}{\detokenize{10.1103/PhysRevD.79.093009}}}.

\bibitem[Murakami and Tait(2015)]{Murakami:2014tna}
Murakami, B.; Tait, T.M.P.
\newblock {Searching for lepton flavor violation at a future high energy e$^{+}$e$^{-}$
  collider}.
\newblock {\em Phys. Rev. D} {\bf 2015}, {\em 91},~015002,
\newblock
  doi:{\changeurlcolor{black}\href{https://doi.org/10.1103/PhysRevD.91.015002}{\detokenize{10.1103/PhysRevD.91.015002}}}.

\bibitem[Calibbi \em{et~al.}(2021)Calibbi, Marcano, and Roy]{Calibbi:2021pyh}
Calibbi, L.; Marcano, X.; Roy, J.
\newblock {Z lepton flavour violation as a probe for new physics at future
  $e^+e^-$ colliders}.
\newblock {\em Eur. Phys. J. C} {\bf 2021}, {\em 81},~1054,
\newblock
  doi:{\changeurlcolor{black}\href{https://doi.org/10.1140/epjc/s10052-021-09777-3}{\detokenize{10.1140/epjc/s10052-021-09777-3}}}.

\bibitem[Hayasaka \em{et~al.}(2010)Hayasaka et~al.]{Hayasaka:2010np}
Hayasaka, K.;  et~al. [Belle collaboration]
\newblock {Search for Lepton Flavor Violating Tau Decays into Three Leptons
  with 719 Million Produced Tau+Tau- Pairs}.
\newblock {\em Phys. Lett. B} {\bf 2010}, {\em 687},~139--143,
\newblock
  doi:{\changeurlcolor{black}\href{https://doi.org/10.1016/j.physletb.2010.03.037}{\detokenize{10.1016/j.physletb.2010.03.037}}}.

\bibitem[Ginzburg \em{et~al.}(1983)Ginzburg, Kotkin, Serbo, and
  Telnov]{Ginzburg:1981vm}
Ginzburg, I.F.; Kotkin, G.L.; Serbo, V.G.; Telnov, V.I.
\newblock {Colliding gamma e and gamma gamma Beams Based on the Single Pass
  Accelerators (of Vlepp Type)}.
\newblock {\em Nucl. Instrum. Meth.} {\bf 1983}, {\em 205},~47--68.
\newblock
  doi:{\changeurlcolor{black}\href{https://doi.org/10.1016/0167-5087(83)90173-4}{\detokenize{10.1016/0167-5087(83)90173-4}}}.

\bibitem[Ginzburg \em{et~al.}(1984)Ginzburg, Kotkin, Panfil, Serbo, and
  Telnov]{Ginzburg:1982yr}
Ginzburg, I.F.; Kotkin, G.L.; Panfil, S.L.; Serbo, V.G.; Telnov, V.I.
\newblock {Colliding gamma e and gamma gamma Beams Based on the Single Pass $e^{+}
  e^{-}$ Accelerators. 2. Polarization Effects. Monochromatization Improvement}.
\newblock {\em Nucl. Instrum. Meth. A} {\bf 1984}, {\em 219},~5--24.
\newblock
  doi:{\changeurlcolor{black}\href{https://doi.org/10.1016/0167-5087(84)90128-5}{\detokenize{10.1016/0167-5087(84)90128-5}}}.

\bibitem[Telnov(1990)]{Telnov:1989sd}
Telnov, V.I.
\newblock {Problems of Obtaining $\gamma \gamma$ and $\gamma \epsilon$
  Colliding Beams at Linear Colliders}.
\newblock {\em Nucl. Instrum. Meth. A} {\bf 1990}, {\em 294},~72--92.
\newblock
  doi:{\changeurlcolor{black}\href{https://doi.org/10.1016/0168-9002(90)91826-W}{\detokenize{10.1016/0168-9002(90)91826-W}}}.

\bibitem[Capdevilla \em{et~al.}(2021)Capdevilla, Curtin, Kahn, and
  Krnjaic]{Capdevilla:2020qel}
Capdevilla, R.; Curtin, D.; Kahn, Y.; Krnjaic, G.
\newblock {Discovering the physics of $(g-2)_\mu$ at future muon colliders}.
\newblock {\em Phys. Rev. D} {\bf 2021}, {\em 103},~075028,
\newblock
  doi:{\changeurlcolor{black}\href{https://doi.org/10.1103/PhysRevD.103.075028}{\detokenize{10.1103/PhysRevD.103.075028}}}.

\bibitem[Iguro \em{et~al.}(2020)Iguro, Mohan, and Yuan]{Iguro:2020qbc}
Iguro, S.; Mohan, K.A.; Yuan, C.P.
\newblock {Detecting a ${\mu}{\tau}$-philic $Z$' boson via photon initiated
  processes at the LHC}.
\newblock {\em Phys. Rev. D} {\bf 2020}, {\em 101},~075011,
\newblock
  doi:{\changeurlcolor{black}\href{https://doi.org/10.1103/PhysRevD.101.075011}{\detokenize{10.1103/PhysRevD.101.075011}}}.

\bibitem[Magg and Wetterich(1980)]{Magg:1980ut}
Magg, M.; Wetterich, C.
\newblock {Neutrino Mass Problem and Gauge Hierarchy}.
\newblock {\em Phys. Lett. B} {\bf 1980}, {\em 94},~61--64.
\newblock
  doi:{\changeurlcolor{black}\href{https://doi.org/10.1016/0370-2693(80)90825-4}{\detokenize{10.1016/0370-2693(80)90825-4}}}.

\bibitem[Schechter and Valle(1980)]{Schechter:1980gr}
Schechter, J.; Valle, J.W.F.
\newblock {Neutrino Masses in SU(2) x U(1) Theories}.
\newblock {\em Phys. Rev. D} {\bf 1980}, {\em 22},~2227.
\newblock
  doi:{\changeurlcolor{black}\href{https://doi.org/10.1103/PhysRevD.22.2227}{\detokenize{10.1103/PhysRevD.22.2227}}}.

\bibitem[Cheng and Li(1980)]{Cheng:1980qt}
Cheng, T.P.; Li, L.F.
\newblock {Neutrino Masses, Mixings and Oscillations in SU(2) x U(1) Models of
  Electroweak Interactions}.
\newblock {\em Phys. Rev. D} {\bf 1980}, {\em 22},~2860.
\newblock
  doi:{\changeurlcolor{black}\href{https://doi.org/10.1103/PhysRevD.22.2860}{\detokenize{10.1103/PhysRevD.22.2860}}}.

\bibitem[Lazarides \em{et~al.}(1981)Lazarides, Shafi, and
  Wetterich]{Lazarides:1980nt}
Lazarides, G.; Shafi, Q.; Wetterich, C.
\newblock {Proton Lifetime and Fermion Masses in an SO(10) Model}.
\newblock {\em Nucl. Phys. B} {\bf 1981}, {\em 181},~287--300.
\newblock
  doi:{\changeurlcolor{black}\href{https://doi.org/10.1016/0550-3213(81)90354-0}{\detokenize{10.1016/0550-3213(81)90354-0}}}.

\bibitem[Mohapatra and Senjanovic(1981)]{Mohapatra:1980yp}
Mohapatra, R.N.; Senjanovic, G.
\newblock {Neutrino Masses and Mixings in Gauge Models with Spontaneous Parity
  Violation}.
\newblock {\em Phys. Rev. D} {\bf 1981}, {\em 23},~165.
\newblock
  doi:{\changeurlcolor{black}\href{https://doi.org/10.1103/PhysRevD.23.165}{\detokenize{10.1103/PhysRevD.23.165}}}.

\bibitem[Pati and Salam(1974)]{Pati:1974yy}
Pati, J.C.; Salam, A.
\newblock {Lepton Number as the Fourth Color}.
\newblock {\em Phys. Rev. D} {\bf 1974}, {\em 10},~275--289.
\newblock Erratum: \emph{Phys. Rev. D} \textbf{1975}, \emph{11}, 703--703, 
  doi:{\changeurlcolor{black}\href{https://doi.org/10.1103/PhysRevD.10.275}{\detokenize{10.1103/PhysRevD.10.275}}}.

\bibitem[Mohapatra and Pati(1975)]{Mohapatra:1974gc}
Mohapatra, R.N.; Pati, J.C.
\newblock {A Natural Left-Right Symmetry}.
\newblock {\em Phys. Rev. D} {\bf 1975}, {\em 11},~2558.
\newblock
  doi:{\changeurlcolor{black}\href{https://doi.org/10.1103/PhysRevD.11.2558}{\detokenize{10.1103/PhysRevD.11.2558}}}.

\bibitem[Senjanovic and Mohapatra(1975)]{Senjanovic:1975rk}
Senjanovic, G.; Mohapatra, R.N.
\newblock {Exact Left-Right Symmetry and Spontaneous Violation of Parity}.
\newblock {\em Phys. Rev. D} {\bf 1975}, {\em 12},~1502.
\newblock
  doi:{\changeurlcolor{black}\href{https://doi.org/10.1103/PhysRevD.12.1502}{\detokenize{10.1103/PhysRevD.12.1502}}}.

\bibitem[Babu(1988)]{Babu:1988ki}
Babu, K.S.
\newblock {Model of 'Calculable' Majorana Neutrino Masses}.
\newblock {\em Phys. Lett. B} {\bf 1988}, {\em 203},~132--136.
\newblock
  doi:{\changeurlcolor{black}\href{https://doi.org/10.1016/0370-2693(88)91584-5}{\detokenize{10.1016/0370-2693(88)91584-5}}}.

\bibitem[Crivellin \em{et~al.}(2019)Crivellin, Ghezzi, Panizzi, Pruna, and
  Signer]{Crivellin:2018ahj}
Crivellin, A.; Ghezzi, M.; Panizzi, L.; Pruna, G.M.; Signer, A.
\newblock {Low- and high-energy phenomenology of a doubly charged scalar}.
\newblock {\em Phys. Rev. D} {\bf 2019}, {\em 99},~035004,
\newblock
  doi:{\changeurlcolor{black}\href{https://doi.org/10.1103/PhysRevD.99.035004}{\detokenize{10.1103/PhysRevD.99.035004}}}.

\bibitem[Aaboud \em{et~al.}(2018)Aaboud et~al.]{ATLAS:2017xqs}
Aaboud, M.;  et~al. [ATLAS Collaboration]
\newblock {Search for doubly charged Higgs boson production in multi-lepton 
  final states with the ATLAS detector using proton\textendash{}proton
  collisions at $\sqrt{s}=13\,\text {TeV}$}.
\newblock {\em Eur. Phys. J. C} {\bf 2018}, {\em 78},~199,
\newblock
  doi:{\changeurlcolor{black}\href{https://doi.org/10.1140/epjc/s10052-018-5661-z}{\detokenize{10.1140/epjc/s10052-018-5661-z}}}.

\bibitem[CMS(2017)]{CMS:2017pet}
{CMS Collaboration. A search for doubly-charged Higgs boson production in three and four lepton
  final states at $\sqrt{s}=13~\mathrm{TeV}$}, CMS-PAS-HIG-16-036. 2017. Available online: \url{http://cds.cern.ch/record/2242956} (accessed on 25 January 2017). 

\bibitem[Aaboud \em{et~al.}(2019)Aaboud et~al.]{ATLAS:2018ceg}
Aaboud, M.;  et~al. [ATLAS Collaboration]
\newblock {Search for doubly charged scalar bosons decaying into same-sign $W$
  boson pairs with the ATLAS detector}.
\newblock {\em Eur. Phys. J. C} {\bf 2019}, {\em 79},~58,
\newblock
  doi:{\changeurlcolor{black}\href{https://doi.org/10.1140/epjc/s10052-018-6500-y}{\detokenize{10.1140/epjc/s10052-018-6500-y}}}.

\bibitem[Aad \em{et~al.}(2021)Aad et~al.]{ATLAS:2021jol}
Aad, G.;  et~al. [ATLAS Collaboration]
\newblock {Search for doubly and singly charged Higgs bosons decaying into
  vector bosons in multi-lepton final states with the ATLAS detector using
  proton-proton collisions at $\sqrt{s}$ = 13 TeV}.
\newblock {\em J. High Energy Phys.} {\bf 2021}, {\em 6},~146,
\newblock
  doi:{\changeurlcolor{black}\href{https://doi.org/10.1007/JHEP06(2021)146}{\detokenize{10.1007/JHEP06(2021)146}}}.

\bibitem[Rizzo(1982)]{Rizzo:1981dla}
Rizzo, T.G.
\newblock {Doubly Charged Higgs Bosons and Lepton Number Violating Processes}.
\newblock {\em Phys. Rev. D} {\bf 1982}, {\em 25},~1355--1364.
\newblock Addendum: \emph{Phys. Rev. D} \textbf{1983}, \emph{27}, 657--659,
  doi:{\changeurlcolor{black}\href{https://doi.org/10.1103/PhysRevD.27.657}{\detokenize{10.1103/PhysRevD.27.657}}}.

\bibitem[Akeroyd and Aoki(2005)]{Akeroyd:2005gt}
Akeroyd, A.G.; Aoki, M.
\newblock {Single and pair production of doubly charged Higgs bosons at hadron
  colliders}.
\newblock {\em Phys. Rev. D} {\bf 2005}, {\em 72},~035011,
\newblock
  doi:{\changeurlcolor{black}\href{https://doi.org/10.1103/PhysRevD.72.035011}{\detokenize{10.1103/PhysRevD.72.035011}}}.

\bibitem[Fileviez~Perez \em{et~al.}(2008)Fileviez~Perez, Han, Huang, Li, and
  Wang]{FileviezPerez:2008jbu}
Fileviez~Perez, P.; Han, T.; Huang, G.y.; Li, T.; Wang, K.
\newblock {Neutrino Masses and the CERN LHC: Testing Type II Seesaw}.
\newblock {\em Phys. Rev. D} {\bf 2008}, {\em 78},~015018,
\newblock
  doi:{\changeurlcolor{black}\href{https://doi.org/10.1103/PhysRevD.78.015018}{\detokenize{10.1103/PhysRevD.78.015018}}}.

\bibitem[Ferreira \em{et~al.}(2019)Ferreira, de~Melo, Kovalenko, Pinheiro, and
  Queiroz]{Ferreira:2019qpf}
Ferreira, M.M.; de~Melo, T.B.; Kovalenko, S.; Pinheiro, P.R.D.; Queiroz, F.S.
\newblock {Lepton Flavor Violation and Collider Searches in a Type I + II
  Seesaw Model}.
\newblock {\em Eur. Phys. J. C} {\bf 2019}, {\em 79},~955,
\newblock
  doi:{\changeurlcolor{black}\href{https://doi.org/10.1140/epjc/s10052-019-7422-z}{\detokenize{10.1140/epjc/s10052-019-7422-z}}}.

\bibitem[de~Melo \em{et~al.}(2019)de~Melo, Queiroz, and
  Villamizar]{deMelo:2019asm}
de~Melo, T.B.; Queiroz, F.S.; Villamizar, Y.
\newblock {Doubly Charged Scalar at the High-Luminosity and High-Energy LHC}.
\newblock {\em Int. J. Mod. Phys. A} {\bf 2019}, {\em 34},~1950157,
\newblock
  doi:{\changeurlcolor{black}\href{https://doi.org/10.1142/S0217751X19501574}{\detokenize{10.1142/S0217751X19501574}}}.

\bibitem[Padhan \em{et~al.}(2020)Padhan, Das, Mitra, and
  Kumar~Nayak]{Padhan:2019jlc}
Padhan, R.; Das, D.; Mitra, M.; Kumar~Nayak, A.
\newblock {Probing doubly and singly charged Higgs bosons at the $pp$ collider
  HE-LHC}.
\newblock {\em Phys. Rev. D} {\bf 2020}, {\em 101},~075050,
\newblock
  doi:{\changeurlcolor{black}\href{https://doi.org/10.1103/PhysRevD.101.075050}{\detokenize{10.1103/PhysRevD.101.075050}}}.

\bibitem[Fuks \em{et~al.}(2020)Fuks, Nemev\v{s}ek, and Ruiz]{Fuks:2019clu}
Fuks, B.; Nemev\v{s}ek, M.; Ruiz, R.
\newblock {Doubly Charged Higgs Boson Production at Hadron Colliders}.
\newblock {\em Phys. Rev. D} {\bf 2020}, {\em 101},~075022,
\newblock
  doi:{\changeurlcolor{black}\href{https://doi.org/10.1103/PhysRevD.101.075022}{\detokenize{10.1103/PhysRevD.101.075022}}}.

\bibitem[Gluza \em{et~al.}(2021)Gluza, Kordiaczynska, and
  Srivastava]{Gluza:2020qrt}
Gluza, J.; Kordiaczynska, M.; Srivastava, T.
\newblock {Discriminating the HTM and MLRSM models in collider studies via
  doubly charged Higgs boson pair production and the subsequent leptonic
  decays}.
\newblock {\em Chin. Phys. C} {\bf 2021}, {\em 45},~073113,
\newblock
  doi:{\changeurlcolor{black}\href{https://doi.org/10.1088/1674-1137/abfe51}{\detokenize{10.1088/1674-1137/abfe51}}}.

\bibitem[Ashanujjaman and Ghosh(2021)]{Ashanujjaman:2021txz}
Ashanujjaman, S.; Ghosh, K.
\newblock {Revisiting Type-II see-saw: Present Limits and Future Prospects at
  LHC} \emph{arXiv} {\bf 2021}, arXiv:hep-ph/2108.10952.

\bibitem[Lusignoli and Petrarca(1989)]{Lusignoli:1989tr}
Lusignoli, M.; Petrarca, S.
\newblock {EXOTIC HIGGS PRODUCTION AT E+ E- COLLIDERS}.
\newblock {\em Phys. Lett. B} {\bf 1989}, {\em 226},~397--400.
\newblock
  doi:{\changeurlcolor{black}\href{https://doi.org/10.1016/0370-2693(89)91218-5}{\detokenize{10.1016/0370-2693(89)91218-5}}}.

\bibitem[Barenboim \em{et~al.}(1997)Barenboim, Huitu, Maalampi, and
  Raidal]{Barenboim:1996pt}
Barenboim, G.; Huitu, K.; Maalampi, J.; Raidal, M.
\newblock {Constraints on doubly charged Higgs interactions at linear
  collider}.
\newblock {\em Phys. Lett. B} {\bf 1997}, {\em 394},~132--138,
\newblock
  doi:{\changeurlcolor{black}\href{https://doi.org/10.1016/S0370-2693(96)01670-X}{\detokenize{10.1016/S0370-2693(96)01670-X}}}.

\bibitem[Kuze and Sirois(2003)]{Kuze:2002vb}
Kuze, M.; Sirois, Y.
\newblock {Search for particles and forces beyond the standard model at HERA ep
  and Tevatron $p \bar{p}$ colliders}.
\newblock {\em Prog. Part. Nucl. Phys.} {\bf 2003}, {\em 50},~1--62,
\newblock Erratum: \emph{Prog. Part. Nucl. Phys.} \textbf{2004}, \emph{53}, 583--677,
  doi:{\changeurlcolor{black}\href{https://doi.org/10.1016/j.ppnp.2004.03.001}{\detokenize{10.1016/j.ppnp.2004.03.001}}}.

\bibitem[Yue and Zhao(2007)]{Yue:2007kv}
Yue, C.X.; Zhao, S.
\newblock {Lepton flavor violating signals of a little Higgs model at the high
  energy linear $e^{+} e^{-}$ colliders}.
\newblock {\em Eur. Phys. J. C} {\bf 2007}, {\em 50},~897--903,
\newblock
  doi:{\changeurlcolor{black}\href{https://doi.org/10.1140/epjc/s10052-007-0234-6}{\detokenize{10.1140/epjc/s10052-007-0234-6}}}.

\bibitem[Yue \em{et~al.}(2007)Yue, Zhao, and Ma]{Yue:2007ym}
Yue, C.X.; Zhao, S.; Ma, W.
\newblock {Single production of the doubly charged scalar in the littlest Higgs
  model}.
\newblock {\em Nucl. Phys. B} {\bf 2007}, {\em 784},~36--48,
\newblock
  doi:{\changeurlcolor{black}\href{https://doi.org/10.1016/j.nuclphysb.2007.06.003}{\detokenize{10.1016/j.nuclphysb.2007.06.003}}}.


\bibitem[Delahaye \em{et~al.}(2019)Delahaye, Diemoz, Long, Mansouli\'e,
  Pastrone, Rivkin, Schulte, Skrinsky, and Wulzer]{Delahaye:2019omf}
Delahaye, J.P.; Diemoz, M.; Long, K.; Mansouli\'e, B.; Pastrone, N.; Rivkin,
  L.; Schulte, D.; Skrinsky, A.; Wulzer, A.
\newblock {Muon Colliders} \emph{arXiv} {\bf 2019}, arXiv:physics.acc-ph/1901.06150.

\bibitem[Bhupal~Dev and Zhang(2018)]{BhupalDev:2018tox}
Bhupal~Dev, P.S.; Zhang, Y.
\newblock {Displaced vertex signatures of doubly charged scalars in the type-II
  seesaw and its left-right extensions}.
\newblock {\em J. High Energy Phys.} {\bf 2018}, {\em 10},~199,
\newblock
  doi:{\changeurlcolor{black}\href{https://doi.org/10.1007/JHEP10(2018)199}{\detokenize{10.1007/JHEP10(2018)199}}}.

\bibitem[Arkani-Hamed \em{et~al.}(2016)Arkani-Hamed, Han, Mangano, and
  Wang]{Arkani-Hamed:2015vfh}
Arkani-Hamed, N.; Han, T.; Mangano, M.; Wang, L.T.
\newblock {Physics opportunities of a 100 TeV proton\textendash{}proton
  collider}.
\newblock {\em Phys. Rept.} {\bf 2016}, {\em 652},~1--49,
\newblock
  doi:{\changeurlcolor{black}\href{https://doi.org/10.1016/j.physrep.2016.07.004}{\detokenize{10.1016/j.physrep.2016.07.004}}}.

\bibitem[Minkowski(1977)]{Minkowski:1977sc}
Minkowski, P.
\newblock {$\mu \to e\gamma$ at a Rate of One Out of $10^{9}$ Muon Decays?}
\newblock {\em Phys. Lett. B} {\bf 1977}, {\em 67},~421--428.
\newblock
  doi:{\changeurlcolor{black}\href{https://doi.org/10.1016/0370-2693(77)90435-X}{\detokenize{10.1016/0370-2693(77)90435-X}}}.

\bibitem[Mohapatra and Senjanovic(1980)]{Mohapatra:1979ia}
Mohapatra, R.N.; Senjanovic, G.
\newblock {Neutrino Mass and Spontaneous Parity Nonconservation}.
\newblock {\em Phys. Rev. Lett.} {\bf 1980}, {\em 44},~912.
\newblock
  doi:{\changeurlcolor{black}\href{https://doi.org/10.1103/PhysRevLett.44.912}{\detokenize{10.1103/PhysRevLett.44.912}}}.

\bibitem[Yanagida(1979)]{Yanagida:1979as}
Yanagida, T.
\newblock {Horizontal gauge symmetry and masses of neutrinos}.
\newblock {\em Conf. Proc. C} {\bf 1979}, {\em 7902131},~95--99.

\bibitem[Gell-Mann \em{et~al.}(1979)Gell-Mann, Ramond, and
  Slansky]{Gell-Mann:1979vob}
Gell-Mann, M.; Ramond, P.; Slansky, R.
\newblock {Complex Spinors and Unified Theories}.
\newblock {\em Conf. Proc. C} {\bf 1979}, {\em 790927},~315--321,

\bibitem[Glashow(1980)]{Glashow:1979nm}
Glashow, S.L.
\newblock {The Future of Elementary Particle Physics}.
\newblock {\em NATO Sci. Ser. B} {\bf 1980}, {\em 61},~687.
\newblock
  doi:{\changeurlcolor{black}\href{https://doi.org/10.1007/978-1-4684-7197-7_15}{\detokenize{10.1007/978-1-4684-7197-7_15}}}.

\bibitem[Mohapatra(1986)]{Mohapatra:1986aw}
Mohapatra, R.N.
\newblock {Mechanism for Understanding Small Neutrino Mass in Superstring
  Theories}.
\newblock {\em Phys. Rev. Lett.} {\bf 1986}, {\em 56},~561--563.
\newblock
  doi:{\changeurlcolor{black}\href{https://doi.org/10.1103/PhysRevLett.56.561}{\detokenize{10.1103/PhysRevLett.56.561}}}.

\bibitem[Mohapatra and Valle(1986)]{Mohapatra:1986bd}
Mohapatra, R.N.; Valle, J.W.F.
\newblock {Neutrino Mass and Baryon Number Nonconservation in Superstring
  Models}.
\newblock {\em Phys. Rev. D} {\bf 1986}, {\em 34},~1642.
\newblock
  doi:{\changeurlcolor{black}\href{https://doi.org/10.1103/PhysRevD.34.1642}{\detokenize{10.1103/PhysRevD.34.1642}}}.

\bibitem[Bernabeu \em{et~al.}(1987)Bernabeu, Santamaria, Vidal, Mendez, and
  Valle]{Bernabeu:1987gr}
Bernabeu, J.; Santamaria, A.; Vidal, J.; Mendez, A.; Valle, J.W.F.
\newblock {Lepton Flavor Nonconservation at High-Energies in a Superstring
  Inspired Standard Model}.
\newblock {\em Phys. Lett. B} {\bf 1987}, {\em 187},~303--308.
\newblock
  doi:{\changeurlcolor{black}\href{https://doi.org/10.1016/0370-2693(87)91100-2}{\detokenize{10.1016/0370-2693(87)91100-2}}}.

\bibitem[Keung and Senjanovic(1983)]{Keung:1983uu}
Keung, W.Y.; Senjanovic, G.
\newblock {Majorana Neutrinos and the Production of the Right-handed Charged
  Gauge Boson}.
\newblock {\em Phys. Rev. Lett.} {\bf 1983}, {\em 50},~1427.
\newblock
  doi:{\changeurlcolor{black}\href{https://doi.org/10.1103/PhysRevLett.50.1427}{\detokenize{10.1103/PhysRevLett.50.1427}}}.

\bibitem[Datta \em{et~al.}(1994)Datta, Guchait, and Pilaftsis]{Datta:1993nm}
Datta, A.; Guchait, M.; Pilaftsis, A.
\newblock {Probing lepton number violation via majorana neutrinos at hadron
  supercolliders}.
\newblock {\em Phys. Rev. D} {\bf 1994}, {\em 50},~3195--3203,
\newblock
  doi:{\changeurlcolor{black}\href{https://doi.org/10.1103/PhysRevD.50.3195}{\detokenize{10.1103/PhysRevD.50.3195}}}.

\bibitem[Dev \em{et~al.}(2014)Dev, Pilaftsis, and Yang]{Dev:2013wba}
Dev, P.S.B.; Pilaftsis, A.; Yang, U.K.
\newblock {New Production Mechanism for Heavy Neutrinos at the LHC}.
\newblock {\em Phys. Rev. Lett.} {\bf 2014}, {\em 112},~081801,
\newblock
  doi:{\changeurlcolor{black}\href{https://doi.org/10.1103/PhysRevLett.112.081801}{\detokenize{10.1103/PhysRevLett.112.081801}}}.

\bibitem[Alva \em{et~al.}(2015)Alva, Han, and Ruiz]{Alva:2014gxa}
Alva, D.; Han, T.; Ruiz, R.
\newblock {Heavy Majorana neutrinos from $W\gamma$ fusion at hadron colliders}.
\newblock {\em J. High Energy Phys.} {\bf 2015}, {\em 2},~072,
\newblock
  doi:{\changeurlcolor{black}\href{https://doi.org/10.1007/JHEP02(2015)072}{\detokenize{10.1007/JHEP02(2015)072}}}.

\bibitem[Degrande \em{et~al.}(2016)Degrande, Mattelaer, Ruiz, and
  Turner]{Degrande:2016aje}
Degrande, C.; Mattelaer, O.; Ruiz, R.; Turner, J.
\newblock {Fully-Automated Precision Predictions for Heavy Neutrino Production
  Mechanisms at Hadron Colliders}.
\newblock {\em Phys. Rev. D} {\bf 2016}, {\em 94},~053002,
\newblock
  doi:{\changeurlcolor{black}\href{https://doi.org/10.1103/PhysRevD.94.053002}{\detokenize{10.1103/PhysRevD.94.053002}}}.

\bibitem[Dicus \em{et~al.}(1991)Dicus, Karatas, and Roy]{Dicus:1991fk}
Dicus, D.A.; Karatas, D.D.; Roy, P.
\newblock {Lepton nonconservation at supercollider energies}.
\newblock {\em Phys. Rev. D} {\bf 1991}, {\em 44},~2033--2037.
\newblock
  doi:{\changeurlcolor{black}\href{https://doi.org/10.1103/PhysRevD.44.2033}{\detokenize{10.1103/PhysRevD.44.2033}}}.

\bibitem[Fuks \em{et~al.}(2021)Fuks, Neundorf, Peters, Ruiz, and
  Saimpert]{Fuks:2020att}
Fuks, B.; Neundorf, J.; Peters, K.; Ruiz, R.; Saimpert, M.
\newblock {Majorana neutrinos in same-sign $W^\pm W^\pm$ scattering at the LHC:
  Breaking the TeV barrier}.
\newblock {\em Phys. Rev. D} {\bf 2021}, {\em 103},~055005,
\newblock
  doi:{\changeurlcolor{black}\href{https://doi.org/10.1103/PhysRevD.103.055005}{\detokenize{10.1103/PhysRevD.103.055005}}}.

\bibitem[Willenbrock and Dicus(1985)]{Willenbrock:1985tj}
Willenbrock, S.S.D.; Dicus, D.A.
\newblock {Production of Heavy Leptons From Gluon Fusion}.
\newblock {\em Phys. Lett. B} {\bf 1985}, {\em 156},~429--433.
\newblock
  doi:{\changeurlcolor{black}\href{https://doi.org/10.1016/0370-2693(85)91638-7}{\detokenize{10.1016/0370-2693(85)91638-7}}}.

\bibitem[Dicus and Roy(1991)]{Dicus:1991wj}
Dicus, D.A.; Roy, P.
\newblock {Supercollider signatures and correlations of heavy neutrinos}.
\newblock {\em Phys. Rev. D} {\bf 1991}, {\em 44},~1593--1596.
\newblock
  doi:{\changeurlcolor{black}\href{https://doi.org/10.1103/PhysRevD.44.1593}{\detokenize{10.1103/PhysRevD.44.1593}}}.

\bibitem[Hessler \em{et~al.}(2015)Hessler, Ibarra, Molinaro, and
  Vogl]{Hessler:2014ssa}
Hessler, A.G.; Ibarra, A.; Molinaro, E.; Vogl, S.
\newblock {Impact of the Higgs boson on the production of exotic particles at
  the LHC}.
\newblock {\em Phys. Rev. D} {\bf 2015}, {\em 91},~115004,
\newblock
  doi:{\changeurlcolor{black}\href{https://doi.org/10.1103/PhysRevD.91.115004}{\detokenize{10.1103/PhysRevD.91.115004}}}.

\bibitem[Ruiz \em{et~al.}(2017)Ruiz, Spannowsky, and Waite]{Ruiz:2017yyf}
Ruiz, R.; Spannowsky, M.; Waite, P.
\newblock {Heavy neutrinos from gluon fusion}.
\newblock {\em Phys. Rev. D} {\bf 2017}, {\em 96},~055042,  doi:10.1103/PhysRevD.96.\linebreak 055042.

\bibitem[Pascoli \em{et~al.}(2019)Pascoli, Ruiz, and Weiland]{Pascoli:2018heg}
Pascoli, S.; Ruiz, R.; Weiland, C.
\newblock {Heavy neutrinos with dynamic jet vetoes: multilepton searches at $
  \sqrt{s}=14 $ , 27, and 100 TeV}.
\newblock {\em J. High Energy Phys.} {\bf 2019}, {\em 06},~049,
\newblock
  doi:{\changeurlcolor{black}\href{https://doi.org/10.1007/JHEP06(2019)049}{\detokenize{10.1007/JHEP06(2019)049}}}.

\bibitem[Arganda \em{et~al.}(2016)Arganda, Herrero, Marcano, and
  Weiland]{Arganda:2015ija}
Arganda, E.; Herrero, M.J.; Marcano, X.; Weiland, C.
\newblock {Exotic \ensuremath{\mu}\ensuremath{\tau}jj events from heavy ISS
  neutrinos at the LHC}.
\newblock {\em Phys. Lett. B} {\bf 2016}, {\em 752},~46--50,
\newblock
  doi:{\changeurlcolor{black}\href{https://doi.org/10.1016/j.physletb.2015.11.013}{\detokenize{10.1016/j.physletb.2015.11.013}}}.

\bibitem[Sirunyan \em{et~al.}(2018)Sirunyan et~al.]{CMS:2018iaf}
Sirunyan, A.M.;  et~al. [CMS Collaboration]
\newblock {Search for heavy neutral leptons in events with three charged
  leptons in proton-proton collisions at $\sqrt{s} =$ 13 TeV}.
\newblock {\em Phys. Rev. Lett.} {\bf 2018}, {\em 120},~221801,
\newblock
  doi:{\changeurlcolor{black}\href{https://doi.org/10.1103/PhysRevLett.120.221801}{\detokenize{10.1103/PhysRevLett.120.221801}}}.

\bibitem[Fernandez-Martinez \em{et~al.}(2016)Fernandez-Martinez,
  Hernandez-Garcia, and Lopez-Pavon]{Fernandez-Martinez:2016lgt}
Fernandez-Martinez, E.; Hernandez-Garcia, J.; Lopez-Pavon, J.
\newblock {Global constraints on heavy neutrino mixing}.
\newblock {\em J. High Energy Phys.} {\bf 2016}, {\em 8},~033,
\newblock
  doi:{\changeurlcolor{black}\href{https://doi.org/10.1007/JHEP08(2016)033}{\detokenize{10.1007/JHEP08(2016)033}}}.

\bibitem[Bolton \em{et~al.}(2020)Bolton, Deppisch, and
  Bhupal~Dev]{Bolton:2019pcu}
Bolton, P.D.; Deppisch, F.F.; Bhupal~Dev, P.S.
\newblock {Neutrinoless double beta decay versus other probes of heavy sterile
  neutrinos}.
\newblock {\em J. High Energy Phys.} {\bf 2020}, {\em 3},~170,
\newblock
  doi:{\changeurlcolor{black}\href{https://doi.org/10.1007/JHEP03(2020)170}{\detokenize{10.1007/JHEP03(2020)170}}}.

\bibitem[Aad \em{et~al.}(2019)Aad et~al.]{ATLAS:2019kpx}
Aad, G.;  et~al. [ATLAS Collaboration]
\newblock {Search for heavy neutral leptons in decays of $W$ bosons produced in
  13 TeV $pp$ collisions using prompt and displaced signatures with the ATLAS
  detector}.
\newblock {\em J. High Energy Phys.} {\bf 2019}, {\em 10},~265,
\newblock
  doi:{\changeurlcolor{black}\href{https://doi.org/10.1007/JHEP10(2019)265}{\detokenize{10.1007/JHEP10(2019)265}}}.

\bibitem[Tumasyan \em{et~al.}(2022)Tumasyan et~al.]{CMS:2022fut}
Tumasyan, A.;  et~al. [CMS Collaboration]
\newblock {Search for long-lived heavy neutral leptons with displaced vertices
  in proton-proton collisions at $\sqrt{s}$ =13 TeV}. \emph{arXiv} {\bf 2022}, arXiv:hep-ex/2201.05578.


\bibitem[Atre \em{et~al.}(2009)Atre, Han, Pascoli, and Zhang]{Atre:2009rg}
Atre, A.; Han, T.; Pascoli, S.; Zhang, B.
\newblock {The Search for Heavy Majorana Neutrinos}.
\newblock {\em J. High Energy Phys.} {\bf 2009}, {\em 5},~030,
\newblock
  doi:{\changeurlcolor{black}\href{https://doi.org/10.1088/1126-6708/2009/05/030}{\detokenize{10.1088/1126-6708/2009/05/030}}}.

\bibitem[Cottin \em{et~al.}(2018)Cottin, Helo, and Hirsch]{Cottin:2018nms}
Cottin, G.; Helo, J.C.; Hirsch, M.
\newblock {Displaced vertices as probes of sterile neutrino mixing at the LHC}.
\newblock {\em Phys. Rev. D} {\bf 2018}, {\em 98},~035012,
\newblock
  doi:{\changeurlcolor{black}\href{https://doi.org/10.1103/PhysRevD.98.035012}{\detokenize{10.1103/PhysRevD.98.035012}}}.

\bibitem[Coloma \em{et~al.}(2021)Coloma, Fern\'andez-Mart\'\i{}nez,
  Gonz\'alez-L\'opez, Hern\'andez-Garc\'\i{}a, and Pavlovic]{Coloma:2020lgy}
Coloma, P.; Fern\'andez-Mart\'\i{}nez, E.; Gonz\'alez-L\'opez, M.;
  Hern\'andez-Garc\'\i{}a, J.; Pavlovic, Z.
\newblock {GeV-scale neutrinos: interactions with mesons and DUNE sensitivity}.
\newblock {\em Eur. Phys. J. C} {\bf 2021}, {\em 81},~78,
\newblock
  doi:{\changeurlcolor{black}\href{https://doi.org/10.1140/epjc/s10052-021-08861-y}{\detokenize{10.1140/epjc/s10052-021-08861-y}}}.

\bibitem[De~Vries \em{et~al.}(2021)De~Vries, Dreiner, G\"unther, Wang, and
  Zhou]{DeVries:2020jbs}
De~Vries, J.; Dreiner, H.K.; G\"unther, J.Y.; Wang, Z.S.; Zhou, G.
\newblock {Long-lived Sterile Neutrinos at the LHC in Effective Field Theory}.
\newblock {\em J. High Energy Phys.} {\bf 2021}, {\em 3},~148,
\newblock
  doi:{\changeurlcolor{black}\href{https://doi.org/10.1007/JHEP03(2021)148}{\detokenize{10.1007/JHEP03(2021)148}}}.

\bibitem[Zhou \em{et~al.}(2021)Zhou, G\"unther, Wang, de~Vries, and
  Dreiner]{Zhou:2021ylt}
Zhou, G.; G\"unther, J.Y.; Wang, Z.S.; de~Vries, J.; Dreiner, H.K.
\newblock {Long-lived Sterile Neutrinos at Belle II in Effective Field Theory}.
 \emph{arXiv} {\bf 2021}, arXiv:hep-ph/2111.04403. 


\bibitem[Hu \em{et~al.}(2019)Hu, Wong, and Xu]{Hu:2019zan}
Hu, S.; Wong, S.M.Y.; Xu, F.
\newblock {Probing Sterile Neutrino via Lepton Flavor Violating Decays of
  Mesons}. \emph{arXiv} {\bf 2019}, arXiv:hep-ph/1904.00568.


\bibitem[Cortina~Gil \em{et~al.}(2018)Cortina~Gil et~al.]{NA62:2017qcd}
Cortina~Gil, E.;  et~al. [NA62 Collaboration]
\newblock {Search for heavy neutral lepton production in $K^+$ decays}.
\newblock {\em Phys. Lett. B} {\bf 2018}, {\em 778},~137--145,
\newblock
  doi:{\changeurlcolor{black}\href{https://doi.org/10.1016/j.physletb.2018.01.031}{\detokenize{10.1016/j.physletb.2018.01.031}}}.

\bibitem[Ilakovac(2000)]{Ilakovac:1999md}
Ilakovac, A.
\newblock {Lepton flavor violation in the standard model extended by heavy
  singlet Dirac neutrinos}.
\newblock {\em Phys. Rev. D} {\bf 2000}, {\em 62},~036010,
\newblock
  doi:{\changeurlcolor{black}\href{https://doi.org/10.1103/PhysRevD.62.036010}{\detokenize{10.1103/PhysRevD.62.036010}}}.

\bibitem[Helo \em{et~al.}(2014)Helo, Hirsch, and Kovalenko]{Helo:2013esa}
Helo, J.C.; Hirsch, M.; Kovalenko, S.
\newblock {Heavy neutrino searches at the LHC with displaced vertices}.
\newblock {\em Phys. Rev. D} {\bf 2014}, {\em 89},~073005,
\newblock Erratum: \emph{Phys. Rev. D} \textbf{2016}, \emph{93}, 099902,
  doi:{\changeurlcolor{black}\href{https://doi.org/10.1103/PhysRevD.89.073005}{\detokenize{10.1103/PhysRevD.89.073005}}}.

\bibitem[Izaguirre and Shuve(2015)]{Izaguirre:2015pga}
Izaguirre, E.; Shuve, B.
\newblock {Multilepton and Lepton Jet Probes of Sub-Weak-Scale Right-Handed
  Neutrinos}.
\newblock {\em Phys. Rev. D} {\bf 2015}, {\em 91},~093010,
\newblock
  doi:{\changeurlcolor{black}\href{https://doi.org/10.1103/PhysRevD.91.093010}{\detokenize{10.1103/PhysRevD.91.093010}}}.

\bibitem[Dube \em{et~al.}(2017)Dube, Gadkari, and Thalapillil]{Dube:2017jgo}
Dube, S.; Gadkari, D.; Thalapillil, A.M.
\newblock {Lepton-Jets and Low-Mass Sterile Neutrinos at Hadron Colliders}.
\newblock {\em Phys. Rev. D} {\bf 2017}, {\em 96},~055031,
\newblock
  doi:{\changeurlcolor{black}\href{https://doi.org/10.1103/PhysRevD.96.055031}{\detokenize{10.1103/PhysRevD.96.055031}}}.

\bibitem[Dib \em{et~al.}(2018)Dib, Kim, Neill, and Yuan]{Dib:2018iyr}
Dib, C.O.; Kim, C.S.; Neill, N.A.; Yuan, X.B.
\newblock {Search for sterile neutrinos decaying into pions at the LHC}.
\newblock {\em Phys. Rev. D} {\bf 2018}, {\em 97},~035022,
\newblock
  doi:{\changeurlcolor{black}\href{https://doi.org/10.1103/PhysRevD.97.035022}{\detokenize{10.1103/PhysRevD.97.035022}}}.

\bibitem[Gago \em{et~al.}(2015)Gago, Hern\'andez, Jones-P\'erez, Losada, and
  Moreno Brice\~no]{Gago:2015vma}
Gago, A.M.; Hern\'andez, P.; Jones-P\'erez, J.; Losada, M.; Moreno Brice\~no,
  A.
\newblock {Probing the Type I Seesaw Mechanism with Displaced Vertices at the
  LHC}.
\newblock {\em Eur. Phys. J. C} {\bf 2015}, {\em 75},~470,
\newblock
  doi:{\changeurlcolor{black}\href{https://doi.org/10.1140/epjc/s10052-015-3693-1}{\detokenize{10.1140/epjc/s10052-015-3693-1}}}.

\bibitem[Accomando \em{et~al.}(2017)Accomando, Delle~Rose, Moretti, Olaiya, and
  Shepherd-Themistocleous]{Accomando:2016rpc}
Accomando, E.; Delle~Rose, L.; Moretti, S.; Olaiya, E.;
  Shepherd-Themistocleous, C.H.
\newblock {Novel SM-like Higgs decay into displaced heavy neutrino pairs in
  U(1)' models}.
\newblock {\em J. High Energy Phys.} {\bf 2017}, {\em 4},~081,
\newblock
  doi:{\changeurlcolor{black}\href{https://doi.org/10.1007/JHEP04(2017)081}{\detokenize{10.1007/JHEP04(2017)081}}}.

\bibitem[Caputo \em{et~al.}(2017)Caputo, Hernandez, Lopez-Pavon, and
  Salvado]{Caputo:2017pit}
Caputo, A.; Hernandez, P.; Lopez-Pavon, J.; Salvado, J.
\newblock {The seesaw portal in testable models of neutrino masses}.
\newblock {\em J. High Energy Phys.} {\bf 2017}, {\em 6},~112,
\newblock
  doi:{\changeurlcolor{black}\href{https://doi.org/10.1007/JHEP06(2017)112}{\detokenize{10.1007/JHEP06(2017)112}}}.

\bibitem[Deppisch \em{et~al.}(2018)Deppisch, Liu, and Mitra]{Deppisch:2018eth}
Deppisch, F.F.; Liu, W.; Mitra, M.
\newblock {Long-lived Heavy Neutrinos from Higgs Decays}.
\newblock {\em J. High Energy Phys.} {\bf 2018}, {\em 8},~181,
\newblock
  doi:{\changeurlcolor{black}\href{https://doi.org/10.1007/JHEP08(2018)181}{\detokenize{10.1007/JHEP08(2018)181}}}.

\bibitem[Liu \em{et~al.}(2019)Liu, Liu, and Wang]{Liu:2018wte}
Liu, J.; Liu, Z.; Wang, L.T.
\newblock {Enhancing Long-Lived Particles Searches at the LHC with Precision
  Timing Information}.
\newblock {\em Phys. Rev. Lett.} {\bf 2019}, {\em 122},~131801,
\newblock
  doi:{\changeurlcolor{black}\href{https://doi.org/10.1103/PhysRevLett.122.131801}{\detokenize{10.1103/PhysRevLett.122.131801}}}.

\bibitem[Antusch \em{et~al.}(2017)Antusch, Cazzato, and
  Fischer]{Antusch:2017hhu}
Antusch, S.; Cazzato, E.; Fischer, O.
\newblock {Sterile neutrino searches via displaced vertices at LHCb}.
\newblock {\em Phys. Lett. B} {\bf 2017}, {\em 774},~114--118,
\newblock
  doi:{\changeurlcolor{black}\href{https://doi.org/10.1016/j.physletb.2017.09.057}{\detokenize{10.1016/j.physletb.2017.09.057}}}.

\bibitem[Kling and Trojanowski(2018)]{Kling:2018wct}
Kling, F.; Trojanowski, S.
\newblock {Heavy Neutral Leptons at FASER}.
\newblock {\em Phys. Rev. D} {\bf 2018}, {\em 97},~095016,
\newblock
  doi:{\changeurlcolor{black}\href{https://doi.org/10.1103/PhysRevD.97.095016}{\detokenize{10.1103/PhysRevD.97.095016}}}.

\bibitem[Helo \em{et~al.}(2018)Helo, Hirsch, and Wang]{Helo:2018qej}
Helo, J.C.; Hirsch, M.; Wang, Z.S.
\newblock {Heavy neutral fermions at the high-luminosity LHC}.
\newblock {\em J. High Energy Phys.} {\bf 2018}, {\em 7},~056,
\newblock
  doi:10.1007/JHEP07\linebreak(2018)056.

\bibitem[Jana \em{et~al.}(2018)Jana, Okada, and Raut]{Jana:2018rdf}
Jana, S.; Okada, N.; Raut, D.
\newblock {Displaced vertex signature of type-I seesaw model}.
\newblock {\em Phys. Rev. D} {\bf 2018}, {\em 98},~035023,
\newblock
  doi:{\changeurlcolor{black}\href{https://doi.org/10.1103/PhysRevD.98.035023}{\detokenize{10.1103/PhysRevD.98.035023}}}.

\bibitem[Caputo \em{et~al.}(2017)Caputo, Hernandez, Kekic, L\'opez-Pav\'on, and
  Salvado]{Caputo:2016ojx}
Caputo, A.; Hernandez, P.; Kekic, M.; L\'opez-Pav\'on, J.; Salvado, J.
\newblock {The seesaw path to leptonic CP violation}.
\newblock {\em Eur. Phys. J. C} {\bf 2017}, {\em 77},~258,
\newblock
  doi:{\changeurlcolor{black}\href{https://doi.org/10.1140/epjc/s10052-017-4823-8}{\detokenize{10.1140/epjc/s10052-017-4823-8}}}.

\bibitem[Blondel \em{et~al.}(2016)Blondel, Graverini, Serra, and
  Shaposhnikov]{Blondel:2014bra}
Blondel, A.; Graverini, E.; Serra, N.; Shaposhnikov, M.
\newblock {Search for Heavy Right Handed Neutrinos at the FCC-ee}.
\newblock {\em Nucl. Part. Phys. Proc.} {\bf 2016}, {\em 273-275},~1883--1890,
\newblock
  doi:{\changeurlcolor{black}\href{https://doi.org/10.1016/j.nuclphysbps.2015.09.304}{\detokenize{10.1016/j.nuclphysbps.2015.09.304}}}.

\bibitem[Antusch \em{et~al.}(2016)Antusch, Cazzato, and
  Fischer]{Antusch:2016vyf}
Antusch, S.; Cazzato, E.; Fischer, O.
\newblock {Displaced vertex searches for sterile neutrinos at future lepton
  colliders}.
\newblock {\em J. High Energy Phys.} {\bf 2016}, {\em 12},~007,
\newblock
  doi:{\changeurlcolor{black}\href{https://doi.org/10.1007/JHEP12(2016)007}{\detokenize{10.1007/JHEP12(2016)007}}}.

\bibitem[Bonivento \em{et~al.}(2013)Bonivento et~al.]{Bonivento:2013jag}
Bonivento, W.;  Boyarsky, A.;  Dijkstra, H.;  Egede, U.;  Ferro-Luzzi, M.; Goddard, B.; Golutvin, A.; Gorbunov, D.; Jacobsson, R.; Panman, J.; et~al.
\newblock {Proposal to Search for Heavy Neutral Leptons at the SPS}. \emph{arXiv} {\bf 2013}, arXiv:hep-ex/1310.1762.


\bibitem[Abada \em{et~al.}(2019)Abada, Bernal, Losada, and
  Marcano]{Abada:2018sfh}
Abada, A.; Bernal, N.; Losada, M.; Marcano, X.
\newblock {Inclusive Displaced Vertex Searches for Heavy Neutral Leptons at the
  LHC}.
\newblock {\em J. High Energy Phys.} {\bf 2019}, {\em 1},~093,
\newblock
  doi:{\changeurlcolor{black}\href{https://doi.org/10.1007/JHEP01(2019)093}{\detokenize{10.1007/JHEP01(2019)093}}}.

\bibitem[Abada \em{et~al.}(2015)Abada, De~Romeri, Monteil, Orloff, and
  Teixeira]{Abada:2014cca}
Abada, A.; De~Romeri, V.; Monteil, S.; Orloff, J.; Teixeira, A.M.
\newblock {Indirect searches for sterile neutrinos at a high-luminosity
  Z-factory}.
\newblock {\em J. High Energy Phys.} {\bf 2015}, {\em 4},~051,
\newblock
  doi:{\changeurlcolor{black}\href{https://doi.org/10.1007/JHEP04(2015)051}{\detokenize{10.1007/JHEP04(2015)051}}}.

\bibitem[Antusch \em{et~al.}(2018)Antusch, Cazzato, Drewes, Fischer, Garbrecht,
  Gueter, and Klaric]{Antusch:2017pkq}
Antusch, S.; Cazzato, E.; Drewes, M.; Fischer, O.; Garbrecht, B.; Gueter, D.;
  Klaric, J.
\newblock {Probing Leptogenesis at Future Colliders}.
\newblock {\em J. High Energy Phys.} {\bf 2018}, {\em 9},~124,
\newblock
  doi:{\changeurlcolor{black}\href{https://doi.org/10.1007/JHEP09(2018)124}{\detokenize{10.1007/JHEP09(2018)124}}}.

\bibitem[Hern\'andez \em{et~al.}(2019)Hern\'andez, Jones-P\'erez, and
  Suarez-Navarro]{Hernandez:2018cgc}
Hern\'andez, P.; Jones-P\'erez, J.; Suarez-Navarro, O.
\newblock {Majorana vs Pseudo-Dirac Neutrinos at the ILC}.
\newblock {\em Eur. Phys. J. C} {\bf 2019}, {\em 79},~220,

\newblock
  doi:{\changeurlcolor{black}\href{https://doi.org/10.1140/epjc/s10052-019-6728-1}{\detokenize{10.1140/epjc/s10052-019-6728-1}}}.

\bibitem[Abada \em{et~al.}(2019)Abada, Arcadi, Domcke, Drewes, Klaric, and
  Lucente]{Abada:2018oly}
Abada, A.; Arcadi, G.; Domcke, V.; Drewes, M.; Klaric, J.; Lucente, M.
\newblock {Low-scale leptogenesis with three heavy neutrinos}.
\newblock {\em J. High Energy Phys.} {\bf 2019}, {\em 1},~164,
\newblock
  doi:{\changeurlcolor{black}\href{https://doi.org/10.1007/JHEP01(2019)164}{\detokenize{10.1007/JHEP01(2019)164}}}.

\bibitem[Boiarska \em{et~al.}(2019)Boiarska, Bondarenko, Boyarsky, Eijima,
  Ovchynnikov, Ruchayskiy, and Timiryasov]{Boiarska:2019jcw}
Boiarska, I.; Bondarenko, K.; Boyarsky, A.; Eijima, S.; Ovchynnikov, M.;
  Ruchayskiy, O.; Timiryasov, I.
\newblock {Probing baryon asymmetry of the Universe at LHC and SHiP}. \emph{arXiv} {\bf
  2019}, arXiv:hep-ph/1902.04535.

\bibitem[Lavignac and Medina(2021)]{Lavignac:2020yld}
Lavignac, S.; Medina, A.D.
\newblock {Displaced Vertex signatures of a pseudo-Goldstone sterile neutrino}.
\newblock {\em J. High Energy Phys.} {\bf 2021}, {\em 1},~151,
\newblock
  doi:{\changeurlcolor{black}\href{https://doi.org/10.1007/JHEP01(2021)151}{\detokenize{10.1007/JHEP01(2021)151}}}.

\bibitem[Dib \em{et~al.}(2020)Dib, Kim, and Tapia~Araya]{Dib:2019ztn}
Dib, C.O.; Kim, C.S.; Tapia~Araya, S.
\newblock {Search for light sterile neutrinos from $W^\pm$ decays at the LHC}.
\newblock {\em Phys. Rev. D} {\bf 2020}, {\em 101},~035022,
\newblock
  doi:{\changeurlcolor{black}\href{https://doi.org/10.1103/PhysRevD.101.035022}{\detokenize{10.1103/PhysRevD.101.035022}}}.

\bibitem[Drewes and Hajer(2020)]{Drewes:2019fou}
Drewes, M.; Hajer, J.
\newblock {Heavy Neutrinos in displaced vertex searches at the LHC and HL-LHC}.
\newblock {\em J. High Energy Phys.} {\bf 2020}, {\em 2},~070,
\newblock
  doi:{\changeurlcolor{black}\href{https://doi.org/10.1007/JHEP02(2020)070}{\detokenize{10.1007/JHEP02(2020)070}}}.

\bibitem[Liu \em{et~al.}(2019)Liu, Liu, Wang, and Wang]{Liu:2019ayx}
Liu, J.; Liu, Z.; Wang, L.T.; Wang, X.P.
\newblock {Seeking for sterile neutrinos with displaced leptons at the LHC}.
\newblock {\em J. High Energy Phys.} {\bf 2019}, {\em 7},~159,
\newblock
  doi:{\changeurlcolor{black}\href{https://doi.org/10.1007/JHEP07(2019)159}{\detokenize{10.1007/JHEP07(2019)159}}}.

\bibitem[Das \em{et~al.}(2019)Das, Dev, and Okada]{Das:2019fee}
Das, A.; Dev, P.S.B.; Okada, N.
\newblock {Long-lived TeV-scale right-handed neutrino production at the LHC in
  gauged $U(1)_X$ model}.
\newblock {\em Phys. Lett. B} {\bf 2019}, {\em 799},~135052,
\newblock
  doi:{\changeurlcolor{black}\href{https://doi.org/10.1016/j.physletb.2019.135052}{\detokenize{10.1016/j.physletb.2019.135052}}}.

\bibitem[Drewes \em{et~al.}(2020)Drewes, Giammanco, Hajer, and
  Lucente]{Drewes:2019vjy}
Drewes, M.; Giammanco, A.; Hajer, J.; Lucente, M.
\newblock {New long-lived particle searches in heavy-ion collisions at the
  LHC}.
\newblock {\em Phys. Rev. D} {\bf 2020}, {\em 101},~055002,
\newblock
  doi:{\changeurlcolor{black}\href{https://doi.org/10.1103/PhysRevD.101.055002}{\detokenize{10.1103/PhysRevD.101.055002}}}.

\bibitem[Chiang \em{et~al.}(2019)Chiang, Cottin, Das, and
  Mandal]{Chiang:2019ajm}
Chiang, C.W.; Cottin, G.; Das, A.; Mandal, S.
\newblock {Displaced heavy neutrinos from $Z'$ decays at the LHC}.
\newblock {\em J. High Energy Phys.} {\bf 2019}, {\em 12},~070,
\newblock
  doi:{\changeurlcolor{black}\href{https://doi.org/10.1007/JHEP12(2019)070}{\detokenize{10.1007/JHEP12(2019)070}}}.

\bibitem[Dib \em{et~al.}(2020)Dib, Helo, Nayak, Neill, Soffer, and
  Zamora-Saa]{Dib:2019tuj}
Dib, C.O.; Helo, J.C.; Nayak, M.; Neill, N.A.; Soffer, A.; Zamora-Saa, J.
\newblock {Searching for a sterile neutrino that mixes predominantly with
  $\nu_\tau$ at $B$ factories}.
\newblock {\em Phys. Rev. D} {\bf 2020}, {\em 101},~093003,
\newblock
  doi:{\changeurlcolor{black}\href{https://doi.org/10.1103/PhysRevD.101.093003}{\detokenize{10.1103/PhysRevD.101.093003}}}.

\bibitem[Jones-P\'erez \em{et~al.}(2020)Jones-P\'erez, Masias, and
  Ruiz-\'Alvarez]{Jones-Perez:2019plk}
Jones-P\'erez, J.; Masias, J.; Ruiz-\'Alvarez, J.D.
\newblock {Search for Long-Lived Heavy Neutrinos at the LHC with a VBF
  Trigger}.
\newblock {\em Eur. Phys. J. C} {\bf 2020}, {\em 80},~642,
\newblock
  doi:{\changeurlcolor{black}\href{https://doi.org/10.1140/epjc/s10052-020-8188-z}{\detokenize{10.1140/epjc/s10052-020-8188-z}}}.

\bibitem[Barducci \em{et~al.}(2021)Barducci, Bertuzzo, Caputo, Hernandez, and
  Mele]{Barducci:2020icf}
Barducci, D.; Bertuzzo, E.; Caputo, A.; Hernandez, P.; Mele, B.
\newblock {The see-saw portal at future Higgs Factories}.
\newblock {\em J. High Energy Phys.} {\bf 2021}, {\em 3},~117,
\newblock
  doi:{\changeurlcolor{black}\href{https://doi.org/10.1007/JHEP03(2021)117}{\detokenize{10.1007/JHEP03(2021)117}}}.

\bibitem[Borsato \em{et~al.}(2021)Borsato et~al.]{Borsato:2021aum}
Borsato, M.; Cid Vidal, X,; Tsai, Y.; Vázquez Sierra, C.; Zurita, J.; Alonso-Álvarez, G.; Boyarsky, A.; Brea Rodríguez, A.; Buarque Franzosi, D.; Cacciapaglia, G.; et~al.
\newblock {Unleashing the full power of LHCb to probe Stealth New Physics}. \emph{arXiv} {\bf
  2021}, arXiv:hep-ph/2105.12668.


\bibitem[Tastet \em{et~al.}(2021)Tastet, Ruchayskiy, and
  Timiryasov]{Tastet:2021vwp}
Tastet, J.L.; Ruchayskiy, O.; Timiryasov, I.
\newblock {Reinterpreting the ATLAS bounds on heavy neutral leptons in a
  realistic neutrino oscillation model}.
\newblock {\em J. High Energy Phys.} {\bf 2021}, {\em 12},~182,
\newblock
  doi:{\changeurlcolor{black}\href{https://doi.org/10.1007/JHEP12(2021)182}{\detokenize{10.1007/JHEP12(2021)182}}}.

\bibitem[Liu \em{et~al.}(2022)Liu, Kulkarni, and Deppisch]{Liu:2022kid}
Liu, W.; Kulkarni, S.; Deppisch, F.F.
\newblock {Heavy Neutrinos at the FCC-hh in the $U(1)_{B-L}$ Model}. \emph{arXiv} {\bf 2022}, arXiv:hep-ph/2202.07310.


\bibitem[Bray \em{et~al.}(2007)Bray, Lee, and Pilaftsis]{Bray:2007ru}
Bray, S.; Lee, J.S.; Pilaftsis, A.
\newblock {Resonant CP violation due to heavy neutrinos at the LHC}.
\newblock {\em Nucl. Phys. B} {\bf 2007}, {\em 786},~95--118,
\newblock
  doi:{\changeurlcolor{black}\href{https://doi.org/10.1016/j.nuclphysb.2007.07.002}{\detokenize{10.1016/j.nuclphysb.2007.07.002}}}.

\bibitem[Tapia \em{et~al.}(2021)Tapia, Vidal-Bravo, and
  Zamora-Saa]{Tapia:2021gne}
Tapia, S.; Vidal-Bravo, M.; Zamora-Saa, J.
\newblock {Discovering heavy neutrino oscillations in rare $B^\pm_c$ meson
  decays at HL-LHCb} \emph{arXiv} {\bf 2021}, arXiv:hep-ph/2109.06027.


\bibitem[Cveti\v{c} \em{et~al.}(2021)Cveti\v{c}, Kim, and
  Zamora-Sa\'a]{Cvetic:2021lmm}
Cveti\v{c}, G.; Kim, C.S.; Zamora-Sa\'a, J.
\newblock {CP violation in the rare Higgs decays via exchange of on-shell
  almost degenerate Majorana neutrinos, $H \to \nu_k N_j \to \nu_k \ell^{-} U
  {\bar D}$ and $H \to \nu_k N_j \to \nu_k \ell^{+} {\bar U} D$}. \emph{arXiv} {\bf 2021}, arXiv:hep-ph/2110.08799.


\bibitem[Abada \em{et~al.}(2019)Abada, Hati, Marcano, and
  Teixeira]{Abada:2019bac}
Abada, A.; Hati, C.; Marcano, X.; Teixeira, A.M.
\newblock {Interference effects in LNV and LFV semileptonic decays: the
  Majorana hypothesis}.
\newblock {\em J. High Energy Phys.} {\bf 2019}, {\em 9},~017,
\newblock
  doi:{\changeurlcolor{black}\href{https://doi.org/10.1007/JHEP09(2019)017}{\detokenize{10.1007/JHEP09(2019)017}}}.

\bibitem[Tumasyan \em{et~al.}(2021)Tumasyan et~al.]{CMS:2021dzb}
Tumasyan, A.;  et~al. [CMS Collaboration]
\newblock {Search for a right-handed W boson and a heavy neutrino in
  proton-proton collisions at $\sqrt{s}$ = 13 TeV}. \emph{arXiv} {\bf 2021}, arXiv:hep-ex/2112.03949.


\bibitem[Nemev\v{s}ek \em{et~al.}(2018)Nemev\v{s}ek, Nesti, and
  Popara]{Nemevsek:2018bbt}
Nemev\v{s}ek, M.; Nesti, F.; Popara, G.
\newblock {Keung-Senjanovi\'c process at the LHC: From lepton number violation
  to displaced vertices to invisible decays}.
\newblock {\em Phys. Rev. D} {\bf 2018}, {\em 97},~115018,
\newblock
  doi:{\changeurlcolor{black}\href{https://doi.org/10.1103/PhysRevD.97.115018}{\detokenize{10.1103/PhysRevD.97.115018}}}.

\bibitem[Chauhan \em{et~al.}(2019)Chauhan, Dev, Mohapatra, and
  Zhang]{Chauhan:2018uuy}
Chauhan, G.; Dev, P.S.B.; Mohapatra, R.N.; Zhang, Y.
\newblock {Perturbativity constraints on $U(1)_{B-L}$ and left-right models and
  implications for heavy gauge boson searches}.
\newblock {\em J. High Energy Phys.} {\bf 2019}, {\em 1},~208,
\newblock
  doi:{\changeurlcolor{black}\href{https://doi.org/10.1007/JHEP01(2019)208}{\detokenize{10.1007/JHEP01(2019)208}}}.

\bibitem[Mitra \em{et~al.}(2016)Mitra, Ruiz, Scott, and
  Spannowsky]{Mitra:2016kov}
Mitra, M.; Ruiz, R.; Scott, D.J.; Spannowsky, M.
\newblock {Neutrino Jets from High-Mass $W_R$ Gauge Bosons in TeV-Scale
  Left-Right Symmetric Models}.
\newblock {\em Phys. Rev. D} {\bf 2016}, {\em 94},~095016,
\newblock
  doi:{\changeurlcolor{black}\href{https://doi.org/10.1103/PhysRevD.94.095016}{\detokenize{10.1103/PhysRevD.94.095016}}}.

\bibitem[Ferrari \em{et~al.}(2000)Ferrari, Collot, Andrieux, Belhorma,
  de~Saintignon, Hostachy, Martin, and Wielers]{Ferrari:2000sp}
Ferrari, A.; Collot, J.; Andrieux, M.L.; Belhorma, B.; de~Saintignon, P.;
  Hostachy, J.Y.; Martin, P.; Wielers, M.
\newblock {Sensitivity study for new gauge bosons and right-handed Majorana
  neutrinos in $p p$ collisions at $s$ = 14-TeV}.
\newblock {\em Phys. Rev. D} {\bf 2000}, {\em 62},~013001.
\newblock
  doi:{\changeurlcolor{black}\href{https://doi.org/10.1103/PhysRevD.62.013001}{\detokenize{10.1103/PhysRevD.62.013001}}}.

\bibitem[Aaboud \em{et~al.}(2019{\natexlab{a}})Aaboud et~al.]{ATLAS:2018dcj}
Aaboud, M.;  et~al. [ATLAS Collaboration]
\newblock {Search for heavy Majorana or Dirac neutrinos and right-handed $W$
  gauge bosons in final states with two charged leptons and two jets at $
  \sqrt{s}=13 $ TeV with the ATLAS detector}.
\newblock {\em J. High Energy Phys.} {\bf 2019}, {\em 1},~016,
\newblock
  doi:{\changeurlcolor{black}\href{https://doi.org/10.1007/JHEP01(2019)016}{\detokenize{10.1007/JHEP01(2019)016}}}.

\bibitem[Aaboud \em{et~al.}(2019{\natexlab{b}})Aaboud et~al.]{ATLAS:2019isd}
Aaboud, M.;  et~al. [ATLAS Collaboration]
\newblock {Search for a right-handed gauge boson decaying into a high-momentum
  heavy neutrino and a charged lepton in $pp$ collisions with the ATLAS
  detector at $\sqrt{s}=13$ TeV}.
\newblock {\em Phys. Lett. B} {\bf 2019}, {\em 798},~134942,
\newblock
  doi:{\changeurlcolor{black}\href{https://doi.org/10.1016/j.physletb.2019.134942}{\detokenize{10.1016/j.physletb.2019.134942}}}.

\bibitem[Maiezza \em{et~al.}(2015)Maiezza, Nemev\v{s}ek, and
  Nesti]{Maiezza:2015lza}
Maiezza, A.; Nemev\v{s}ek, M.; Nesti, F.
\newblock {Lepton Number Violation in Higgs Decay at LHC}.
\newblock {\em Phys. Rev. Lett.} {\bf 2015}, {\em 115},~081802,
\newblock
  doi:{\changeurlcolor{black}\href{https://doi.org/10.1103/PhysRevLett.115.081802}{\detokenize{10.1103/PhysRevLett.115.081802}}}.

\bibitem[Nemev\v{s}ek \em{et~al.}(2017)Nemev\v{s}ek, Nesti, and
  Vasquez]{Nemevsek:2016enw}
Nemev\v{s}ek, M.; Nesti, F.; Vasquez, J.C.
\newblock {Majorana Higgses at colliders}.
\newblock {\em J. High Energy Phys.} {\bf 2017}, {\em 4},~114,
\newblock
  doi:{\changeurlcolor{black}\href{https://doi.org/10.1007/JHEP04(2017)114}{\detokenize{10.1007/JHEP04(2017)114}}}.

\bibitem[Cottin \em{et~al.}(2018)Cottin, Helo, and Hirsch]{Cottin:2018kmq}
Cottin, G.; Helo, J.C.; Hirsch, M.
\newblock {Searches for light sterile neutrinos with multitrack displaced
  vertices}.
\newblock {\em Phys. Rev. D} {\bf 2018}, {\em 97},~055025,
\newblock
  doi:{\changeurlcolor{black}\href{https://doi.org/10.1103/PhysRevD.97.055025}{\detokenize{10.1103/PhysRevD.97.055025}}}.

\bibitem[Cottin \em{et~al.}(2019)Cottin, Helo, Hirsch, and
  Silva]{Cottin:2019drg}
Cottin, G.; Helo, J.C.; Hirsch, M.; Silva, D.
\newblock {Revisiting the LHC reach in the displaced region of the minimal
  left-right symmetric model}.
\newblock {\em Phys. Rev. D} {\bf 2019}, {\em 99},~115013,
\newblock
  doi:{\changeurlcolor{black}\href{https://doi.org/10.1103/PhysRevD.99.115013}{\detokenize{10.1103/PhysRevD.99.115013}}}.

\bibitem[Langacker and Plumacher(2000)]{Langacker:2000ju}
Langacker, P.; Plumacher, M.
\newblock {Flavor changing effects in theories with a heavy $Z^\prime$ boson
  with family nonuniversal couplings}.
\newblock {\em Phys. Rev. D} {\bf 2000}, {\em 62},~013006,
\newblock
  doi:{\changeurlcolor{black}\href{https://doi.org/10.1103/PhysRevD.62.013006}{\detokenize{10.1103/PhysRevD.62.013006}}}.

\bibitem[Langacker(2009)]{Langacker:2008yv}
Langacker, P.
\newblock {The Physics of Heavy $Z^\prime$ Gauge Bosons}.
\newblock {\em Rev. Mod. Phys.} {\bf 2009}, {\em 81},~1199--1228,
\newblock
  doi:{\changeurlcolor{black}\href{https://doi.org/10.1103/RevModPhys.81.1199}{\detokenize{10.1103/RevModPhys.81.1199}}}.

\bibitem[del Aguila \em{et~al.}(2010)del Aguila, de~Blas, and
  Perez-Victoria]{delAguila:2010mx}
del Aguila, F.; de~Blas, J.; Perez-Victoria, M.
\newblock {Electroweak Limits on General New Vector Bosons}.
\newblock {\em J. High Energy Phys.} {\bf 2010}, {\em 9},~033,
\newblock
  doi:{\changeurlcolor{black}\href{https://doi.org/10.1007/JHEP09(2010)033}{\detokenize{10.1007/JHEP09(2010)033}}}.

\bibitem[de~Blas \em{et~al.}(2013)de~Blas, Lizana, and
  Perez-Victoria]{deBlas:2012qp}
de~Blas, J.; Lizana, J.M.; Perez-Victoria, M.
\newblock {Combining searches of Z' and W' bosons}.
\newblock {\em J. High Energy Phys.} {\bf 2013}, {\em 1},~166,
\newblock
  doi:{\changeurlcolor{black}\href{https://doi.org/10.1007/JHEP01(2013)166}{\detokenize{10.1007/JHEP01(2013)166}}}.

\bibitem[Accomando \em{et~al.}(2013)Accomando, Becciolini, Belyaev, Moretti,
  and Shepherd-Themistocleous]{Accomando:2013sfa}
Accomando, E.; Becciolini, D.; Belyaev, A.; Moretti, S.;
  Shepherd-Themistocleous, C.
\newblock {$Z'$ at the LHC: Interference and Finite Width Effects in
  Drell-Yan}.
\newblock {\em J. High Energy Phys.} {\bf 2013}, {\em 10},~153,
\newblock
  doi:{\changeurlcolor{black}\href{https://doi.org/10.1007/JHEP10(2013)153}{\detokenize{10.1007/JHEP10(2013)153}}}.

\bibitem[Jezo \em{et~al.}(2014)Jezo, Klasen, Lamprea, Lyonnet, and
  Schienbein]{Jezo:2014wra}
Jezo, T.; Klasen, M.; Lamprea, D.R.; Lyonnet, F.; Schienbein, I.
\newblock {NLO+NLL limits on $W'$ and $Z'$ gauge boson masses in general
  extensions of the Standard Model}.
\newblock {\em J. High Energy Phys.} {\bf 2014}, {\em 12},~092,
\newblock
  doi:{\changeurlcolor{black}\href{https://doi.org/10.1007/JHEP12(2014)092}{\detokenize{10.1007/JHEP12(2014)092}}}.

\bibitem[Accomando \em{et~al.}(2016)Accomando, Coriano, Delle~Rose, Fiaschi,
  Marzo, and Moretti]{Accomando:2016sge}
Accomando, E.; Coriano, C.; Delle~Rose, L.; Fiaschi, J.; Marzo, C.; Moretti, S.
\newblock {$Z'$, Higgses and heavy neutrinos in $U(1)'$ models: from the LHC to
  the GUT scale}.
\newblock {\em J. High Energy Phys.} {\bf 2016}, {\em 7},~086,
\newblock
  doi:{\changeurlcolor{black}\href{https://doi.org/10.1007/JHEP07(2016)086}{\detokenize{10.1007/JHEP07(2016)086}}}.

\bibitem[Deppisch \em{et~al.}(2019)Deppisch, Kulkarni, and
  Liu]{Deppisch:2019kvs}
Deppisch, F.; Kulkarni, S.; Liu, W.
\newblock {Heavy neutrino production via $Z'$ at the lifetime frontier}.
\newblock {\em Phys. Rev. D} {\bf 2019}, {\em 100},~035005,
\newblock
  doi:{\changeurlcolor{black}\href{https://doi.org/10.1103/PhysRevD.100.035005}{\detokenize{10.1103/PhysRevD.100.035005}}}.

\bibitem[Buras \em{et~al.}(2021)Buras, Crivellin, Kirk, Manzari, and
  Montull]{Buras:2021btx}
Buras, A.J.; Crivellin, A.; Kirk, F.; Manzari, C.A.; Montull, M.
\newblock {Global analysis of leptophilic Z' bosons}.
\newblock {\em J. High Energy Phys.} {\bf 2021}, {\em 6},~068,
\newblock
  doi:{\changeurlcolor{black}\href{https://doi.org/10.1007/JHEP06(2021)068}{\detokenize{10.1007/JHEP06(2021)068}}}.

\bibitem[Bossi and Ciafaloni(2020)]{Bossi:2020yne}
Bossi, F.; Ciafaloni, P.
\newblock {Lepton Flavor Violation at muon-electron colliders}.
\newblock {\em J. High Energy Phys.} {\bf 2020}, {\em 10},~033,
\newblock
  doi:{\changeurlcolor{black}\href{https://doi.org/10.1007/JHEP10(2020)033}{\detokenize{10.1007/JHEP10(2020)033}}}.

\bibitem[Wetterich(1981)]{Wetterich:1981bx}
Wetterich, C.
\newblock {Neutrino Masses and the Scale of B-L Violation}.
\newblock {\em Nucl. Phys. B} {\bf 1981}, {\em 187},~343--375.
\newblock
  doi:{\changeurlcolor{black}\href{https://doi.org/10.1016/0550-3213(81)90279-0}{\detokenize{10.1016/0550-3213(81)90279-0}}}.

\bibitem[Buchmuller \em{et~al.}(1991)Buchmuller, Greub, and
  Minkowski]{Buchmuller:1991ce}
Buchmuller, W.; Greub, C.; Minkowski, P.
\newblock {Neutrino masses, neutral vector bosons and the scale of B-L
  breaking}.
\newblock {\em Phys. Lett. B} {\bf 1991}, {\em 267},~395--399.
\newblock
  doi:{\changeurlcolor{black}\href{https://doi.org/10.1016/0370-2693(91)90952-M}{\detokenize{10.1016/0370-2693(91)90952-M}}}.

\bibitem[Emam and Khalil(2007)]{Emam:2007dy}
Emam, W.; Khalil, S.
\newblock {Higgs and Z-prime phenomenology in B-L extension of the standard
  model at LHC}.
\newblock {\em Eur. Phys. J. C} {\bf 2007}, {\em 52},~625--633,
\newblock
  doi:{\changeurlcolor{black}\href{https://doi.org/10.1140/epjc/s10052-007-0411-7}{\detokenize{10.1140/epjc/s10052-007-0411-7}}}.

\bibitem[Basso \em{et~al.}(2009)Basso, Belyaev, Moretti, and
  Shepherd-Themistocleous]{Basso:2008iv}
Basso, L.; Belyaev, A.; Moretti, S.; Shepherd-Themistocleous, C.H.
\newblock {Phenomenology of the minimal B-L extension of the Standard model: Z'
  and neutrinos}.
\newblock {\em Phys. Rev. D} {\bf 2009}, {\em 80},~055030,
\newblock
  doi:{\changeurlcolor{black}\href{https://doi.org/10.1103/PhysRevD.80.055030}{\detokenize{10.1103/PhysRevD.80.055030}}}.

\bibitem[Fileviez~Perez \em{et~al.}(2009)Fileviez~Perez, Han, and
  Li]{FileviezPerez:2009hdc}
Fileviez~Perez, P.; Han, T.; Li, T.
\newblock {Testability of Type I Seesaw at the CERN LHC: Revealing the
  Existence of the B-L Symmetry}.
\newblock {\em Phys. Rev. D} {\bf 2009}, {\em 80},~073015,
\newblock
  doi:{\changeurlcolor{black}\href{https://doi.org/10.1103/PhysRevD.80.073015}{\detokenize{10.1103/PhysRevD.80.073015}}}.

\bibitem[Heeck(2014)]{Heeck:2014zfa}
Heeck, J.
\newblock {Unbroken B \textendash{} L symmetry}.
\newblock {\em Phys. Lett. B} {\bf 2014}, {\em 739},~256--262,
\newblock
  doi:{\changeurlcolor{black}\href{https://doi.org/10.1016/j.physletb.2014.10.067}{\detokenize{10.1016/j.physletb.2014.10.067}}}.

\bibitem[Khalil(2008)]{Khalil:2006yi}
Khalil, S.
\newblock {Low scale $B$ - L extension of the Standard Model at the LHC}.
\newblock {\em J. Phys. G} {\bf 2008}, {\em 35},~055001,
\newblock
  doi:{\changeurlcolor{black}\href{https://doi.org/10.1088/0954-3899/35/5/055001}{\detokenize{10.1088/0954-3899/35/5/055001}}}.

\bibitem[Huitu \em{et~al.}(2008)Huitu, Khalil, Okada, and Rai]{Huitu:2008gf}
Huitu, K.; Khalil, S.; Okada, H.; Rai, S.K.
\newblock {Signatures for right-handed neutrinos at the Large Hadron Collider}.
\newblock {\em Phys. Rev. Lett.} {\bf 2008}, {\em 101},~181802,
\newblock
  doi:{\changeurlcolor{black}\href{https://doi.org/10.1103/PhysRevLett.101.181802}{\detokenize{10.1103/PhysRevLett.101.181802}}}.

\bibitem[Accomando \em{et~al.}(2018)Accomando, Delle~Rose, Moretti, Olaiya, and
  Shepherd-Themistocleous]{Accomando:2017qcs}
Accomando, E.; Delle~Rose, L.; Moretti, S.; Olaiya, E.;
  Shepherd-Themistocleous, C.H.
\newblock {Extra Higgs boson and $Z'$ as portals to signatures of heavy
  neutrinos at the LHC}.
\newblock {\em J. High Energy Phys.} {\bf 2018}, {\em 2},~109,
\newblock
  doi:{\changeurlcolor{black}\href{https://doi.org/10.1007/JHEP02(2018)109}{\detokenize{10.1007/JHEP02(2018)109}}}.

\bibitem[Dev \em{et~al.}(2018)Dev, Mohapatra, and Zhang]{Dev:2017xry}
Dev, P.S.B.; Mohapatra, R.N.; Zhang, Y.
\newblock {Leptogenesis constraints on $B-L$ breaking Higgs boson in TeV scale
  seesaw models}.
\newblock {\em J. High Energy Phys.} {\bf 2018}, {\em 3},~122,
\newblock
  doi:{\changeurlcolor{black}\href{https://doi.org/10.1007/JHEP03(2018)122}{\detokenize{10.1007/JHEP03(2018)122}}}.

\bibitem[Davidson \em{et~al.}(2012)Davidson, Lacroix, and
  Verdier]{Davidson:2012wn}
Davidson, S.; Lacroix, S.; Verdier, P.
\newblock {LHC sensitivity to lepton flavour violating Z boson decays}.
\newblock {\em J. High Energy Phys.} {\bf 2012}, {\em 9},~092,
\newblock
  doi:{\changeurlcolor{black}\href{https://doi.org/10.1007/JHEP09(2012)092}{\detokenize{10.1007/JHEP09(2012)092}}}.

\bibitem[Abada \em{et~al.}(2015)Abada, Be\v{c}irevi\'c, Lucente, and
  Sumensari]{Abada:2015zea}
Abada, A.; Be\v{c}irevi\'c, D.; Lucente, M.; Sumensari, O.
\newblock {Lepton flavor violating decays of vector quarkonia and of the $Z$
  boson}.
\newblock {\em Phys. Rev. D} {\bf 2015}, {\em 91},~113013,
\newblock
  doi:{\changeurlcolor{black}\href{https://doi.org/10.1103/PhysRevD.91.113013}{\detokenize{10.1103/PhysRevD.91.113013}}}.

\bibitem[De~Romeri \em{et~al.}(2017)De~Romeri, Herrero, Marcano, and
  Scarcella]{DeRomeri:2016gum}
De~Romeri, V.; Herrero, M.J.; Marcano, X.; Scarcella, F.
\newblock {Lepton flavor violating Z decays: A promising window to low scale
  seesaw neutrinos}.
\newblock {\em Phys. Rev. D} {\bf 2017}, {\em 95},~075028,
\newblock
  doi:{\changeurlcolor{black}\href{https://doi.org/10.1103/PhysRevD.95.075028}{\detokenize{10.1103/PhysRevD.95.075028}}}.

\bibitem[Herrero \em{et~al.}(2018)Herrero, Marcano, Morales, and
  Szynkman]{Herrero:2018luu}
Herrero, M.J.; Marcano, X.; Morales, R.; Szynkman, A.
\newblock {One-loop effective LFV $Zl_kl_m$ vertex from heavy neutrinos within
  the mass insertion approximation}.
\newblock {\em Eur. Phys. J. C} {\bf 2018}, {\em 78},~815,
\newblock
  doi:{\changeurlcolor{black}\href{https://doi.org/10.1140/epjc/s10052-018-6281-3}{\detokenize{10.1140/epjc/s10052-018-6281-3}}}.

\bibitem[Langacker and London(1988)]{Langacker:1988ur}
Langacker, P.; London, D.
\newblock {Mixing Between Ordinary and Exotic Fermions}.
\newblock {\em Phys. Rev. D} {\bf 1988}, {\em 38},~886.
\newblock
  doi:{\changeurlcolor{black}\href{https://doi.org/10.1103/PhysRevD.38.886}{\detokenize{10.1103/PhysRevD.38.886}}}.

\bibitem[Altmannshofer \em{et~al.}(2016)Altmannshofer, Chen, Bhupal~Dev, and
  Soni]{Altmannshofer:2016brv}
Altmannshofer, W.; Chen, C.Y.; Bhupal~Dev, P.S.; Soni, A.
\newblock {Lepton flavor violating Z' explanation of the muon anomalous
  magnetic moment}.
\newblock {\em Phys. Lett. B} {\bf 2016}, {\em 762},~389--398,
\newblock
  doi:{\changeurlcolor{black}\href{https://doi.org/10.1016/j.physletb.2016.09.046}{\detokenize{10.1016/j.physletb.2016.09.046}}}.

\bibitem[CMS(2021)]{CMS:2021tau}
CMS Collaboration. {Search for heavy resonances and quantum black holes in e$\mu$, e$\tau$, and
  $\mu\tau$ final states in proton-proton collisions at
  $\sqrt{s}=13~\mathrm{TeV}$}, CMS-PAS-EXO-19-014. 2021. Available online: \url{http://cds.cern.ch/record/2779023}  (accessed on 22 August 2021 date month year). 

\bibitem[Dev \em{et~al.}(2020)Dev, Rodejohann, Xu, and Zhang]{Dev:2020drf}
Dev, P.S.B.; Rodejohann, W.; Xu, X.J.; Zhang, Y.
\newblock {MUonE sensitivity to new physics explanations of the muon anomalous
  magnetic moment}.
\newblock {\em J. High Energy Phys.} {\bf 2020}, {\em 5},~053,
\newblock
  doi:{\changeurlcolor{black}\href{https://doi.org/10.1007/JHEP05(2020)053}{\detokenize{10.1007/JHEP05(2020)053}}}.

\bibitem[Masiero \em{et~al.}(2020)Masiero, Paradisi, and
  Passera]{Masiero:2020vxk}
Masiero, A.; Paradisi, P.; Passera, M.
\newblock {New physics at the MUonE experiment at CERN}.
\newblock {\em Phys. Rev. D} {\bf 2020}, {\em 102},~075013,
\newblock
  doi:{\changeurlcolor{black}\href{https://doi.org/10.1103/PhysRevD.102.075013}{\detokenize{10.1103/PhysRevD.102.075013}}}.

\bibitem{CarloniCalame:2015obs}
Carloni Calame, C.M.; Passera, M.;
Trentadue, L.; Venanzoni, G.
\newblock{A new approach to evaluate the leading hadronic corrections to the muon $g$-2}. 
\newblock{ \em Phys. Lett. B} {\bf 2015}, {\em 746}, 325-329, 
\newblock
  doi:{\changeurlcolor{black}\href{https://doi.org/10.1016/j.physletb.2015.05.020}{\detokenize{10.1016/j.physletb.2015.05.020}}}.

\bibitem[Abbiendi \em{et~al.}(2017)Abbiendi et~al.]{Abbiendi:2016xup}
Abbiendi, G.; Carloni Calame, C.M.; Marconi, U.; Matteuzzi, C.; Montagna, G.; Nicrosini, O.; Passera, M.; Piccinini, F.; Tenchini, R.; Trentadue, L.; et~al.
\newblock {Measuring the leading hadronic contribution to the muon g-2 via $\mu
  e$ scattering}.
\newblock {\em Eur. Phys. J. C} {\bf 2017}, {\em 77},~139,
\newblock
  doi:{\changeurlcolor{black}\href{https://doi.org/10.1140/epjc/s10052-017-4633-z}{\detokenize{10.1140/epjc/s10052-017-4633-z}}}.





\end{thebibliography}
\end{document}